\documentclass[12pt]{article}
\usepackage
[left=2.40cm, right=2.40cm, top=2.40cm, bottom=2.40cm]
{geometry}
\usepackage{setspace}
\usepackage[bbgreekl]{mathbbol}
\usepackage{amsmath,amssymb,amsfonts,amsxtra,mathrsfs,graphics,graphicx,amsthm,epsfig,ytableau,bm,longtable,float,color,tikz,mathtools,xfrac,footnote,rotating,lscape}
\usetikzlibrary{calc,positioning,arrows.meta}
\usepackage[debug,pageanchor=false]{hyperref}
\definecolor{link}{rgb}{0., 0.1, 0.5}
\hypersetup{colorlinks=true,linkcolor=link,citecolor=link,urlcolor=link,linktocpage}

\usepackage[titles]{tocloft}

\restylefloat{table}
\usepackage{dsfont}
\usepackage{multicol}
\usepackage{lipsum}
\usepackage{textgreek}
\usetikzlibrary{decorations.pathmorphing}
\usetikzlibrary{decorations.markings}
\usetikzlibrary{quotes,arrows.meta}
\usetikzlibrary{arrows, decorations.markings, calc, fadings, decorations.pathreplacing, patterns, decorations.pathmorphing, positioning}
\usepackage{tikz-cd}

\usepackage{fixmath} % for \mathbold
\usepackage{scalerel}
\newlength\bshft
\def\fakebold#1{\ThisStyle{\ooalign{$\SavedStyle#1$\cr%
  \kern-\bshft$\SavedStyle#1$\cr%
  \kern\bshft$\SavedStyle#1$}}}

\usetikzlibrary{positioning,shapes}
\usetikzlibrary{chains}
\usetikzlibrary{arrows,fit,decorations.pathreplacing}
\tikzstyle{every picture}+=[remember picture]
\tikzstyle{na} = [baseline=-.5ex]

\usepackage{empheq}
\usepackage{cite}
\usepackage{multirow}
\usepackage{booktabs}
\usepackage[american]{babel}

\usepackage[latin1]{inputenc}

\usepackage{array,booktabs}

\makeatletter
\newcommand{\vast}{\bBigg@{1}}
\newcommand{\Vast}{\bBigg@{5}}
\makeatother

\numberwithin{equation}{section}

\makeatletter
\@addtoreset{equation}{section}
\makeatother

\newcommand{\bea}{\begin{equation} \begin{aligned}} \newcommand{\eea}{\end{aligned} \end{equation}}

\newcommand{\ii}{\mathrm{i}}

\newcommand{\p}[1]{{\left({#1}\right)}}

\newcommand{\dd}{{\rm d}}

\newcommand{\fks}{\sigma}

\def\U{\mathrm{U}}

\newcommand{\wb}{\overline}
\newcommand{\wt}{\widetilde}

\newcommand{\cB}{\mathcal{B}}
\newcommand{\cC}{\mathcal{C}}
\newcommand{\cD}{\mathcal{D}}

\newcommand{\cI}{\mathcal{I}}

\newcommand{\cL}{\mathcal{L}}

\newcommand{\cN}{\mathcal{N}}
\newcommand{\cO}{\mathcal{O}}

\newcommand{\cR}{\mathcal{R}}
\newcommand{\cS}{\mathcal{S}}
\newcommand{\cT}{\mathcal{T}}

\newcommand{\cV}{\mathcal{V}}

\newcommand{\cX}{\mathcal{X}}

\newcommand{\bC}{\mathbb{C}}

\newcommand{\bN}{\mathbb{N}}

\newcommand{\bQ}{\mathbb{Q}}
\newcommand{\bR}{\mathbb{R}}
\newcommand{\bT}{\mathbb{T}}
\newcommand{\bZ}{\mathbb{Z}}

\newcommand{\fn}{\mathfrak{n}}

\usepackage{bbold}

\usepackage[Symbol]{upgreek}
\usepackage{bm}

\usepackage{calligra}
\DeclareMathAlphabet{\mathcalligra}{T1}{calligra}{m}{n}

\makeatletter
\g@addto@macro\bfseries{\boldmath}
\makeatother

\makeatletter
\newcommand*{\rom}[1]{\expandafter\@slowromancap\romannumeral #1@}
\makeatother

\usepackage{bookmark}
\usepackage{cleveref}
\usepackage{physics}
\usepackage{tensor}
\usepackage{color}
\usepackage{enumerate}

\AtBeginEnvironment{pmatrix}{\everymath{\displaystyle}}
\AtBeginEnvironment{bmatrix}{\everymath{\displaystyle}}
\DeclareMathAlphabet{\mathpzc}{OT1}{pzc}{m}{it}

\newcommand{\iu}{\ensuremath{\mathrm{i}}}
\newcommand{\eu}{\ensuremath{\mathrm{e}}}
\newcommand{\curly}[1]{\ensuremath{\mathpzc{#1}}}

\usepackage{empheq}
\usepackage[most]{tcolorbox}

\newtcbox{\mymath}[1][]{%
    nobeforeafter, math upper, tcbox raise base,
    enhanced, colframe=blue!30!black,
    colback=blue!30, boxrule=1pt,
    #1}

\newcommand{\X}{\mathcal X}
\newcommand{\func}{\cV}
\newcommand{\scal}{X}
\newcommand{\V}{V}
\newcommand{\selfdual}{\Theta}
\newcommand{\cpot}{\gamma}
\newcommand{\Pot}{\Phi}
\newcommand{\etacoho}{\alpha}
\newcommand{\flux}{\mathfrak n}
\newcommand{\gaugeint}{m}
\newcommand{\gaugereal}{M}
\newcommand{\torusint}{\cT}
\newcommand{\gauge}{\nu}
\newcommand{\nv}{n_V}
\newcommand{\holonomy}{\varphi}

\begin{document}

%*******************************************************************************
%
%     Title page
%
%*******************************************************************************
	\begin{titlepage}
	~\\
		\begin{center}

		   %    	{\Large \bf{Patch-wise localization in 5D gauged supergravity}}
		{\Large \bf{Patch-wise localization with Chern-Simons forms\\[3mm]
		in five dimensional supergravity}}
			\vskip 5mm 
				\bigskip
				
					Edoardo Colombo,$^{a,b}$ Vasil Dimitrov$^{a,b}$, Dario Martelli,$^{a,b}$
					and Alberto Zaffaroni$^{c,d}$\\
					
				\bigskip
				\vskip 9mm 

{\footnotesize
{\it
 $^a$Dipartimento di Matematica,
% ``Giuseppe Peano'',
 Universit\`a di Torino, Via Carlo Alberto 10, 10123 Torino, Italy \\
$^b$INFN, Sezione di Torino, Via Pietro Giuria 1, 10125 Torino, Italy \\
$^c$Dipartimento di Fisica, Universit\`a di Milano-Bicocca, I-20126 Milano, Italy \\
$^d$INFN, sezione di Milano-Bicocca, I-20126 Milano, Italy
}
}

       \vskip .2in
       
{\footnotesize
       {\it E-mail:}
       \href{edoardo.colombo@unito.it}{\texttt{edoardo.colombo@unito.it}},
       \href{vasilradoslavov.dimitrov@unito.it}{\texttt{vasilradoslavov.dimitrov@unito.it}},
       \href{dario.martelli@unito.it}{\texttt{dario.martelli@unito.it}},
       \href{alberto.zaffaroni@mib.infn.it}{\texttt{alberto.zaffaroni@mib.infn.it}}
}       
       \vskip .5in
       {\bf Abstract } 
       \vskip .1in
       \end{center}

In this paper, using equivariant localization for foliations, we compute the on-shell action of a general class of supersymmetric solutions of five-dimensional gauged supergravity with vector multiplets. Unlike previous literature, we also allow for a non-trivial topology for the space-time as well as  for the gauge fields. In practice, we achieve this by covering the spacetime manifold with patches and localizing in each patch. We derive a general formula for the on-shell action that depends on topological data only and can be used without a detailed knowledge of the solution. Our final result is relevant for the physically interesting examples of multi-center black holes, black rings and black lenses,  topological solitons and Euclidean black saddles.  We also show how to connect the topological data with the thermodynamic data: electrostatic potential, the magnetic fluxes and the possible flat connections of the solutions. Our formula for the on-shell action is derived in the context of gauged supergravity, but it is straightforward to take the ungauged limit. Thus, we reproduce known results for a large class of explicit asymptotically flat supersymmetric solutions, and we provide predictions for AdS solutions still to be found.

       \end{titlepage}

%*******************************************************************************
%
%     TOC
%
%*******************************************************************************

\setcounter{tocdepth}{2}

\hrule

\tableofcontents

\vspace{0.65cm}
\hrule

%*******************************************************************************
%
%     Main body
%
%*******************************************************************************

\section{Introduction}
\label{sect:intro}
 
 Equivariant localization  has been recently applied to the calculation of on-shell actions in supergravity and their applications to holography, in particular to the physics of supersymmetric black holes in Anti-de-Sitter (AdS). In the approach  introduced in \cite{BenettiGenolini:2023kxp}, the action is written as an integral of an equivariantly closed form constructed with Killing spinor bilinears and the result evaluated using the BVAB formula \cite{Duistermaat,berline1982classes,ATIYAH19841}. Most of these results can be reinterpreted in terms  of the equivariant volume of the internal manifold \cite{Martelli:2023oqk,Colombo:2023fhu}, which seems to play  an universal role in extremization problems in holography and in black hole physics. Many results have been obtained with these alternative approaches by applying equivariant localization to even-dimensional manifolds or orbifolds. 
 Less has been done for odd-dimensional spaces, where the fixed loci of the torus action are themselves odd-dimensional. In \cite{Colombo:2025ihp} we considered an alternative to the BVAB formula which has been  proposed in \cite{Goertsches:2015vga} and applies to odd-dimensional foliations. We generalized the theorem of \cite{Goertsches:2015vga} to toric manifolds with boundary, following a similar strategy to \cite{Couzens:2024vbn}, and applied it to the calculation of the action of supersymmetric solutions  to five dimensional minimal gauged supergravity. In particular, we derived the on-shell action  of the asymptotically AdS$_5$ supersymmetric black holes with equal electric charges without explicit knowledge of the solution found in \cite{Chong:2005hr}, reproducing for the first time using localization the known result in the literature \cite{Hosseini:2017mds,Hosseini:2018dob,Cabo-Bizet:2018ehj,Cassani:2019mms}. We also computed the action for a class of Euclidean saddles with trivial topology for the gauge field. Some other results for five-dimensional geometries have been obtained in 
 \cite{Cassani:2024kjn}, where the on-shell action of ungauged  supergravity is computed using the BVAB formula with respect to a subgroup of the isometry group, and also in \cite{BenettiGenolini:2025icr} by dimensionally reducing the supergravity solution to four dimensions.
All these results assumed trivial gauge configurations for simplicity,
thus excluding many interesting solutions.

In this paper we extend the results in \cite{Colombo:2025ihp} to cover solutions in an arbitrary five-dimensional gauged supergravity with  vector multiplets and with a non-trivial topology for the gauge fields. This will allow us to cover the physically interesting examples of multi-center black holes, black rings and black lenses,  topological solitons and Euclidean black saddles. We derive a general formula for the on-shell action that depends on topological data only and can be used without a detailed knowledge of the solution. We will also show how to connect the topological data with the thermodynamic potentials, the magnetic fluxes and the possible flat connections of the solutions. Our formula applies to gauged supergravity but we can easily take the ungauged limit to cover asymptotically flat solutions too. 

We show that, for Euclidean supersymmetric solutions with five dimensional manifold $M$, the
on-shell Lagrangian can be written as 
\begin{equation}\label{Iaction}
%16\pi \iu G_5\cdot I[M]
	\cL\,\big|_{\text{on-shell}}=\frac16C_{IJK} %\int_M
		\:\eta^I\wedge\dd\eta^J\wedge\dd\eta^K+%\int_{\partial M}\alpha(M)\:,
	 \dd[]{\alpha}\:,
\end{equation}
where $C_{IJK}$ are the coefficients of the Chern-Simons form in the supergravity action, $\eta^I$ are one-forms, $\dd\eta^I$ is basic with respect to the supersymmetric Killing vector $\xi$, and $\dd[]{\alpha}$ is an exact form. 
We will compute the integral over $M$ of the above Lagrangian for solutions with a torus $\mathbb{T}=\U(1)^3$ effective action by equivariantizing  with  respect to a generic vector field $X$ in $\mathbb{T}$ and using the odd-dimensional localization introduced in \cite{Goertsches:2015vga} and generalised to manifolds with boundary in \cite{Colombo:2025ihp}. One complication compared to \cite{Colombo:2025ihp} is that, when the gauge fields have   non-trivial topology, the one-forms $\eta^I$ are not globally defined and multiple patches are needed. The final result  consists of a bulk contribution $\widehat \cI$ coming from the closed orbits of the $\mathbb{T}$ action and interface integrals among the patches, and a boundary contribution $\cB$. Quite remarkably, both the bulk and boundary contribution from the localization formula  are separately independent of the auxiliary vector $X$.
Moreover, we will show that the total boundary term, obtained as: $\mathcal{B} = \int_{\partial M} \qty[\alpha + \text{localization boundary contribution} + \text{GHY} + \text{standard counterterms}]$, depends only on the leading order asymptotic boundary data and, whenever we can apply boundary subtraction, will cancel between the solution and the subtraction manifold.
The  final  result $\widehat \cI$ can be expressed as in formula \eqref{OSA_localized} and depends only on topological data.

One of the main challenges of the present paper is to deal with multiple patches to account for the non-trivial topology of the gauge fields. In addition to being a technical problem, the introduction of patches allows one to clarify the topology of the solutions which is quite rich, possibly allowing for magnetic fluxes and flat connections.  In particular, every non-trivial two-cycle implies a quantization condition for the parameters in the solution which gives rise to a magnetic flux. Using an UV/IR relation, we will be able to identify the topological data of the solutions with the thermodynamic potentials that enters in the quantum statistical relation and determine the thermodynamics and  the entropy of the black object. 
Our final result for the on-shell action $\widehat\cI$ is gauge-invariant modulo $2\pi \ii \mathbb{Z}$, after we 
take into account the quantization conditions for the fluxes and the quantization of the Chern-Simons level. 

We will present several examples of evaluation of the on-shell action for multi-center black holes, black rings and black lenses,  topological solitons and Euclidean black saddles. Since very few of such solutions are known in AdS, we will mostly compare our results with asymptotically flat solutions which exist in abundance. In particular, we will reproduce the recent results for the on-shell action of a large class of asymptotically flat solutions obtained in \cite{Cassani:2025iix}.
In AdS we can reproduce the physics of Kerr-Newman black holes \cite{Gutowski:2004ez,Gutowski:2004yv,Chong:2005da,Chong:2005hr,Kunduri:2006ek}, black saddles \cite{Aharony:2021zkr} and topological solitons \cite{Chong:2005hr,Cassani:2015upa,Durgut:2021bpk}, and more importantly we provide a prediction for the on-shell action for AdS solutions still to be found.

Our general formula \eqref{OSA_localized} depends explicitly on the topological data of the solution. These can be expressed in terms of the thermodynamic potentials but the explicit mapping is non-trivial. We have been able to fully re-express \eqref{OSA_localized} in terms of the thermodynamic potentials for a general class of solutions that could arise as a continuation of  Lorentzian multi-center black holes. In doing so, we observed that the analytical form of the on-shell action as a function of the thermodynamic potentials can be re-interpreted in terms of the equivariant volume of the six-dimensional Euclidean geometry associated to the same toric data. We plan to explore further 
this relation and find a general expression of  \eqref{OSA_localized} in terms of the thermodynamic potentials for generic solutions in the near future.

It would be also interesting to fully understand the cancellation of boundary terms  using  holographic renormalization and supersymmetric counter-terms without resorting to background subtraction. In this paper we just comment about subtleties arising for black holes with squashed boundaries and leave the full analysis for the future.

 The paper is organized as follows. In section \ref{sect:sugra} we  rewrite the action of  five-dimensional gauged supergravity with an arbitrary number of vector multiplets as the integral of an equivariant form for a general supersymmetric solution with a time-like Killing vector. We also discuss the toric geometry structure and the topological properties of such solutions.
 In section \ref{Patch-wise localization of the on-shell action} we discuss the patch-wise approach to the localization of the on-shell action for a generic solution of five-dimensional $\cN=2$ Fayet-Iliopoulos gauged supergravity with U(1)$^3$ isometry. The main result is formula  \eqref{OSA_localized}. In  section \ref{Topology, thermodynamics and UV-IR relations}  we discuss  how to rewrite the topological data in \eqref{OSA_localized}
in terms of more physically relevant quantities, such as magnetic fluxes, flat connections and thermodynamic potentials. We will also discuss
 the gauge invariance (modulo $2\pi\ii\,\bZ$) of the on-shell action. In  section \ref{sect:examples} we describe  the examples of Kerr-Newman black holes, black rings and black lenses and topological solitons. For black rings and black lenses
we reproduce known results in the asymptotically flat limit and we provide a formula for the on-shell action and the entropy of AdS solutions still to be found. We conclude by discussing a fairly general class of fans where we can simplify our general formula  \eqref{OSA_localized} and we rewrite the result in terms of the equivariant volume of an associated geometry in the spirit of \cite{Martelli:2023oqk}.  In appendix \ref{Gauge invariance} we discuss technical  aspects for a correct definition of the Chern-Simon functional with our choice of patches and we prove the gauge invariance of \eqref{OSA_localized} under large gauge transformation. Finally in appendix \ref{app:renormalization} we discuss some preliminary results about the physical meaning of the boundary terms when no background subtraction method is employed.

\section{Supersymmetric solutions of 5D gauged supergravity}
\label{sect:sugra}

We consider five-dimensional $\cN=2$ gauged supergravity with an arbitrary number of vector multiplets (but no hypermultiplets), whose bosonic sector
is described by the Lagrangian \cite{Gunaydin:1984ak}
\begin{equation}
\label{Lagrangian}
	\cL=(R_g - 2\mathcal{V}) *_g1 - \, Q_{IJ} \dd[]{X^I} \wedge *_g \dd X^J - Q_{IJ} F^I \wedge *_g F^J -
		\frac{1}{6} C_{IJK} F^I \wedge F^J \wedge A^K\:.
\end{equation}
The indices $I,J,K$ run from 1 to $\nv+1$, where $\nv$ is the number of vector multiplets, and Einstein summation is implied.
The symmetric tensor of constants $C_{IJK}$ determines the coupling of the Chern-Simons terms, which mix the $\nv+1$ gauge fields $A^I$.
The scalars $\scal^I$ are subject to the constraint\footnote{Note that the symbol $X$ is being overloaded. $X^I$ (with index) signifies the scalars in the gauged supergravity, $X$ (no index) signifies the vector with respect to which we localize.}
\begin{equation}
\label{scal_constraint}
	\scal^I\scal_I=1\:,\qquad\scal_I\equiv\frac16C_{IJK}\scal^J\scal^K\:.
\end{equation}
In particular there are only $\nv$ independent real scalar fields, as expected from a theory with $\nv$ vector multiplets;
it is however more convenient to work with the $\nv+1$ constrained scalars $\scal^I$.
These scalars are coupled to the Maxwell terms via the matrix $Q_{IJ}$, which is defined by
\begin{equation}
\label{Qdef}
	Q_{IJ}=\frac92\scal_I\scal_J-\frac12C_{IJK}X^K
\end{equation}
and is assumed to be invertible, with inverse $Q^{IJ}$.
The scalar potential is then defined as
\begin{equation}
	\mathcal{V} = \frac{9}{2} V_I V_J \qty(Q^{IJ} - 2 X^I X^J) \:, 
\end{equation}
where $\V_I$ are the Fayet-Iliopoulos (FI) gauging constants.
The potential $\cV$ is assumed to have a minimum; for solutions that are asymptotically-locally-AdS, the value of the minumum of the potential is related to the
AdS radius $\ell$ by
\begin{equation}
	\cV\,\big|_{\text{minimum}}=-\frac6{\ell^2}\:.
\end{equation}
Notice that in our conventions the FI parameters scale as $\V_I\sim\ell^{-1}$.
If one were to set $\V_I=0$ (and thus $\cV=0$), the resulting supergravity would be \emph{ungauged}.
Ungauged supergravities are the models to consider if one is interested in asymptotically-flat solutions;
in section \ref{The limit to ungauged supergravity} we will explain how our results can also be extended to the ungauged case,
and some of the examples that we will discuss in section \ref{sect:examples} are asymptotically-flat.

We are using a $(-,+,+,+,+)$ signature for the space-time metric $g$.
$R_g$ is the scalar curvature of $g$, and $*_g$ the Hodge star. The action of the theory can formally be written as
\begin{equation}
	S=\frac1{16\pi G}\int\cL\:+\:(\text{boundary terms})\:,
\end{equation}
where $G$ is the five-dimensional Newton constant. Since the theory contains Chern-Simons terms that are not locally gauge invariant,
the above definition of the action strictly speaking only applies to cases where there is a single globally defined gauge for the $A^I$.
More generally, the definition of the action involves extending the gauge fields to a six-dimensional manifold. Furthermore, the Lorentzian action
is only defined modulo $2\pi\bZ$. We will discuss all of these subtleties in detail later on.

The solutions that we are interested in are (complex) Euclidean rather than Lorentzian;
most of the analysis in this section will be done in Lorentzian signature, and we will analytically continue to Euclidean later.

\paragraph{Symmetric scalar cosets.}

Among the general Fayet-Iliopoulos supergravities that we consider in this paper, there is a subclass of theories that enjoys special properties,
which are associated to a symmetric scalar coset. 
This subclass will be relevant in some of the examples that we discuss in section \ref{sect:examples},
even if the main result of this paper is valid more in general.
For these supergravities there exist a symmetric tensor of constants $C^{IJK}$ that satisfies the following property:
\footnote{In the literature it is often also demanded that $C^{IJK}=C_{IJK}$. In the discussion here we will not demand it.}
\begin{equation}
\label{CIJK_inverse}
	C^{IJK}\,C_{J(LM}\,C_{NP)K}=\frac43\,\delta^I_{(L}C_{MNP)}\:.
\end{equation}
Using $C^{IJK}$ it is possible to give an explicit expression for the inverse matrix $Q^{IJ}$,
\begin{equation}
	  Q^{IJ} = 2 X^I X^J - 6\,C^{IJK} X_K \,, 
\end{equation}
which can be used to simplify the expression for the scalar potential to
\begin{align}\label{eq:V_symmetric}
  \mathcal{V} ={}& -27\,C^{IJK}\,V_I V_J X_K \,. 
\end{align}
If we denote as $\bar\scal^I$ and $\bar\scal_I$ the constant values of $\scal^I$ and $\scal_I$ at the minimum of $\cV$, it is easy to see that
\begin{equation}
\label{AdS_radius_symmetric}
	\bar\scal_I=\ell\,\V_I\:,\qquad\ell^{-3}=\frac92\,C^{IJK}\,\V_I\V_J\V_K\:.
\end{equation}

\paragraph{STU model.}
\label{STU_model}

A particularly interesting model within the class of Fayet-Iliopoulos supergravities with a symmetric scalar coset is the STU model,
which is also called U(1)$^3$ gauged supergravity.
It has two vector multiplets, $\nv=2$, and it is described by the following choice of constants:
\begin{equation}
\label{STU}
	C_{IJK}=C^{IJK}=|\epsilon_{IJK}|=\begin{cases}1&\quad I\ne J\ne K\ne I\\0&\quad\text{otherwise}\end{cases}\:,
		\qquad \V_I=\frac1{3\ell}\:,\qquad I=1,2,3\:.
\end{equation}
This model uplifts to type IIB supergravity on $S^5$, and thus it has a holographic dual description in terms of $\cN=4$ super-Yang-Mills.

\paragraph{Minimal supergravity.} 

Minimal supergravity only has a single gauge field, the graviphoton, and no scalar fields.
Formally, it is possible to obtain it from the Lagrangian \eqref{Lagrangian} by setting $\nv=0$.
Minimal supergravity is either gauged or ungauged depending on the presence or absence of a $*_g({12}/{\ell^2})$ term in the Lagrangian.
Minimal supergravity is a consistent truncation of the STU model and many other supergravities with a holographic dual.
From the STU model, the truncation to minimal supergravity sets $A^1=A^2=A^3\equiv A$ and $X^1=X^2=X^3=1$.
The Lagrangian of minimal gauged supergravity is then
\begin{align}
  \cL ={}&\qty(R_g + \frac{12}{\ell^2}) *_g - \, \frac{3}{2} F \wedge *_g F - F \wedge F \wedge A \,. 
\end{align}
In \cite{Colombo:2025ihp} we rewrote the above Lagrangian,%
\footnote{Note that $\curly{x}^{\text{there}} = \sqrt{3}$.
		Also, compared to \cite{Colombo:2025ihp} we have chosen a different convention for some orientation signs.}
evaluated on supersymmetric Euclidean solutions, in a way that is suitable for localization techniques.
In section \ref{The on-shell Lagrangian} we will generalize this rewriting to Fayet-Iliopoulos supergravities.

\subsection{Supersymmetric solutions in the time-like class}
\label{Supersymmetric solutions in the time-like class}

Supersymmetric solutions of Fayet-Iliopoulos supergravities have been classified \cite{Gutowski:2004yv,Gutowski:2005id} in terms of Killing spinor bilinears
satisfying certain algebraic and differential relations.
In this section we will review the key aspects of this classification that will be relevant for our localization computation.

Bosonic supersymmetric solutions of Fayet-Iliopoulos supergravities have a Killing spinor $\epsilon^\alpha$, $\alpha=1,2$, which obeys
Killing spinor equations that are equivalent to setting the supersymmetric gravitino and dilatino variations to zero.
A Killing vector $\xi$, called the supersymmetric Killing vector, can be constructed as a bilinear of the Killing spinor as
\begin{equation}
	\xi^\mu=\bar\epsilon^1\Gamma^\mu\epsilon^2\:,
\end{equation}
where the $\Gamma_\mu$ are five-dimensional gamma matrices.
We will focus on solutions in the ``time-like class", for which $\lVert\xi\rVert^2$ is non-vanishing in some region of space-time where is then possible
to write the metric as
\begin{equation}
	  g = - f^2\,\big(\dd[]{y} + \omega_m\dd x^m\big)^2 + f^{-1}\,\gamma_{mn}\,\dd x^m\,\dd x^n \,.
\end{equation}
Here we are using coordinates $(y,x^m)$ such that $\xi=\partial_y$ and the
$x^m$ parametrize a local four-dimensional base with an induced conformally-K\"ahler metric $f^{-1}\gamma_{mn}$,
where $\gamma_{mn}$ is K\"ahler, and $f=\bar\epsilon^1\epsilon^2$ gives the conformal factor.
Since $\xi=\partial_y$ is a Killing vector, $f$, $\omega_m$ and $\gamma_{mn}$ do not depend on the coordinate $y$, and only depend on $x^m$.

The K\"ahler form of $\gamma_{mn}$, which we denote as $\cX^1$, is found as a bilinear of the Killing spinor, together with two other two-forms
$\cX^2$ and $\cX^3$, as follows:
\begin{equation}
	B^{(\alpha\beta)}_{\mu\nu}=\bar\epsilon^\alpha\Gamma_{\mu\nu}\epsilon^\beta\:,\quad B^{(11)}=\cX^1+\ii\cX^2\:,\quad
		B^{(22)}=\cX^1-\ii\cX^2\:,\quad B^{(12)}=-\ii\cX^3\:.
\end{equation}
Note that $\cX^{2}$ and $\cX^3$ are not gauge-invariant, they transform under gauge shifts of the linear combination $\V_IA^I$
under which the Killing spinor is charged. For ungauged supergravities $\V_I=0$ and thus $\cX^{2}$, $\cX^3$ are gauge invariant;
furthermore it is possible to show that in the ungauged case $\cX^{2}$, $\cX^3$ become K\"ahler forms and also satisfy the required conditions
for giving $\gamma_{mn}$ an hyper-K\"ahler structure. More generally, when the supergravity is gauged, one can define the form
\begin{equation}
\label{Ricci_potential}
	P_m=3\V_I(A^I_m-f\scal^I\omega_m)\:,
\end{equation}
and verify that it gives the Ricci potential for $\gamma_{mn}$, i.e.\ its differential is the Ricci form $\cR$:
\begin{equation}
\label{Ricci_form}
	\dd P=\cR\:,\qquad\cR_{mn}= \frac{1}{2} (R_\gamma)_{mnpq} (\mathcal{X}^1)^{pq}\:.
\end{equation}
The forms $\cX^2$, $\cX^3$ are not closed, instead they define a complex form $\Omega$ of type $(2,0)$ satisfying
\begin{equation}\label{differentgauge}
	\dd\Omega+\ii P\wedge\Omega=0\:,\qquad\Omega\,\equiv\,(\cX^2+\ii\cX^3)|_{\text{gauge fixed so that }\iota_\xi A^I=f\scal^I}\:.
\end{equation}
One can show that the $\cX^{1,2,3}$ (when the gauge is fixed as above) give $\gamma_{mn}$ an almost-hyper-K\"ahler structure.
Notice that we will be using a different gauge in this paper.

We can define a vielbein $\{e^a\}_{a=0}^4$ for the five-dimensional metric $g$ from a vielbein $\{e_\gamma^a\}_{a=1}^4$ for the metric $\gamma_{mn}$
as follows:
\begin{equation}
\label{vielbein}
	e^0=f\,\big(\dd{y} + \omega_m\,\dd x^m\big)\:,\qquad e^a=f^{-\frac12}\,e^a_\gamma\:.
\end{equation}
Then we can express the volume forms as
\begin{equation}
\label{Hodge_conventions}
	*_g1=f^{-2}\,e^0\wedge*_\gamma1\:,\qquad*_\gamma1=-\frac12\cX^1\wedge\cX^1\:,
\end{equation}
and we can write the very useful identities
\begin{equation}
\label{Hodge_identities}
  *_g \qty(e^0 \wedge \theta_{(1)}) = -  f^{-1} *_\gamma \theta_{(1)} \,, \qquad
 	 *_g \theta_{(2)} =  e^0 \wedge *_\gamma \theta_{(2)} \,,
\end{equation}
for any one-form $\theta_{(1)}$ and any two-form $\theta_{(2)}$ such that $\iota_\xi\theta_{(1)}=0=\iota_\xi\theta_{(2)}$.
It is also worth noting that the two-form $\cX^1$ (and $\cX^{2,3}$ as well) satisfies
\begin{equation}
	*_\gamma\cX^1=-\cX^1\:,
\end{equation}
which makes it \emph{anti-self-dual} with respect to $\gamma_{mn}$.

Now that we have reviewed the Killing bilinears and the properties of the metric, let us discuss some of the constraints that supersymmetry imposes on the gauge
fields and scalars.
The field-strengths $F^I=\dd A^I$ can be written as
\begin{equation}
\label{field_strength}
	F^I=\dd\left(\scal^Ie^0\right)+\Theta^I+\func^I\cX^1\:,
\end{equation}
where we have defined
\begin{equation}
\label{func_def}
  \func^I = \frac{3}{2} f^{-1} \qty(Q^{IJ} - 2 X^I X^J) V_J \,, 
\end{equation}
and the two-forms $\Theta^I$ satisfy
\begin{equation}
\label{selfdual_relations}
	\iota_\xi\Theta^I=0\:,\qquad\cL_\xi\Theta^I=0\:,\qquad*_\gamma\Theta^I=\Theta^I\:,\qquad\scal_I\Theta^I=-\frac23G^+\:,\qquad
		V_I \Theta^I = \frac{1}{3} \qty(\mathcal{R} - \frac{1}{4} R_\gamma\mathcal{X}^1)\:.
\end{equation}
The first two properties are equivalent to the statement that $\Theta^I$ is \emph{basic}, the third condition means that it is also \emph{self-dual}
(with respect to $\gamma_{mn}$), while the fourth involves the self-dual form $G^+$, which together with the anti-self-dual $G^-$ is defined as
\begin{equation}
	G^\pm=\frac f2\,(\dd\omega\pm*_\gamma\dd\omega)\:,\qquad\omega\equiv\omega_m\dd x^m\:,\qquad*_\gamma G^\pm=\pm\,G^\pm\:.
\end{equation}
Lastly, the fifth relation in \eqref{selfdual_relations} is related to the already mentioned fact that the Ricci potential $P$ and $\V_IA^I$ are connected by
\eqref{Ricci_potential}. Indeed, the scalar curvature $R_\gamma$ of $\gamma_{mn}$ is related to $f$ and the scalar potential by
\begin{equation}
\label{scalar_curvature_base}
	R_\gamma=4f^{-1}\cV\:,
\end{equation}
which implies that
\begin{equation}
\label{V_func}
	\V_I\func^I=\frac13f^{-1}\cV=\frac1{12}R_\gamma\:.
\end{equation}
We note that in addition to the aforementioned properties, $\Theta^I$ must also satisfy the Bianchi identity
$0=\dd F^I=\dd\left(\selfdual^I+\func^I\X^1\right)$.

Using the self-duality of $\Theta^I$ and the anti-self-duality of $\cX^1$ together
with the relations \eqref{Hodge_identities}, we can write $*_gF^I$ as
\begin{equation}
\label{star_field_strength}
	*_gF^I=f^{-2}*_\gamma\dd\left(f\scal^I\right)+e^0\wedge\left(\scal^I(G^+-G^-)+\selfdual^I-\func^I\X^1\right)\:,
\end{equation}
which can be used to rewrite the equation of motion for the gauge fields,
\begin{equation}
\label{F_eq}
	\dd\left(Q_{IJ}*_gF^J\right)+\frac14C_{IJK}F^J\wedge F^K=0\:,
\end{equation}
as follows \cite{Gutowski:2005id}:
\begin{equation}
\label{F_eq2}
\begin{aligned}
	0=&-\dd*_\gamma\dd\left(f^{-1}\scal_I\right)+\frac16C_{IJK}\selfdual^J\wedge\selfdual^K-2f^{-1}\V_I\,G^-\wedge\X^1+\\
	&+\frac38f^{-2}\left(C_{IJK}Q^{JM}Q^{KN}\V_M\V_N+8\V_I\V_M\scal^M\right)\X^1\wedge\X^1\:.
\end{aligned}
\end{equation}

\subsection{The on-shell Lagrangian}
\label{The on-shell Lagrangian}

In this section we simplify the Lagrangian \eqref{Lagrangian} when evaluated on a supersymmetric solution
by using the equations of motions and the various identities required by supersymmetry that we reviewed in the previous section.
Let us start from the Einstein equation:
\begin{equation}
\label{eq:einstein}
	(R_g)_{\mu\nu} - Q_{IJ}\partial_\mu X^I \partial_\nu X^J- Q_{IJ}F\indices{^I_\mu^\rho} F\indices{^J_\nu_\rho}+
		\frac{1}{6}\,g_{\mu\nu}\left(Q_{IJ} F^{I\rho\sigma} F\indices{^J_\rho_\sigma} - 4\mathcal{V}\right) =0\:.
\end{equation}
If we contract the above with $g^{\mu\nu}$ we get an expression for the scalar curvature $R_g$ that we can substitute in the Lagrangian \eqref{Lagrangian},
obtaining
\begin{equation}
\label{Lagrangian_on_shell1}
	\cL\,\big|_{\text{on-shell}}=\frac43\cV*_g1-\frac23Q_{IJ}F^I\wedge*_gF^J-\frac16C_{IJK}F^I\wedge F^J \wedge A^K\:.
\end{equation}
The $*_g1$ term can be rewritten as
\begin{equation}
	\frac43\cV*_g1=-\frac23f^{-2}\cV\,e^0\wedge(\cX^1)^2
\end{equation}
where we have used \eqref{Hodge_conventions}. Next, we will turn our attention to the Maxwell terms.

Using \eqref{field_strength}, \eqref{star_field_strength}, $f\dd(f^{-1}e^0)=\dd\omega=G^++G^-$,
and the fact that any five-form that vanishes when contracted with
$\iota_\xi$ must be identically zero, we can rewrite the Maxwell terms as follows:
\begin{equation}
\label{Maxell}
\begin{aligned}
	&\,-\frac23Q_{IJ}F^I\wedge*_gF^J=\frac23f^{-3}e^0\wedge Q_{IJ}\,\dd\left(f\scal^I\right)\wedge*_\gamma\dd\left(f\scal^J\right)+\\
	&\,\qquad+\frac23e^0Q_{IJ}\left(\scal^I(G^++G^-)+\selfdual^I+\func^I\X^1\right)\wedge\left(\scal^J(G^--G^+)-\selfdual^J+\func^J\X^1\right)\:.
\end{aligned}
\end{equation}
The definition of $Q_{IJ}$ \eqref{Qdef} and the constraint $\scal^I\scal_I=1$ \eqref{scal_constraint} imply various useful algebraic relations
that we will frequently use in this section:
\begin{equation}
\label{X_Q_contractions}
	X^IQ_{IJ}=\frac32X_J\:,\qquad Q^{IJ}X_J=\frac23X^I\:,\qquad X^IQ_{IJ}X^J=\frac32\:,
\end{equation}
and they also imply that
\begin{equation}
	Q_{IJ}\left(\scal^J(G^--G^+)-\selfdual^J+\func^J\X^1\right)=\frac32X_I(G^++G^-)+\frac12C_{IJK}X^J\selfdual^K+Q_{IJ}\func^J\X^1\:,
\end{equation}
thus simplifying the bottom line of \eqref{Maxell}.
It is also useful to remember that wedging a self-dual form with an anti-self-dual one always yields zero, which in particular means that
\begin{equation}
\label{G_cancellations}
	G^+\wedge G^-=0\:,\qquad G^+\wedge\cX^1=0\:,\qquad\Theta^I\wedge G^-=0\:,\qquad\Theta^I\wedge\cX^1=0\:.
\end{equation}
Using the above identities many terms coming from the bottom line of \eqref{Maxell} can be simplified.
\footnote{Specifically, using \eqref{X_Q_contractions}, \eqref{G_cancellations}, \eqref{scal_constraint}, \eqref{func_def}, \eqref{selfdual_relations}
		we have that
\begin{align}
\label{Maxwell_simplifications}\nonumber
	\,&\frac23\scal^I(G^++G^-)\wedge\frac32\scal_I(G^++G^-)=(G^+)^2+(G^-)^2\:,\:\:
		&&\frac23\scal^I(G^++G^-)\wedge\frac12C_{IJK}\scal^J\selfdual^K=-\frac43(G^+)^2\:,\\\nonumber
	\,&\frac23\scal^I(G^++G^-)\wedge Q_{IJ}\func^J\X^1=-2f^{-1}(\V_I\scal^I)G^-\wedge\X^1\:,\:\:
		&&\frac23\selfdual^I\wedge\frac32\scal_I(G^++G^-)=-\frac23(G^+)^2\:,\\\nonumber
	\,&\frac23\func^I\X^1\wedge\frac32\scal_I(G^++G^-)=-2f^{-1}(\V_I\scal^I)G^-\wedge\X^1\:,\:\:
		&&\frac23\func^I\X^1\wedge Q_{IJ}\func^J\X^1=f^{-2}\left(18(\V_I\scal^I)^2+\cV\right)(\X^1)^2\:,\\
	\,&\frac23\selfdual^I\wedge Q_{IJ}\func^J\X^1+\frac23\func^I\X^1\wedge\frac12C_{IJK}\scal^J\selfdual^K=0\:.
\end{align}}

In order to fully simplify the Maxwell terms we will also need to use the equation of motion for the gauge fields \eqref{F_eq2} contracted with $X^I$,
which can be written as
\begin{equation}
\begin{aligned}
\label{F_eq3}
	0=&-\scal^I\dd*_\gamma\dd\left(f^{-1}\scal_I\right)+\frac16C_{IJK}\scal^I\selfdual^J\wedge\selfdual^K
		-2f^{-1}\left(\V_I\scal^I\right)G^-\wedge\X^1+\\
	&+f^{-2}\Big(3\left(\V_I\scal^I\right)^2-\frac16\cV\Big)\X^1\wedge\X^1\:,
\end{aligned}
\end{equation}
where $X^IC_{IJK}=9X_JX_K-2Q_{JK}$ has been used to simplify some terms.
It is convenient to rewrite
\begin{equation}
	\dd\left(f^{-1}\scal_I\right)=\frac23\,\dd\left(f^{-2}Q_{IJ}\cdot f\scal^J\right)=-\frac23f^{-2}Q_{IJ}\,\dd\left(f\scal^J\right)\: ,
\end{equation}
where the second step is a consequence of the fact that $X^IQ_{IJ}X^J$ is a constant and thus
\begin{equation}
	0=\dd\left(f^{-2}Q_{IJ}\cdot f\scal^I\cdot f\scal^J\right)=f\scal^I\left[f\scal^J\dd\left(f^{-2}Q_{IJ}\right)+2f^{-2}Q_{IJ}\,\dd\left(f\scal^I\right)\right]\:,
\end{equation}
Then the following combination of terms is a total derivative,
\begin{equation}
\begin{aligned}
	\,&\frac23f^{-3}e^0\wedge Q_{IJ}\,\dd\left(f\scal^I\right)\wedge*_\gamma\dd\left(f\scal^J\right)
		-\scal^Ie^0\wedge\dd*_\gamma\dd\left(f^{-1}\scal_I\right)=\\
	\,&\qquad=\frac23f^{-1}e^0\wedge\left(f^{-2}Q_{IJ}\cdot f\scal^J*_\gamma\dd\left(f\scal^I\right)\right)=
		\dd\left(-f^{-2}\scal_I\,e^0\wedge*_\gamma\dd\left(f\scal^I\right)\right)\:,
\end{aligned}
\end{equation}
and we can use equation \eqref{F_eq3} to simplify the Maxwell term as follows:
\begin{align}
\label{Maxwell2}
	\,&-\frac23Q_{IJ}F^I\wedge*_gF^J=\eqref{Maxell}\big|_{\eqref{Maxwell_simplifications}}+e^0\wedge\eqref{F_eq3}=\\\nonumber
	\,&\qquad=\dd\left(-f^{-2}\scal_I\,e^0\wedge*_\gamma\dd\left(f\scal^I\right)\right)+\frac12C_{IJK}\scal^Ie^0\wedge\selfdual^J\wedge\selfdual^K
		+e^0\wedge\left((G^-)^2-(G^+)^2\right)+\\\nonumber
	\,&\qquad\quad+f^{-2}\left(9\left(\V_I\scal^I\right)^2+\frac16\cV\right)e^0\wedge(\X^1)^2
		-4\left(\V_I\scal^I\right)e^0\wedge G^-\wedge\X^1\:.
\end{align}

In order to simplify the Chern-Simons terms, let us first define the form
\begin{empheq}[box={\mymath[colback=white, colframe=black]}]{equation}
\label{eta_def}
	\eta^I=\scal^Ie^0-A^I\:.
\end{empheq}
Its differential is a \emph{basic form} (i.e.\ $\iota_\xi\dd\eta=0=\cL_\xi\dd\eta$), and it is given by
\begin{equation}
\label{dd_eta}
	\dd\eta^I=-\selfdual^I-\func^I\X^1\:.
\end{equation}
Then we can write the Chern-Simons term as
\begin{equation}
	-\frac16C_{IJK}A^I\wedge F^J \wedge F^K=-\frac16C_{IJK}A^I\wedge\left(\dd\left(\scal^Je^0\right)-\dd\eta^J\right)
		\wedge\left(\dd\left(\scal^Ke^0\right)-\dd\eta^K\right)\:,
\end{equation}
and we have the simplifications
\begin{align}
{}&\begin{aligned}
	\frac13C_{IJK}A^I\wedge\dd\eta^J \wedge\dd\left(\scal^Ke^0\right)=&\,2e^0\wedge\left(\scal_I\dd\eta^I\wedge f\dd\omega\right)
		-\frac13C_{IJK}\scal^Ie^0\wedge\dd\eta^J\wedge\dd\eta^K+\\
	&+\dd\left(-\frac13C_{IJK}\scal^Ie^0\wedge\eta^J\wedge\dd\eta^K\right)\:,
\end{aligned}\\
{}&\begin{aligned}
	-\frac16C_{IJK}A^I\wedge\dd\left(\scal^Je^0\right)\wedge\dd\left(\scal^Ke^0\right)=&\,
		-e^0\wedge(f\dd\omega)^2+2e^0\wedge\left(\scal_I\dd\eta^I\wedge f\dd\omega\right)+\\
	&+\dd\left(\frac16C_{IJK}\scal^Ie^0\wedge\eta^J\wedge\dd\left(\scal^Ke^0\right)\right)\:,
\end{aligned}
\end{align}
where in both cases we have used $A^I=X^Ie^0-\eta^I$, we have applied Leibniz identity to the piece with $\eta^I$, and used
$\dd(\scal^Ke^0)=f\scal^K\dd\omega+e^0\wedge(\ldots)$ together with the definition \eqref{scal_constraint} of $X_I$ to simplify the piece with $X^Ie^0$.
Then we can use $f\dd\omega=G^++G^-$, \eqref{dd_eta} and the usual identities to write
\begin{equation}
	\scal_I\dd\eta^I\wedge f\dd\omega  = \frac23(G^+)^2+2f^{-1}\left(\V_I\scal^I\right)G^-\wedge\X^1\:.
\end{equation}

Putting all the pieces together, the on-shell Lagrangian \eqref{Lagrangian_on_shell1} can be written as
\begin{align}\nonumber
\label{Lagrangian_on_shell2}
	\cL\,\big|_{\text{on-shell}}=&\,\dd\left(-f^{-2}\scal_I\,e^0\wedge*_\gamma\dd\left(f\scal^I\right)+\frac16C_{IJK}\scal^Ie^0\wedge\eta^J\wedge
		\dd\left(\scal^Ke^0-2\eta^K\right)\right)-\\
	&-\frac16C_{IJK}A^I\wedge\dd\eta^J\wedge\dd\eta^K
		+\frac16C_{IJK}\scal^Ie^0\wedge\left(3\selfdual^J\wedge\selfdual^K-2\dd\eta^J\wedge\dd\eta^K\right)+\\\nonumber
	&+\frac12f^{-2}\left(18\left(\V_I\scal^I\right)^2-\cV\right)e^0\wedge(\X^1)^2\:.
\end{align}
Using that
\begin{equation}
	C_{IJK}\scal^I\func^J\func^K=\left(9X_JX_K-2Q_{JK}\right)\func^J\func^K=f^{-2}\left(18\left(\V_I\scal^I\right)^2-\cV\right)
\end{equation}
and also \eqref{dd_eta}, \eqref{G_cancellations}, we can rewrite \eqref{Lagrangian_on_shell2} simply as
\begin{empheq}[box={\mymath[colback=white, colframe=black]}]{equation}
\label{Lagrangian_rewritten}
	\cL\,\big|_{\text{on-shell}}=\frac16C_{IJK}\:\eta^I\wedge\dd\eta^J\wedge\dd\eta^K+\dd\alpha\:,
\end{empheq}
where we have defined the four-form $\alpha$ as
\begin{equation}
\label{alpha}
	\alpha=-f^{-2}\scal_I\,e^0\wedge*_\gamma\dd\left(f\scal^I\right)+\frac16C_{IJK}\scal^Ie^0\wedge\eta^J\wedge\dd\left(\scal^Ke^0-2\eta^K\right)\:.
\end{equation}
We stress that $\alpha$ is a regular and smooth form whenever the one-forms $A^I$ and $e^0$ are regular and smooth.

So far we have worked in Lorentzian signature, but we are interested in complex Euclidean solutions.
We can use analytic continuation, and the rewriting of the on-shell action \eqref{Lagrangian_rewritten} is still valid, but the one-forms $\eta^I$
and $e^0$ will be complex forms.
Continuous parameters that the solution depends on, such as the components of the Killing vector $\xi$ and others that will be introduced later,
will always be assumed complex unless specified otherwise, in the same spirit as \cite{Cabo-Bizet:2018ehj}.
The Euclidean solutions that we consider have the time direction compactified to a circle, which may shrink in the bulk,
and we can assume that $e^0$ is well defined everywhere.\footnote{In Lorentzian signature $f$ vanishes at the horizon and possibly at {\it evanescent ergosurfaces}. For Euclidean saddles, on the other hand, the solutions cap off smoothly at the horizon with non-vanishing $f$ while, generically, the evanescent ergosurfaces are eliminated by complexifying the metric (see for example the asymptotically flat saddles discussed in  \cite{Cassani:2025iix}, which for generic complex values of the parameters have no vanishing $f$).}
When there is an everywhere regular gauge for $A^I$ for all $I$, then the Euclidean on-shell action is found from the on-shell Lagrangian by
\begin{equation}
	\cI=\frac1{16\pi\ii G}\int\cL\,\big|_{\text{on-shell}}\:+\:(\text{boundary terms})\:.
\end{equation}
In section \ref{The on-shell action with patches and interfaces} we will provide a more rigorous definition of $\cI$.
From now on it will always be implicit that the solutions that we consider are (complex) Euclidean.

\subsection{Geometry of solutions with U(1)$^3$ symmetry}
\label{Geometry of solutions}

We will be interested in solutions with U(1)$^3$ symmetry. In this section we describe the geometry and the main topological properties of such solutions.

We can describe a five-dimensional (smooth) geometry $M$ with a $\bT=$U(1)$^3$ symmetry in terms of information about the loci where $\bT$
has a non-trivial isotropy group. Following the conventions in \cite{Colombo:2025ihp}, we denote with $\cD_a$ the three-dimensional submanifolds
where $\bT$ has a U(1) isotropy subgroup.
\footnote{More precisely, we define the $\cD_a$ to be the closures of the connected components of the loci where $\bT$
		has isotropy subgroup isomorphic to U(1).}
Throughout this paper, we will primarily use a basis of independent $2\pi$-periodic angular coordinates
$(\phi_0,\phi_1,\phi_2)$ rotated by the torus $\bT$. In these
coordinates the action of the U(1) isotropy group at the locus $\cD_a$ can be described 
in terms of three integers $V^a_0,V^a_1,V^a_2\in\bZ$ as follows:
\begin{equation}
\label{U(1)_isotropy}
	\phi_i\:\longrightarrow\:\phi_i+V_i^a\,\varphi\:,\qquad\varphi\in\bR/2\pi\bZ\:.
\end{equation}
Thus the three-dimensional vector $V^a\in\bZ^3$ specifies the angle that is degenerating at $\cD_a$. For the $\bT$ action to be effective we need gcd$(V^a_0,V^a_1,V^a_2)=1$.  The overall sign of $V^a\in\bZ^3$ is not fixed by the above definition,
but it will be fixed in section \eqref{Cohomology of the base}.
Under fairly general assumptions, we can order the $\cD_a$ with respect to the index $a=0,\ldots,d$ in such a way that
\begin{equation}
	\cD_{a-1}\cap\cD_a=L_a\,\cong\,S^1\:,\qquad\cD_b\cap\cD_c=\emptyset\quad\text{if }\:\:|b-c|>1\:,
\end{equation}
where the \{$L_a\}_{a=1}^d$ are the loci where the $\bT$ action has U(1)$^2$ isotropy. When we will discuss the localization formula in section
\ref{The localization formula} the $L_a$ will play the important role of being the \emph{localization loci}.
Notice the absence of $L_0$: since the solutions of our interest have a boundary $\partial M$, the loci $\cD_0$ and $\cD_d$ do not intersect each other but rather
they intersect $\partial M$, and no other $\cD_a$ intersects $\partial M$.

All the relevant topological information about these geometries can be derived from the collection of vectors $V^0,V^1,\ldots,V^d$,
where $V^a\in\bZ^3$ for each $a$. With some abuse of language, we will refer to this (ordered) list of vectors as the \emph{fan} of $M$.
\footnote{This terminology is taken from toric geometry; if the cone over $M$ admits a symplectic form such that the U(1)$^3$ action is Hamiltonian,
		then the geometry of $M$ would be describable by a toric fan generated by the lattice vectors $V^a$. However the cone over a generic 
		supergravity solution $M$ with U(1)$^3$ symmetry may not admit such a symplectic form.}
In the rest of this section we will discuss how to extract from the fan 
the behavior of a smooth metric near the circles $L_a$, the topology of the three-dimensional $\cD_a$ and the topology of the boundary $\partial M$.

Given \emph{any} smooth $\bT$-invariant metric $g$ on $M$, sufficiently close to each $L_a$ we can write
\begin{equation}
\label{flat_metric_5D}
	g\big|_{\text{near }L_a}\approx\big(R_a\,\dd\varphi_a^0\big)^2+\sum_{i=1,2}
		\Big[\big(\dd r_a^i\big)^2+(r_a^i)^2\big(\dd\varphi_a^i+c_a^i\dd\varphi_a^0\big)^2\Big]\:,
\end{equation}
where $R_a$ and $c_a^i$ are constants, the angular coordinates $(\varphi^0_a,\varphi^1_a,\varphi^2_a)$ differ from the angles $(\phi_0,\phi_1,\phi_2)$
by an SL(3,$\bZ$) transformation (and possibly a change of sign), and the $r^i_a$ are two radial coordinates in the planes orthogonal to $L_a$.
By convention we require that the ordering $(\varphi_a^0,r_a^1,\varphi_a^1,r_a^2,\varphi_a^2)$ is aligned with the orientation of $M$.
The coordinates $\varphi^i_a$ can be determined from the fan by imposing that
\begin{equation}
\label{adapted_coord_condition}
\begin{aligned}
	g(V^{a},V^{a})\,\big|_{\cD_{a}}=0\quad&\implies\quad
		\iota_{V^{a}}\,\dd\varphi_a^0=0=\iota_{V^{a}}\big(\dd\varphi_a^1+c_a^1\dd\varphi_a^0\big)\:,\\
	g(V^{a-1},V^{a-1})\,\big|_{\cD_{a-1}}=0\quad&\implies\quad
		\iota_{V^{a-1}}\,\dd\varphi_a^0=0=\iota_{V^{a-1}}\big(\dd\varphi_a^2+c_a^2\dd\varphi_a^0\big)\:,
\end{aligned}
\end{equation}
where with some abuse of language we are denoting the Killing vector $V^a=\sum_iV^a_i\,\partial_{\phi_i}$ that vanishes at $\cD_a$
with the same letter as the fan vector $V^a=(V^a_0,V^a_1,V^a_2)\in\bZ^3$.
It is possible to solve the above system provided that there exist a choice of lattice vector $w^a\in\bZ^3$ such that
\begin{equation}
\label{wa_def}
	1=(V^{a-1},V^a,w^a)\,\equiv\,
	\det \, \begin{pmatrix}
		V^{a-1}_0&\:\:V^{a-1}_1&\:\:V^{a-1}_2\\
		V^a_0&V^a_1&V^a_2\\
		w^a_0&w^a_1&w^a_2
	\end{pmatrix}\:.
\end{equation}
The vector $w^a$ is not unique, being defined only up to the shift  $w^a\to w^a+n_1V^{a-1}+n_2V^a$ where $n_{1,2}\in\bZ$.
If a lattice vector $w^a$ satisfying the above condition does not exist, the solution has orbifold singularities.
For simplicity, we will only consider smooth solutions that do admit such a $w^a$.
Then the solution to the system \eqref{adapted_coord_condition} is given by
\begin{equation}
\label{phiLa}
	\varphi_a^0=\mathfrak s_a\cdot(\phi,V^{a-1},V^a)\:,\quad\varphi_a^1=(\phi,V^a,w^a)\:,\quad\varphi_a^2=(\phi,w^a,V^{a-1})\:,
\end{equation}
where $\phi\equiv (\phi_0,\phi_1,\phi_2)$,  for some sign $\mathfrak s_a$, which can be fixed by imposing the correct orientation.
These signs are all equal and we can take them to be $\mathfrak s_a=-1$, see footnote \ref{orientation_signs}. 
From \eqref{phiLa} it is immediate to find the components of the Killing vector $\xi$ (or any Killing vector) in the basis of  angles $\varphi_a^i$.
Defining $\xi=\sum_i\xi_a^i\,\partial_{\varphi^i_a}$, then we have
\begin{equation}
\label{xiLa_determinant}
	\xi_a^0=\mathfrak s_a\cdot(\xi,V^{a-1},V^a)\:,\quad\xi_a^1=(\xi,V^a,w^a)\:,\quad\xi_a^2=(\xi,w^a,V^{a-1})\:,
\end{equation}
where by convention the above three-by-three determinants have the components of the Killing vector in the base $\partial_{\phi_i}$ in the first row.

Let us discuss how the topology of the three-dimensional locus $\cD_a$ can be determined from the vectors of the fan.
We focus on the case $a\ne0,d$, for which $\cD_a$ is compact.
The SL(3,$\bZ$) transformation 
\begin{equation}
	\begin{pmatrix}
		V^a_0&V^a_1&V^a_2\\
		w^a_0&w^a_1&w^a_2\\
		V^{a-1}_0&\:\:V^{a-1}_1&\:\:V^{a-1}_2
	\end{pmatrix}^{-1}
\end{equation}
sends the vectors $V^{a-1}$, $V^a$, $V^{a+1}$ into 
\begin{equation}
\begin{aligned}
	\,&\wb V^{a-1}\!\!\!\!\!&&=(0,0,1)\:,\\
	\,&\wb V^a\!\!\!\!\!&&=(1,0,0)\:,\\
	\,&\wb V^{a+1}\!\!\!\!\!&&=\big((V^{a-1},V^{a+1},w^a),\:(V^{a-1},V^a,V^{a+1}),
		\,-(V^{a},V^{a+1},w^a)\big)\:,
\end{aligned}
\end{equation}
and defines a new basis of $2\pi$-periodic angles $\wb\phi_0,\wb\phi_1,\wb\phi_2$
such that the angle $\wb\phi_0$ degenerates at $\cD_a$, whereas $\wb\phi_1,\wb\phi_2$ can be used to parametrize the angular coordinates of $\cD_a$ itself.
In particular $\cD_a$ is a three-dimensional manifold with a U(1)$^2$ symmetry, described by the fan obtained by ignoring the $\partial_{\,\wb\phi_0}$
components of $\wb V^{a-1}$ and $\wb V^{a+1}$:
\begin{equation}
	(0,1)\:,\qquad\big((V^{a-1},V^a,V^{a+1}),\,-(V^{a},V^{a+1},w^a)\big)\:.
\end{equation}
From the above fan it is then easy to extract the topology of $\cD_a$,
\footnote{The fan $(0,1)$, $(p,-q)$ corresponds to a smooth geometry only if gcd$(p,q)=1$. An SL(2,$\bZ$) transformation can be used to shift $q$ by any
		integer multiple of $p$.
		When $p=0$, 
		the topology of $\cD_a$ is a fibration of $S^1\times S^1$, where the $S^1$ parametrized by $\wb\phi_1$ never shrinks, whereas
		the  $S^1$ parametrized by $\wb\phi_2$ shrinks smoothly at both endpoints, hence $\cD_a\,\cong\,S^1\times S^2$.
		When $p\ne0$, we can assume that $p>0$ without loss of generality, since the overall sign of these fan vectors does not
		matter for this discussion and $L(p,q)\cong L(p,-q)$. In order to  see that $\cD_a$ has the topology of $L(p,q)\,\cong\,S^3/\bZ_p$ (and thus $S^3$ for $p=1$), we can 
		parametrize $S^3$ as $\big\{\big(\rho^2\,e^{\ii\,\varphi_1},(1-\rho^2)\,e^{\ii\,\varphi_2}\big)\:|\:\varphi_{1,2}\in\bR/2\pi\bZ,\:\rho\in[0,1]\big\}$,
		so that $L(p,q)$ is obtained by taking the quotient with respect to the action 
		$\bZ_p\ni n:(\varphi_1,\varphi_2)\mapsto(\varphi_1+\frac{2\pi}pn,\,\varphi_2+\frac{2\pi q}pn)$.
		In the angular coordinates $(\wb\phi_1,\wb\phi_2)=(p\,\varphi_1,\,\varphi_2-q\,\varphi_1)$ with periodicity $\wb\phi_1\sim\wb\phi_1+p\,2\pi$ and
		$\wb\phi_2\sim\wb\phi_2+2\pi$, the $\bZ_p$ shifts $\wb\phi_1$ by multiples of $2\pi$, and thus taking the quotient by $\bZ_p$
		reduces the periodicity of $\wb\phi_1$ to $2\pi$. The resulting geometry is $L(p,q)$, and the vectors
		$\partial_{\varphi_1}=(p,-q)\cdot(\partial_{\wb\phi_1},\partial_{\wb\phi_2})$ and
		$\partial_{\varphi_2}=(0,1)\cdot(\partial_{\wb\phi_1},\partial_{\wb\phi_2})$ vanish at $\rho=0$ and $\rho=1$ respectively, reproducing the fan
		$(0,1)$, $(p,-q)$.}
\begin{equation}
\label{cD_a_topology}
	\cD_a\,\cong\,
		\begin{cases}
			\:S^1\times S^2&\qquad\text{if}\quad(V^{a-1},V^a,V^{a+1})=0\:,\\[1mm]
			\:S^3&\qquad\text{if}\quad\big|(V^{a-1},V^a,V^{a+1})\big|=1\:,\\[1mm]
			\:L(p,q)&\qquad\text{otherwise}\:,
		\end{cases}
\end{equation}
where the coprime integers $p$, $q$ are given by
\footnote{Notice that $q$ does not depend on the specific choice of $w^a$. Indeed, if we redefined $w^a$ by $w^a\to w^a+n_1V^{a-1}+n_2V^a$
		for some integers $n_{1,2}$, then we would have $(V^{a},V^{a+1},w^a)\to(V^{a},V^{a+1},w^a)+n_1(V^{a-1},V^a,V^{a+1})=
		(V^{a},V^{a+1},w^a)\mod p$.}
\begin{equation}
\label{p,q_fmla}
	p=\big|(V^{a-1},V^a,V^{a+1})\big|\:,\qquad q=(V^{a},V^{a+1},w^a)\mod p\:.
\end{equation}

The lattice vectors $V^0,V^d$ together determine the topology of the boundary $\partial M$. Up to a redefinition of the angles,
we can assume without loss of generality that they are given by
\begin{equation}
\label{V^0,V^d}
	V^0=(0,0,1)\:,\qquad V^d=(0,p,-q)\:,
\end{equation}
where $p\geq0$ and $q$ are coprime and $0\leq q<p$ when $p\ne0$.
In all the examples throughout this paper we will adopt the above convention for $V^0$ and $V^d$.
Then the angular coordinate $\phi_0$ never shrinks in $\partial M$, and it parametrizes a $S^1$ factor of the geometry.
The remaining coordinates parametrize a three-dimensional manifold with U(1)$^2$ symmetry and fan given by $(0,1)$, $(p,-q)$.
With a similar logic as \eqref{cD_a_topology}, the possible boundary topologies are then
\begin{equation}
\label{boundary_topology}
	\partial M\,\cong\,
		\begin{cases}
			T^2\times S^2&\qquad\text{if}\quad p=0\:,\\[1mm]
			S^1\times S^3&\qquad\text{if}\quad p=1\:,\\[1mm]
			S^1\times L(p,q)&\qquad\text{otherwise}\:.
		\end{cases}
\end{equation}
In this paper we will not consider the case $\partial M\,\cong\,T^2\times S^2$  as it requires the introduction of two patches in the boundary. Instead, we will mostly focus on
$S^1\times S^3$ boundaries and make some comments on the $S^1\times L(p,q)$ case, leaving the extension to a systematic treatment of multiple boundary patches for future work.

\subsubsection{Cohomology of $\dd\eta^I$}
\label{Cohomology of the base}

Unlike the case of minimally gauged supergravity discussed in \cite{Colombo:2025ihp}, the two-forms $\dd\eta^I$ are not proportional to the Ricci form $\cR$.
Their expansion in cohomology classes of the base will play an important role in fixing the gauge in each patch, and in the application of the localization formula.
We will also see in section \ref{alpha vs varphi}  that the coefficients of this expansion are related to the
thermodynamic potentials.

Pragmatically, we can define a basic one-form $\wt\eta^I$ on $M\smallsetminus\bigcup_a\cD_a$ by
\begin{equation}
  \label{wteta_pragmatic}
  \wt\eta^I=\left(fX^I\omega_m-A^I_m\right)\dd x^m
\end{equation}
where $x^m$ are the coordinates on the base, 
for some choice of gauge $A^I$ that is regular over $M\smallsetminus\bigcup_a\cD_a$, but might be irregular at the loci $\cD_a$ since it could have
a non-zero component along the angular coordinate that degenerates at $\cD_a$. Such a choice of $A^I$ always exists, and the above form by construction
satisfies $\cL_\xi\wt\eta^I=0=\iota_\xi\wt\eta^I$ and
\begin{equation}
  \dd\wt\eta^I=\dd\eta^I\:,
\end{equation}
which follows simply by comparing \eqref{wteta_pragmatic} with \eqref{eta_def} and using the fact that $\dd\eta^I$ is basic.
In particular, $\wt\eta^I$ and $\eta^I$ always differ by a U(1)$^3$-invariant gauge transformation.
Using that $\cL_{V^a}\wt\eta^I=0$ and that $\dd\eta^I$ is regular over all $M$, we find
\begin{equation}
  \label{weta_constant}
  \dd\left(\iota_{V^a}\wt\eta^I\right)\big|_{\cD_a}=-\iota_{V^a}\dd\eta^I\,\big|_{\cD_a}=0
\end{equation}
since the Killing vector $V^a$ vanishes at $\cD_a$. In particular, the above equation implies that $\iota_{V^a}\wt\eta^I$ is constant over $\cD_a$,
and we can define a constant $\etacoho^I_a$ such that
\begin{equation}
  \label{etacoho_def_simple}
  \iota_{V^a}\wt\eta^I\,\big|_{\cD_a}\equiv-\etacoho^I_a\:.
\end{equation}
As we will explain shortly, these constants determine the coefficients of the expansion of $\dd\eta^I$ in basic cohomological classes.

The form $\wt\eta^I$ will be important in the manipulations of section \ref{Patch-wise localization of the on-shell action}.
Let us notice that by simple comparison between \eqref{wteta_pragmatic} and \eqref{Ricci_potential} we find that 
a linear combination of the $\wt\eta^I$ gives the Ricci potential $P$:
\begin{equation}
  \label{wteta_P}
  V_I\,\wt\eta^I=-\frac13P\:,
\end{equation}
which will imply a linear relation among the $\etacoho^I_a$, as we will discuss at the end of this section.

\paragraph{Toric geometry on the base and basic cohomology.}

Let $\mathbb{B}$ denote the four-dimensional base of $M$ with respect to the fibration given by $\xi$. We can restrict ourselves to the values of $\xi$
that make $\mathbb{B}$ an orbifold, and extend our results to generic values of $\xi$ by analytic continuation. In particular if $\xi = c_\xi n_i \partial_{\phi_i}$ with $c_\xi\in \mathbb{R}$ and $n_i\in \mathbb{Z}$,
 we can define an $SL(3,\mathbb{Z})$ transformation such that
\begin{equation}
\label{base_toric_data}
	B_{ij}n_j=\delta_{0i}\:,\qquad\phi_1^B =B_{1i} \phi_i\:\qquad \phi_2^B =B_{2i} \phi_i\:,\qquad v_i^a= B_{ij} V_j^a
\end{equation}
where $\phi^B_{1,2}$ are $2\pi$-periodic angles on the base, and the two-component vectors $v^a$ give the toric data of the base \cite{Colombo:2025ihp}.
Notice that toric geometry fixes the sign of the vectors $v^a$, which fixes the sign ambiguity of the $V^a$.
Then since $\dd\eta^I$ is basic we can expand it in the cohomology classes of the base $\mathbb{B}$ as
\begin{equation}
\label{ddeta_cohomology}
	\big[\dd\eta^I\big]=2\pi\sum_{a=0}^d\etacoho^I_a\,c_1(\cL_a)\:,
\end{equation}
where $\cL_a$ is the line bundle associated to the divisor $D_a=\cD_a\cap \mathbb{B}$ and $[\:\:]$ is the equivalence class of forms differing by an exact form on $\mathbb{B}$.
It will soon be clear why the coefficients of the above expansion are the same as the constants defined in \eqref{etacoho_def_simple}.
We have the following cohomological relations between the Chern classes on the base:
\begin{equation}
	\sum_{a=0}^dv^a_i\,c_1(\cL_a)=0\:,\qquad i=1,2\:.
\end{equation}
Due to these relations the definition of the $\etacoho^I_a$ is ambiguous, and the $\etacoho^I_a$ enjoy the following ``gauge transformations":
\begin{equation}
\label{etacoho_redef}
	\etacoho^I_a\:\longrightarrow\:\etacoho^I_a+\beta_i^I\,v_i^a\:,\qquad\beta_{1,2}^I\text{ arbitrary}.
\end{equation}
It is especially convenient to fix the ``gauge" choice for the $\etacoho^I_a$ that makes
\begin{equation}
	\etacoho_0^I=0=\etacoho_d^I\:.       
\end{equation}
Notice that this is possible since we are assuming that $V^0$ and $V^d$ are not collinear, since this corresponds to $T^2\times S^2$ boundaries that we do
not discuss in this paper.
In terms of five-dimensional data the gauge transformation would read
\begin{equation}
\label{etacoho_redef2}
	\etacoho^I_a\:\longrightarrow\:\etacoho^I_a+\wt\beta_i^I\,V_i^a\:,\qquad \xi_i \wt\beta_{i}^I=0.
\end{equation}
where now $i=0,1,2$.

As discussed in \cite{Colombo:2025ihp}, to which we refer for details, a representative of the  Chern classes is expressible in terms of moment maps $\mu_i^a$,
\begin{equation}
\label{Chern_class}
c_1(\cL_a) = \sum_{i=1}^2 \dd (\mu_i^a \dd\phi_i^B ) \, ,
\end{equation}
where the moment maps restricted to the fixed points of the toric action $p_a=D_{a-1} \cap D_a$ satisfy
\begin{equation}\label{fp}
 \mu_i^a v^a_k \Big |_{p_b} = - \frac{\delta_{ik}}{2\pi} \delta_{ab} \, .
\end{equation}
Then, using the same logic as in section 4.1 of \cite{Colombo:2025ihp},
we can write the one-form $\wt\eta^I$ defined in \eqref{wteta_pragmatic} as
\begin{equation}
\label{eta_tilde_explicit_def}
\wt\eta^I = 2\pi \sum_{a=1}^d\sum_{i=1}^2 \alpha^I_a \, \mu_i^a \, \dd \phi_i^B + (\text{regular\, one-form})\, .\
\end{equation}
Indeed, from \eqref{eta_tilde_explicit_def} we can immediately derive
\begin{equation}
\label{eta_tilde_def}
	\iota_\xi\wt\eta^I=0\:,\qquad\dd\wt\eta^I=\dd\eta^I\:,\qquad\iota_{V^a}\wt\eta^I\big|_{\cD_a}=-\etacoho_a^I\:,
\end{equation}
where the last identity follows from the fact that \eqref{eta_tilde_explicit_def} 
is constant on $\cD_a$
by the same logic as \eqref{weta_constant}, and then we can evaluate it at $p_a\in\cD_a$ where we can use \eqref{fp}.
The first two properties in \eqref{eta_tilde_def} imply that \eqref{wteta_pragmatic} and \eqref{eta_tilde_explicit_def}
are the same (up to a basic gauge transformation),
and the last one implies that the $\etacoho^I_a$ defined in \eqref{etacoho_def_simple} are the same as the coefficients of the cohomological expansion
\eqref{ddeta_cohomology}.

\paragraph{A linear relation between the $\etacoho^I_a$.}

Even if each two-form $\dd\eta^I$ is unrelated to the Ricci form $\cR$, there is however
a linear combination  that is proportional to $\cR$. 
If we take the differential of \eqref{wteta_P} we find that
\footnote{Alternatively one can also use \eqref{dd_eta}, \eqref{selfdual_relations} and \eqref{V_func}.}
\begin{equation}
\label{Ricci_vs_ddeta}
	\V_I\,\dd\eta^I=-\frac13\cR\:.
\end{equation}
Using the above relation and the fact that we can expand $\cR$ in basic cohomological classes as
\begin{equation}
	\big[\cR\big]=2\pi\sum_{a=0}^dc_1(\cL_a)\:,
\end{equation}
comparison with \eqref{ddeta_cohomology} leads to
\begin{equation}
\label{FI_and_etacoho}
	\V_I\,\etacoho_a^I=-\frac13+\V_I\,\wt\beta_i^I\,V_i^a\:,
\end{equation}
where we are also taking in account the possibility of redefinition of the form \eqref{etacoho_redef2} with $\xi_i\wt\beta_i^I=0$.
Then the ``gauge" choice $\etacoho^I_0=0=\etacoho^I_d$ corresponds to setting
\begin{equation}
	\V_I\,\wt\beta^I_i=\frac13\,\frac{\epsilon_{ijk}\,\xi_j\left(V^0_k-V^d_k\right)}{(\xi,V^0,V^d)}\:,
\end{equation}
which leads to the linear relation
\begin{equation}
\label{etacoho_linear_relation}
		\V_I\,\etacoho_a^I=\frac13\,\frac{(\xi,V^0,V^a)+(\xi,V^a,V^d)-(\xi,V^0,V^d)}{(\xi,V^0,V^d)}\:,\qquad\text{with }
	\:\etacoho^I_0=0=\etacoho^I_d\:.
\end{equation}
As we will see in section \ref{Topology, thermodynamics and UV-IR relations}, this relation implies constraints among the thermodynamic potentials and magnetic fluxes of the solution.
In the special case of a electrically charged, rotating black hole solution of the STU model \eqref{etacoho_linear_relation} reduces to the familiar
\begin{equation}
  \text{const} \times \qty(\varphi^1+\varphi^2+\varphi^3) = \omega_1+\omega_2-2\pi\ii\qquad(\text{STU model, Kerr-Newman})\:.
\end{equation}
We will discuss this in more detail in \cref{example BH}. \eqref{etacoho_linear_relation} will also impose constraints on the topology of asymptotically-flat solutions as we now discuss. 

\subsection{The limit to ungauged supergravity}
\label{The limit to ungauged supergravity}

At the level of the Lagrangian \eqref{Lagrangian}, ungauged supergravities can be obtained by setting $\V_I=0$;
indeed, the Fayet-Iliopoulos constants $\V_I\sim\ell^{-1}$ parametrize the gauging of the theory.
The equation \eqref{Lagrangian_rewritten} for the on-shell Lagrangian can be derived in the same manner even when $\V_I=0$,
meaning that the same expression for the on-shell $\cL$ can be used for both gauged and ungauged supergravities.
The localization computation that we will present in section \ref{Patch-wise localization of the on-shell action} will then be valid for both cases,
with only some technical differences in the handling of the boundary terms.
Remarkably, if a suitable renormalization scheme is chosen for the on shell action $\cI$, we will find that the expression for
$\cI$ is formally identical for solutions in gauged and ungauged supergravities that have the same topology.%
\footnote{Although, it often happens that topologies that can support solutions in one case cannot support solutions in the other.}
For our purposes, the key difference between the two cases is the following: the relation \eqref{selfdual_relations}, which is valid for gauged supergravities
and rewrites $\V_I\Theta^I$ in terms of the curvature invariants on the base, becomes simply $\cR=0$
in the ungauged limit.
As a consequence, the linear relation \eqref{etacoho_linear_relation} which, in gauged supergravities fixes $\V_I\,\etacoho^I_a$ in terms of the fan
and $\xi$, no longer involves the $\etacoho^I_a$ in the ungauged case, becoming
\begin{equation}
	(\xi,V^0,V^a)+(\xi,V^a,V^d)=(\xi,V^0,V^d)\qquad\forall\,a\:.
\end{equation}
Using the expression \eqref{V^0,V^d} for $V^0,V^d$, the above conditions can be written as
\begin{equation}
\label{ungauged_constraint}
	-\frac{(1+q)\,\xi_1+p\,\xi_2}{p\,\xi_0}\:V^a_0+\frac{1+q}p\:V^a_1+V^a_2\,=\,1\qquad\forall\,a\:,
\end{equation}
which takes the  form $\nu\cdot V^a=1$ for a particular $\nu\in\bQ^3$.

Focusing on the case $\partial M\,\cong\,S^1\times S^3$ (and thus $p=1$, $q=0$), supersymmetric boundary conditions impose that $\xi$ can be written as
\cite{Cabo-Bizet:2018ehj,Cassani:2024kjn,Cassani:2025iix}
\begin{equation}
	\xi=\frac1\beta\left(2\pi\ii\,\partial_{\phi_0}-\omega_1\partial_{\phi_1}-\omega_2\partial_{\phi_2}\right)\:,
\end{equation}
where the $\omega_{1,2}\in\bC$ are chemical potentials for the angular momenta $J_{1,2}$ which further satisfy
\begin{equation}
\label{AF_constraint}
	\omega_1+\omega_2=2\pi\ii\:,
\end{equation}
while $\beta$ is the radius of the $S^1$ at the boundary. Then \eqref{ungauged_constraint} becomes
\begin{equation}\label{CY}
	V^a_0+V^a_1+V^a_2=1\qquad\forall\,a\:,
\end{equation}
and is solved by setting
\begin{equation}
\label{flat_fan}
	V^a = (n_a,p_a,1-p_a-n_a)\qquad\forall\,a\:
\end{equation}
for some integers $n_a$, $p_a$. The above form for the $V^a$ matches the one presented in the classification of asymptotically
flat solutions in minimal ungauged supergravity done in \cite{Cassani:2025iix}.
Furthermore, from \eqref{cD_a_topology} and \eqref{p,q_fmla} we can reproduce the the analysis of the topology of the $\cD_a$
done in \cite{Cassani:2025iix}, assuming absence of orbifold singularities.%
\footnote{From \eqref{flat_fan} and \eqref{p,q_fmla} we find $p=|n_{a-1}(p_{a}-p_{a+1})+n_a(p_{a+1}-p_{a-1})+n_{a+1}(p_{a-1}-p_a)|$
		for the topology of $\cD_a$, matching section 3.3.2 of \cite{Cassani:2025iix} in all cases. For $q$, we need a case-by-case analysis:
\begin{itemize}
\item If $n_{a-1}=n_a=n_{a+1}=0$ we find $p=0$ and thus $\cD_a\,\cong\,S^1\times S^2$, no further analysis is needed.
\item If $n_{a-1}=n_a=0$ but $n_{a+1}\ne0$, then $1=(V^{a-1},V^a,w^a)=w^a_0\,(p_{a-1}-p_a)$ is solvable only if $|p_{a-1}-p_a|=1$.
	We can set $w^a=(p_{a-1}-p_a,0,0)$ and find $q=(V^a,V^{a+1},w^a)\mod p=(p_{a-1}-p_a)(p_{a+1}-p_a)$,
	and thus $\cD_a\,\cong\,L(|n_{a+1}|,p_{a+1}-p_a)$ (since $L(p,q)\,\cong\,L(p,-q)$).
\item If $n_{a-1}=n_{a+1}=0$ but $n_{a}\ne0$, we may set $w^a=(-\mathfrak a,\mathfrak b(p_{a+1}-p_{a-1}),\mathfrak{a-b}(p_{a+1}-p_{a-1}))$,
	then the condition $1=(V^{a-1},V^a,w^a)=\mathfrak a(p_a-p_{a-1})+\mathfrak b\,n_a(p_{a+1}-p_{a-1})$ defines $\mathfrak{a,b}\in\bZ$,
	and we find $\cD_a\,\cong\,L(|n_{a+1}(p_{a-1}-p_a)|,1+\mathfrak a(p_{a-1}-p_{a+1}))$, reproducing \cite{Cassani:2025iix} with ``$a+1$" and
	``$a-1$" swapped. This derivation assumes gcd$(p_{a+1}-p_{a-1},p_a-p_{a-1})=1$; we find that this assumption is not necessary, if we simply take
	$w^a=(\mathfrak{-c,d,c-d})$ with $\mathfrak{c,d}\in\bZ$ defined by $\mathfrak c(p_a-p_{a-1})+\mathfrak d\,n_a=1$, giving
	$\cD_a\,\cong\, L(|n_{a+1}(p_{a-1}-p_a)|,1+\mathfrak c(p_{a-1}-p_{a+1}))$.
\item If $n_a=0$ but $n_{a-1}\ne0$ and $n_{a+1}\ne0$, we may set $w^a=(\mathfrak{-a+b}(n_{a-1}-n_{a+1}),\mathfrak b(p_{a-1}-p_{a+1}),
	\mathfrak{a+b}(n_{a+1}-n_{a-1}+p_{a+1}-p_{a-1}))$ with $\mathfrak{a,b}\in\bZ$ defined by
	$\mathfrak a(p_a-p_{a-1})+\mathfrak b(n_{a+1}(p_a-p_{a-1})+n_{a-1}(p_{a+1}-p_a))=1$, and we find
	$\cD_a\,\cong\,L(|n_{a-1}(p_{a}-p_{a+1})+n_{a+1}(p_{a-1}-p_a)|,1+\mathfrak a(p_{a-1}-p_{a+1}))$,
	reproducing \cite{Cassani:2025iix} with ``$a+1$" and ``$a-1$" swapped. A simpler derivation with minimal assumptions involves taking
	$w^a=(\mathfrak{-c,d,c-d})$ with $\mathfrak{c,d}\in\bZ$ defined by $\mathfrak c(p_a-p_{a-1})-\mathfrak d\,n_{a-1}=1$, giving
	$\cD_a\,\cong\, L(|n_{a-1}(p_{a}-p_{a+1})+n_{a+1}(p_{a-1}-p_a)|,\mathfrak c(p_{a+1}-p_{a-1})+\mathfrak d\,n_{a+1})$.
\end{itemize}
}
In the classification of \cite{Cassani:2025iix}, the $\cD_a$ with $n_a=0$ are called ``bubbles", whereas the $\cD_a$ with $n_a\ne0$ are called ``horizons",
even when the Euclidean saddle may not be the analytic continuation of a Lorentzian solution. In this paper we will often follow the same convention.

Notice that \eqref{CY} implies that the vectors of the fan lie on a plane. This is a generalized Calabi-Yau condition, and it is the same as the one we imposed
in \cite{Colombo:2025ihp} in a slightly different context.
If our geometry were compact and we could uplift it to a toric symplectic  cone over $M$ this would be a CY three-fold.

\section{Patch-wise localization of the on-shell action}
\label{Patch-wise localization of the on-shell action}

\begin{figure}[ht]
  \centering
\begin{tikzpicture}[scale=1, every node/.style={font=\sffamily}]

  % colors
  \definecolor{myteal}{RGB}{0,140,115}
  \definecolor{myred}{RGB}{200,40,40}
  \definecolor{mygray}{RGB}{40,40,40}

  % Top boundary
  \coordinate (BTleft)  at (-5.6,3.6);
  \coordinate (BTright) at ( 5.6,3.6);
  \draw[line width=1.0pt,myteal] (BTleft) -- (BTright);

  % Bottom polyline vertices
  \coordinate (L1) at (-4.0,-1.9);
  \coordinate (L2) at (-1.6,-2.7);
  \coordinate (L3) at ( 1.6,-2.7);
  \coordinate (L4) at ( 4.2,-1.9);

  % Polyline connecting to boundary
  \draw[line width=1.2pt,mygray] (BTleft) -- (L1) -- (L2) -- (L3) -- (L4) -- (BTright);

  % Black nodes with labels
  \foreach \pt/\lab in {L1/{L$_1$}, L2/{L$_2$}, L3/{L$_3$}, L4/{L$_4$}}
    \filldraw[black] (\pt) circle (3.2pt) node[below=6pt,font=\small] {\lab};

  % Central red dot
  \coordinate (C) at (0,0.55);
  \filldraw[myred] (C) circle (3.6pt);

  % Midpoints
  \coordinate (M12) at ($0.5*(L1)+0.5*(L2)$);
  \coordinate (M23) at ($0.5*(L2)+0.5*(L3)$);
  \coordinate (M34) at ($0.5*(L3)+0.5*(L4)$);

  % External midpoints (between boundary and first/last vertex)
  \coordinate (ExtL) at ($0.5*(BTleft)+0.5*(L1)$);
  \coordinate (ExtR) at ($0.5*(BTright)+0.5*(L4)$);

  % Red dashed rays
  \foreach \P in {ExtL,M12,M23,M34,ExtR} {
    \draw[myred,dashed,line width=0.95pt] (C) -- (\P);
  }

  % Labels for dashed lines: just below endpoints, with shifts
  \node[myred,font=\small,below=2pt,left=3pt] at (ExtL) {$U_{(01)}$};
   \node[black,font=\small, above=25pt] at (ExtL) {$\cD_0$};
  \node[myred,font=\small,below=2pt] at (M12) {$U_{(12)}$};
  \node[black,font=\small, above left=5pt] at (M12) {$\cD_1$};
  
  \node[myred,font=\small,below=2pt] at (M23) {$U_{(23)}$};
   \node[black,font=\small, above left=1pt] at (M23) {$\cD_2$};
  
  \node[myred,font=\small,below=2pt] at (M34) {$U_{(34)}$};
  \node[black,font=\small, above right =5pt] at (M34) {$\cD_3$};
  
  \node[myred,font=\small,below=2pt,right=3pt] at (ExtR) {$U_{(40)}$};
\node[black,font=\small, above =25pt] at (ExtR) {$\cD_4$};

  % Region labels (U_(i)) placed in wedges
  \path let \p1=($(ExtL)!0.5!(M12)$), \p2=($(C)!0.65!(\p1)$)
    in node[myred,font=\small] at (\p2) {$U_{(1)}$};

  \path let \p1=($(M12)!0.5!(M23)$), \p2=($(C)!0.65!(\p1)$)
    in node[myred,font=\small] at (\p2) {$U_{(2)}$};

  \path let \p1=($(M23)!0.5!(M34)$), \p2=($(C)!0.65!(\p1)$)
    in node[myred,font=\small] at (\p2) {$U_{(3)}$};

  \path let \p1=($(M34)!0.5!(ExtR)$), \p2=($(C)!0.65!(\p1)$)
    in node[myred,font=\small] at (\p2) {$U_{(4)}$};

  % Central label moved up
  \node[myred,font=\small,above=4pt] at (C) {$U_{(0,...,d)}$};
  
   \node[myred,font=\small,above=40pt] at (C) {$U_{(0)}$};

  % Boundary label
  \node[myteal,font=\large] at ($(BTleft)!0.5!(BTright)+(0,0.25)$) {BOUNDARY};

\end{tikzpicture}
\caption{The disposition of the patches $U_{(a)}$, interfaces $U_{(ab)}$ and intersection of interfaces $U_{(0,\ldots,d)}$ in the toric diagram for the
		four-dimensional base of a generic solution with $d=4$.}
\label{fig:Patches}

\end{figure}
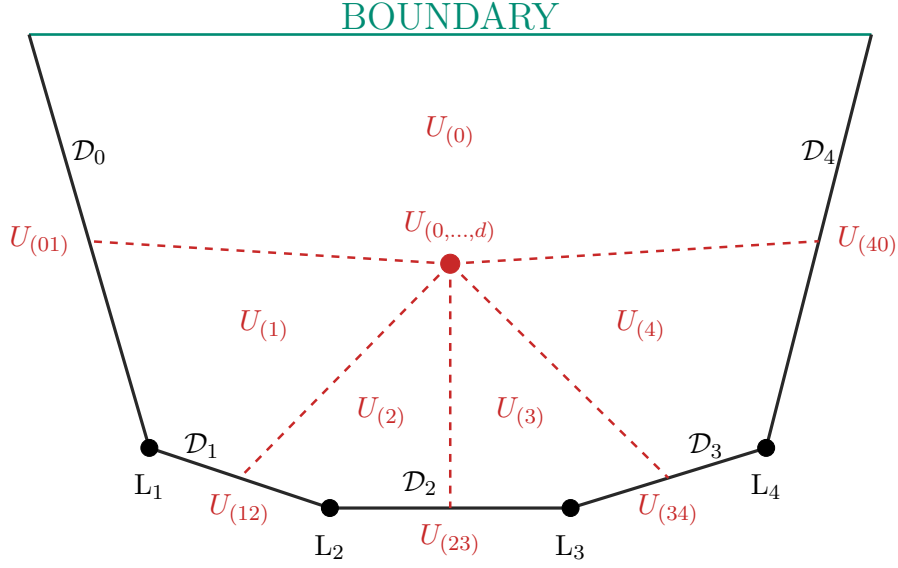

In this section we will compute the on-shell action of a generic solution of five-dimensional $\cN=2$ Fayet-Iliopoulos gauged supergravity with U(1)$^3$ isometry.
In doing so we will also extend the results for minimal gauged supergravity obtained in \cite{Colombo:2025ihp} which assume a globally defined gauge field and cannot be applied to solutions with non trivial topology.
Generically the gauge fields will not be defined globally, but only in patches $U_{(a)}$,
with a gauge transformation required to go from one patch to the other.
We choose our five-dimensional patches $U_{(a)}$ so that the overlaps have null measure; the overlaps are four-dimensional submanifolds
that we call interfaces. Furthermore, the disposition of patches that we will use is such that each patch is invariant under the U(1)$^3$ isometry
and the patch $U_{(a)}$ intersects the loci $\cD_{a-1}$ and $\cD_a$ but no other $\cD_b$.
The notation that we use is
\begin{center}
\begin{tabular}{lll}
	Patches (5D) & $\longrightarrow\qquad U_{(a)}\:,$ & with $a=0,\ldots,d\:,$\\
	Interfaces (4D) & $\longrightarrow\qquad U_{(a,a+1)}\:,$ & such that $\quad\partial U_{(a)}=U_{(a,a+1)}-U_{(a-1,a)}\:,$\\
	Interface intersection (3D) & $\longrightarrow\qquad U_{(0,\ldots,d)}\:,$ & such that $\quad\partial U_{(a,a+1)}=U_{(0,\ldots,d)}\:.$
\end{tabular}
\end{center}
\vspace{3mm}
Diagrammatically the patches are thus disposed as in figure \ref{fig:Patches}. Notice that we have a single patch around the boundary, $U_{(0)}$.
This will simplify the discussion in this section and the boundary analysis of section \ref{Boundary analysis},
at the cost of not being able to treat $T^2\times S^2$ solutions, which can support magnetic fluxes on the $S^2$ at the boundary,
and also will pose some limitation for asymptotically $S^1\times L(p,q)$ solutions, which may have flat connections in the $L(p,q)$ at the boundary
as we discuss in section \ref{An example of a solution with torsion}.

The gauge field $A^I_{(a)}$ in the patch $U_{(a)}$ must then satisfy the regularity conditions
\begin{equation}
\label{explicit_gauge_regularity}
	\iota_{V^{a-1}}\,A^I_{(a)}\,\big|_{\cD_{a-1}}=\,0\,=\,\iota_{V^a}\,A^I_{(a)}\,\big|_{\cD_a}\:,
\end{equation}
which follow from $V^{a-1}\,\big|_{\cD_{a-1}}=0=V^a\,\big|_{\cD_a}$ the fact that $A^I_{(a)}$ is regular everywhere on $U_{(a)}$
which intersects $\cD_{a-1}$ and $\cD_a$.
The forms $\eta^I$ defined by \eqref{eta_def} will also assume a different value $\eta^I_{(a)}$ in each patch,
and are subject to a similar regularity condition:
\begin{equation}
\label{eta_regularity}
	\iota_{V^{a-1}}\,\eta^I_{(a)}\,\big|_{\cD_{a-1}}=\,0\,=\,\iota_{V^a}\,\eta^I_{(a)}\,\big|_{\cD_a}\:.
\end{equation}
Given that $\scal^Ie^0$ is a regular one-form on the whole manifold, $\eta^I_{(a)}$ is regular on the patch $U_{(a)}$ if and only if
the gauge fields $A_{(a)}^I$ are regular on $U_{(a)}$, and thus imposing the condition \eqref{eta_regularity} is equivalent to imposing
\eqref{explicit_gauge_regularity}.

Since each $\eta^I_{(a)}$ differs from the basic (non-regular) one-form $\wt\eta^I$ defined in \eqref{eta_tilde_def}
by a U(1)$^3$\,-\,invariant gauge transformation, we can introduce constants $\left(\gauge^I_a\right)_i$ such that
\footnote{ Since $\dd \eta^I_{(a)}\equiv \dd \eta^I=\dd \wt\eta^I$ this gauge transformation can only be constant.}

\begin{equation}
\label{gauge_constants_def}
	\eta^I_{(a)}-\wt\eta^I\,=\,\gauge^I_a\cdot\dd\phi\:\equiv\sum_{i=0,1,2}\left(\gauge^I_a\right)_i\,\dd\phi_i\:.
\end{equation}
Since $\wt\eta^I$ is basic, we observe that
\begin{equation}
\label{iota_xi_eta}
	\iota_\xi\,\eta^I_{(a)}=\gauge^I_a\cdot\xi\:\equiv\:\text{constant on each patch }U_{(a)}\:.
\end{equation}
The fact that $\iota_\xi\,\eta^I_{(a)}$ is a constant can also be derived by observing that $\iota_\xi\dd\eta^I=0$ and $\cL_\xi\,\eta^I_{(a)}=0$.
We anticipate that the constant $\gauge^I_a\cdot\xi$ will play an important role in the localization of the on-shell action.
It is important to observe that due to \eqref{eta_tilde_def}, the condition \eqref{eta_regularity} for the regularity of $\eta^I_{(a)}$
(or equivalently $A^I$) can be rewritten as
\begin{equation}
\label{gauge_regularity}
	\gauge^I_a\cdot V^{a-1}=\etacoho^I_{a-1}\:,\qquad\gauge^I_a\cdot V^{a}=\etacoho^I_{a}\:.
\end{equation}

\subsection{The on-shell action with patches and interfaces}
\label{The on-shell action with patches and interfaces}

It is well known that a proper gauge invariant definition of the Chern-Simons functional requires  introducing a 6D manifold $\wb M$ such that $\partial\wb M=M$. 
\footnote{There are subtleties due to the fact that $M$ also has a boundary $\partial M$.
		Appropriate boundary conditions must be imposed; when the boundary is $S^1\times S^3$
		with a round $S^3$ metric, it is appropriate to choose Dirichlet boundary conditions for which $F^I|_{\partial M}=0$
		and $M$ can be considered as being effectively compact. Where the boundary conditions are not Dirichlet there is generically  no guarantee that
		the action will be invariant under gauge transformation at the boundary, but the interface contributions that we include in this section will
		be enough to guarantee gauge invariance of the bulk contribution $\widehat\cI$ to the total on-shell action.}
Although in six-dimensions the Chern-Simons integral  depends  only on curvature fields that are gauge-invariant and globally defined,
when we rewrite the functional in terms of five-dimensional fields 
we need to apply Stokes patch-by-patch. The result of this procedure is a Chern-Simons functional that involves not only sum
of integrals on the patches, but also integrals on interfaces and intersections thereof.
The details of this procedure are discussed in appendix \ref{Interface integrals for a well defined Chern-Simons action}. The result is
\begin{align}\nonumber
\label{CS_interfaces}
	16\pi\ii G\cdot\text{(Chern-Simons)}=-\frac16C_{IJK}\sum_{a=0}^d\Bigg[&\int_{U_{(a)}} A_{(a)}^I\wedge F^J\wedge F^K
		+\int_{U_{(a,a+1)}}\!\!\!\!\!\!\!\!\!\!\dd\Lambda^I_{a,a+1}\wedge A^J_{(a)}\wedge F^K-\\
	&-\int_{U_{(0,\ldots,d)}}\!\!\!\!\!\!\!\!\!\!\dd\Lambda^J_{(a,a+1)}\wedge\dd\Lambda^I_{a,0}\wedge A^K_{(a+1)}\Bigg]\:,
\end{align}
where $\dd\Lambda^I_{a,b}=A_{(a)}^I-A_{(b)}^I$.
Notice that the five-dimensional integrals over the patches $U_{(a)}$ involve the same term that appears in the Lagrangian \eqref{Lagrangian},
with $A^I\to A^I_{(a)}$. In particular, the appropriate definition of the on-shell action $\cI$ involves integrating the Lagrangian in each patch separately
and also adding the interface integrals coming from \eqref{CS_interfaces}:
\begin{align}
\label{OSA_proper}
	16\pi\ii G\cdot\cI=\sum_{a=0}^d\Bigg[{}&\int_{U_{(a)}}\cL\,\big|_{A^I\to A^I_{(a)}}
		-\frac16C_{IJK}\int_{U_{(a,a+1)}}\!\!\!\!\!\!\!\!\!\!\dd\Lambda^I_{a,a+1}\wedge A^J_{(a)}\wedge F^K+\\\nonumber
	&+\frac16C_{IJK}\int_{U_{(0,\ldots,d)}}\!\!\!\!\!\!\!\!\!\!\dd\Lambda^J_{(a,a+1)}\wedge\dd\Lambda^I_{a,0}\wedge A^K_{(a+1)}\Bigg]
		+\int_{\partial M}\text{GHY + counterterms}\:.
\end{align}
Notice that we have also (schematically) added the Gibbons-Hawking-York (GHY) boundary term and holographic renormalization counterterms,
which are both needed for a well-defined on-shell action.
We will give the precise formula for these terms in section \ref{Boundary analysis}, when we will analyze carefully all the boundary contributions.

If we combine the rewriting \eqref{Lagrangian_rewritten} of the Lagrangian with 
\eqref{OSA_proper}, we obtain the following expression for the on-shell action:
\begin{align}\nonumber
\label{OSA_initial}
	16\pi\ii G\cdot\cI=&\:\frac16C_{IJK}\sum_{a=0}^d\Bigg[\int_{U_{(a)}}\eta_{(a)}^I\wedge\dd\eta^J\wedge\dd\eta^K+
		\int_{U_{(a,a+1)}}\!\!\!\!\!\!\!\!\!\!\dd\Lambda_{a,a+1}^I\wedge\scal^Je^0\wedge\dd\left(\scal^Ke^0-2\eta^K\right)-\\\nonumber
	&\qquad\qquad\qquad-\int_{U_{(a,a+1)}}\!\!\!\!\!\!\!\!\!\!\dd\Lambda_{a,a+1}^I\wedge A_{(a)}^J\wedge F^K
		+\int_{U_{(0,\ldots,d)}}\!\!\!\!\!\!\!\!\!\!\dd\Lambda_{a,a+1}^I\wedge\dd\Lambda_{a,0}^I\wedge A_{(a+1)}^J\Bigg]+\\\nonumber
	&+\int_{\partial M}\bigg(-f^{-2}\scal_I\,e^0\wedge*_\gamma\dd\left(f\scal^I\right)+
		\frac16C_{IJK}\scal^Ie^0\wedge\eta_{(0)}^J\wedge\dd\left(\scal^Ke^0-2\eta^K\right)\bigg)+\\
	&+\int_{\partial M}\text{GHY + counterterms}\:.
\end{align}
In the above expression we have used Stokes to rewrite the exact ``$\dd\alpha$" term in the on-shell Lagrangian as
integrals of $\alpha$ over the boundaries of the patches, thus obtaining a new $\partial M$ term and new interface integrals.
\footnote{The four-form $\alpha$, defined in \eqref{alpha}, is also dependent on the choice of gauge in each patch, and thus we get new interface integrals
		involving integration over the \emph{difference} between the value of $\alpha$ at the two sides of the interface, which is non-zero.}
In the following sections we will localize the above action. First, we will review the localization formula and show how to simplify the
boundary contributions coming from it.  The reader not interested in the details of the derivation can jump to section \ref{The formula for the localized on-shell action} where the final result is summarized.

\subsection{The localization formula}
\label{The localization formula}

Let us review the localization formula of \cite{Goertsches:2015vga}, valid for odd-dimensional manifolds,
which was generalized to the case of manifolds with boundary in \cite{Colombo:2025ihp}.
We consider a compact orientable manifold $M$ with boundary $\partial M$ and dimension $\,\text{dim}\,M=D\in2\,\bN+1$,
together with the action of a torus $\bT=U(1)^{\frac{D+1}2}$ on $M$.
We will assume that this action is effective and has no fixed points at the boundary $\partial M$.
Just like in the discussion of section \ref{Geometry of solutions}, we will denote the submanifolds of $M$ where the torus $\bT$
action has a U(1)$^{\frac{D-1}2}$ isotropy group as $L_a$, with $a=0,\ldots,d$.
In particular each $L_a$ is diffeomorphic to a circle, $L_a\,\cong\,S^1$, and the torus action simply rotates its points within the circle itself.
As already discussed, any arbitrary $\bT$-invariant Riemannian metric $g_R$ in the vicinity of $L_a$ can be written as
\begin{equation}
\label{flat_metric}
	g_R\big|_{\text{near }L_a}\approx\big(R_a\,\dd\varphi_a^0\big)^2+\sum_{i=1}^{\frac{D-1}2}
		\Big[\big(\dd r_a^i\big)^2+(r_a^i)^2\big(\dd\varphi_a^i+c_a^i\dd\varphi_a^0\big)^2\Big]\:,
\end{equation}
for some real constants $R_a,c_a^i$. Here the $\varphi^i_a$ form a basis of independent $2\pi$-periodic angles that the action of the torus $\bT$ rotates
while leaving the radii $r_a^i$ fixed. By convention, the ordering $(\varphi_a^0,r_a^1,\varphi_a^1,r_a^2,\varphi_a^2,\ldots)$
is aligned with the orientation of $M$.
In the following discussion the adapted coordinates $\varphi^i_a$ will play an important role, whereas the metric $g_R$ will not;
the localization formula only deals with integrals of differential forms, we simply introduced the metric $g_R$ as a shortcut to define the adapted coordinates.
\footnote{A choice of Riemannian metric $g_R$ is also convenient in order to provide a proof of the localization formula, which can be found in section
2.2 of \cite{Colombo:2025ihp}.}

The localization formula  computes integrals of the form
\begin{equation}
\label{localizand}
	\int_M\eta\wedge\Phi_{D-1}\:,
\end{equation}
where $\eta$ is a one-form on $M$ satisfying
\begin{equation}
\label{eta_condition}
	\iota_\xi\,\dd\eta=0
\end{equation}
with respect to a given vector field $\xi$ generated by an element of the Lie algebra of $\bT$. 
On the other hand, the $(D-1)$-form $\Phi_{D-1}$ must be basic with respect to the foliation whose leaves are the orbits of $\xi$;
explicitly, this means that it must satisfy
\begin{equation}
	\iota_\xi\,\Phi_{D-1}=0=\cL_\xi\,\Phi_{D-1}\:.
\end{equation}
Furthermore, we assume that it is possible to define a polyform
\begin{equation}
	\Phi=\Phi_0+\Phi_2+\ldots+\Phi_{D-1}
\end{equation}
such that $\Phi_{2k}$ is a basic $2k$-form (with respect to the foliation generated by $\xi$) and
\begin{equation}
	\dd_X\,\Phi=0\:,\qquad\dd_X\,\equiv\,\dd-\iota_X\:,\qquad\cL_X\,\Phi=0\, , 
\end{equation}
for a generic 
 vector field $X$ generated by an element of the Lie algebra of $\bT$.
The vector $X$ is assumed to be non-collinear with $\xi$ on all points of $M\smallsetminus\bigcup_aL_a$.
Notice that the \emph{equivariant differential} operator $\dd_X$ defines a cohomology on polyforms that have vanishing Lie derivative $\cL_X$,
since $(\dd_X)^2=-\cL_X$. In particular, $\Phi$ is said to be an \emph{equivariantly-closed basic polyform}.
Let us denote the components of the vectors $\xi$, $X$ in the basis of angular coordinates adapted to $L_a$ as $\xi^i_a$, $X_a^i$:
\begin{equation}
	\xi={\sum}_{i=0}^{\frac{D-1}2}\xi_a^i\,\partial_{\varphi^i_a}\:,\qquad X={\sum}_{i=0}^{\frac{D-1}2}X_a^i\,\partial_{\varphi^i_a}\:.
\end{equation}
Then the localization formula recasts the integral \eqref{localizand} as a sum over contributions localized at each $L_a$, plus a boundary term,
as follows:
\begin{empheq}[box={\mymath[colback=white, colframe=black]}]{equation}
\label{localization_formula}
	\int_M\!\eta\wedge\Phi_{D-1}=(-2\pi)^{\frac{D-1}2}\sum_a\frac{\Phi_0\big|_{L_a}\cdot\int_{L_a}\eta}
		{\,{\displaystyle\prod}_{\,i=1}^{\,\frac{D-1}2}\!\left(\frac{X_a^0}{\xi_a^0}\:\xi_a^i-X_a^i\right)}
		-\int_{\partial M}\!\eta\wedge\vartheta\wedge\big(\dd_X\vartheta\big)^{-1}\wedge\Phi,
\end{empheq}
where $\vartheta$ is an arbitrary $\bT$-invariant basic one-form on $\partial M$ such that $\iota_X\vartheta$ never vanishes.
The integrand of the boundary integral can also be written in a more explicit manner if one uses that
\begin{equation}
\label{Theta_def}
	\vartheta\wedge\left(\dd_X\vartheta\right)^{-1}=-\sum_{j=0}^{\frac{D-3}2}\Theta\wedge\left(\dd\Theta\right)^j\:,\qquad
		\Theta=\frac{\vartheta}{\iota_X\vartheta}\:.
\end{equation}
It is possible to further simplify the boundary contribution if  we assume that
the polyform $\Phi$ can be written as $\Phi=\dd_X\Psi$ for a (not necessarily regular) basic polyform $\Psi$,
which will be the case for all the applications that we will consider.
We will discuss this simplification in section \ref{Simplification of the boundary integral}.

In this paper we will be interested in the case $D=5$.
In section \ref{Geometry of solutions} we have already discussed the geometry of five-dimensional manifolds with boundary and a $\bT=$U(1)$^3$ symmetry.
All the relevant topological information can be obtained from the fan vectors $V^a$ defined in \eqref{U(1)_isotropy}.
In particular it is possible to extract an expression for the components of the vectors $\xi$, $X$ in the adapted coordinates around $L_a$,
see \eqref{xiLa_determinant}. This immediately implies the following expression for the value of the weights at the denominator of the localization formula:
\footnote{This is a consequence of the determinant formula
\begin{align}
\label{determinant_product_formula}
	(w^1,w^2,w^3) (w^0,w^3,w^4) + (w^2,w^0,w^3) (w^1,w^3,w^4) + (w^0,w^1,w^3) (w^2,w^3,w^4)=0\:,
\end{align}
valid for any choice of vectors $w^i\in\bC^3$, which follows from the more general identity
\begin{align}
\label{determinant_product_formula_origin}
	(w^1,w^2,w^3)\,w^0-(w^0,w^2,w^3)\,w^1+(w^0,w^1,w^3)\,w^2-(w^0,w^1,w^2)\,w^3=0\:.
\end{align}
}
\begin{equation}\label{weights}
	\frac{X_a^0}{\xi_a^0}\,\xi_a^1-X_a^1=-\frac{(\xi,X,V^a)}{(\xi,V^{a-1},V^a)}\:,\qquad
		\frac{X_a^0}{\xi_a^0}\,\xi_a^2-X_a^2=\frac{(\xi,X,V^{a-1})}{(\xi,V^{a-1},V^a)}\:.
\end{equation}
If one further assumes that $\cL_\xi\eta=0$ (as will be the case for the applications in this paper), then this condition together with \eqref{eta_condition}
imply that $\iota_\xi\eta$ is a constant and thus
\begin{equation}\label{inteta} 
	\int_{L_a}\eta\,=\,2\pi\,\frac{\iota_\xi\eta}{\xi_a^0}\,=\,\frac{\mathfrak s_a\:2\pi\,\iota_\xi\eta}{(\xi,V^{a-1},V^a)}\:.
\end{equation}
The only term in the $L_a$ contribution to the localization formula \eqref{localization_formula} that is not fixed by the above formulas
is the evaluation of the function $\Phi_0$ at $L_a$, which we will explain how to compute in each case of interest.

\subsubsection{A physically relevant example: the localization of the on-shell action of Kerr-Newman black holes}
\label{BH single patch}

As a simple example of the application of the localization formula \eqref{localization_formula}, we can show how to derive the
on-shell action of Euclidean Kerr-Newman black holes, obtained by analytically continuing the supersymmetric Lorenztian solutions of
\cite{Gutowski:2004ez,Gutowski:2004yv,Chong:2005da,Chong:2005hr,Kunduri:2006ek},
without using any explicit information about the solution itself other than its topology.

According to the rules and conventions that we discussed in section \ref{Geometry of solutions}, the fan
\begin{equation}
\label{BH_fan}
	V^0=(0,0,1)\:,\qquad V^1=(1,0,0)\:,\qquad V^2=(0,1,0) \:,
\end{equation}
corresponds to a geometry where the boundary is homeomorphic to $S^1\times S^3$, 
and in the bulk the $2\pi$-periodic coordinate $\phi_0$ that parametrizes the asymptotic $S^1$ degenerates at the locus $\cD_1\,\cong\,S^3$
where the Killing vector $V^1$ vanishes.
This is precisely the topology expected for a black hole solution with horizon topology $S^3$ and no other topologically non-trivial features,
after analytic continuation to a non-extremal complex Euclidean saddle.
Indeed, this analytic continuation procedure involves (among other things) the compactification of the time coordinate to an $S^1$ circle,
the removal of the space-time region inside the horizon and the shrinking of the time circle at the horizon \cite{Cabo-Bizet:2018ehj}.
The black hole horizon thus becomes the locus $\cD_1$, and the Killing vector that was null at the horizon now vanishes at $\cD_1$,
and it matches $V^1$. For more details on the geometry described by the fan \eqref{BH_fan} we refer to section 4.2 of \cite{Colombo:2025ihp}.

The vectors \eqref{BH_fan}, together with the components of the supersymmetric Killing vector $\xi$,
contain all the information required for computing the on-shell action of Kerr-Newman black holes by means of localization.
First, let us parametrize $\xi$ as
\begin{equation}
	\xi=\frac1\beta\left(2\pi\ii\,\partial_{\phi_0}-\omega_1\partial_{\phi_1}-\omega_2\partial_{\phi_2}\right)\:,
\end{equation}
where $\omega_{1,2}$ are complex.
Although  for a generic solution multiple patches are required in order to describe the gauge field,  for the fan \eqref{BH_fan} a single global patch is sufficient,
as it can be seen from the fact that setting
\begin{equation}
\label{BH_single_gauge}
	\left(\gauge^I_0\right)_i=\left(\gauge^I_1\right)_i=\left(\gauge^I_2\right)_i=\etacoho^I_{a\,=\,(i+1\text{ mod }3)}\,\equiv\,\left(\gauge^I\right)_i
\end{equation}
satisfies the regularity condition \eqref{gauge_regularity} for the gauge field,
and any other choice of the  $\gauge^I_a$ that also satisfies \eqref{gauge_regularity}
differs from the above  by unphysical gauge  parameters.%
\footnote{Indeed, this geometry cannot support magnetic fluxes or flat connections, as it will be clarified by the discussion of sections
		\ref{Topological constraints on the thermodynamic potentials} and \ref{Flat connections and gauge invariance}.}
Since we are using a single patch, all the interface terms in the on-shell action \eqref{OSA_initial} can be ignored.
We will employ a background subtraction scheme for renormalising the on-shell action, and subtract from the on-shell action of the black hole
the on-shell action of (global Euclidean) AdS$_5$, whose fan is missing $V^1$:
\begin{equation}
	V^0=(0,0,1)\:,\qquad V^2=(0,1,0)\qquad\qquad(\text{fan of AdS}_5)\:.
\end{equation}
We will ignore all boundary terms in this discussion, and show that the bulk terms are sufficient to reproduce the known results for the on-shell action
of these black holes. 
In \cite{Colombo:2025ihp} we have already shown (in the context of minimal gauged supergravity) that with AdS-subtraction all boundary terms cancel,
provided that the same global gauge $\gauge^I$ and the same arbitrary vector $X$ are picked for both the solution $M$ and AdS$_5$;
in section \ref{Boundary analysis} we will show that the same is true for the gauged supergravities of interest in this paper,
and we will  discuss the holographic renormalization scheme as well.

Let us use the localization formula \eqref{localization_formula} to compute the integral
\begin{equation}
	16\pi\ii G\cdot\cI=\frac16C_{IJK}\Bigg(\int_M\eta^I\wedge\dd\eta^J\wedge\dd\eta^K-\int_{\text{AdS}}\eta^I\wedge\dd\eta^J\wedge\dd\eta^K\bigg)
		+(\text{boundary terms})\:.
\end{equation}
First, we construct the basic polyform
\begin{equation}
\label{BH_first_polyform}
	\Phi^{JK}=\dd_X\eta^J\wedge\dd_X\eta^K=\dd\eta^J\wedge\dd\eta^K
		-\left(\iota_X\eta^J\,\dd\eta^K+\iota_X\eta^K\,\dd\eta^J\right)+\iota_X\eta^J\,\iota_X\eta^K\:,
\end{equation}
which satisfies $0=\dd_X\Phi^{JK}=\cL_X\Phi^{JK}=\cL_\xi\Phi^{JK}$ and
\begin{equation}
	\frac16C_{IJK}\int_{M\text{ (or AdS)}}\eta^I\wedge\dd\eta^J\wedge\dd\eta^K=\frac16C_{IJK}\int_{M\text{ (or AdS)}}\eta^I\wedge\Phi^{JK}\:.
\end{equation}
In order to apply the localization formula \eqref{localization_formula}, we are only missing the following piece:
\begin{equation}
	\Phi_0^{JK}\big|_{L_a}=\left(\iota_X\eta^J\,\iota_X\eta^K\right)\big|_{L_a}=
		\left(\frac{X^0_a}{\xi^0_a}\right)^2\left(\iota_\xi\eta^J\right)\left(\iota_\xi\eta^K\right)=
		\frac{(X,V^{a-1},V^a)^2}{(\xi,V^{a-1},V^a)^2}\left(\iota_\xi\eta^J\right)\left(\iota_\xi\eta^K\right)\:,
\end{equation}
where the second step follows from the fact that since $\eta^I$ is a regular form
\begin{equation}
\label{eta_restrictions}
	\left(X-\frac{X^0_a}{\xi^0_a}\,\xi\right)\bigg|_{L_a}=0\quad\implies\quad
		\left(\iota_X\eta^I-\frac{X^0_a}{\xi^0_a}\:\iota_\xi\eta^I\right)\bigg|_{L_a}=0\:.
\end{equation}
We can now apply the localization formula \eqref{localization_formula} to the black hole solution, which receives two distinct bulk contibutions
$L_1$ and $L_2$ corresponding to the North and South poles of the horizon $\cD_1$:
\begin{equation}
\label{BH_localization}
	\frac16C_{IJK}\int_{M}\eta^I\wedge\Phi^{JK}=-\frac{(2\pi)^3}6C_{IJK}\sum_{a=1,2}
		\frac{\left(\iota_\xi\eta^I\right)\left(\iota_\xi\eta^J\right)\left(\iota_\xi\eta^K\right)(X,V^{a-1},V^a)^2}
		{(\xi,V^{a-1},V^a)(\xi,X,V^a)(\xi,V^{a-1},X)}+(\text{boundary})\:,
\end{equation}
where we have also used \eqref{weights} and \eqref{inteta}.
A similar computation is also performed for AdS$_5$, whose only bulk contribution comes from the AdS center $L_0$:%
\footnote{The orientation signs $\mathfrak s_a$ are taken to be $-1$ for the black hole's $L_1$ and $L_2$ contributions, and $+1$ for the $L_0$
		contribution of AdS$_5$. This is the opposite sign convention compared to \cite{Colombo:2025ihp}.}
\begin{equation}
\label{AdS_localization}
	\frac16C_{IJK}\int_{\text{AdS}}\eta^I\wedge\Phi^{JK}=\frac{(2\pi)^3}6C_{IJK}
		\frac{\left(\iota_\xi\eta^I\right)\left(\iota_\xi\eta^J\right)\left(\iota_\xi\eta^K\right)(X,V^{2},V^0)^2}
		{(\xi,V^{2},V^0)(\xi,X,V^0)(\xi,V^{2},X)}+(\text{boundary})\:.
\end{equation}
We can see that the $X$-dependence completely drops out when taking the difference of the bulk contributions of \eqref{BH_localization} and
\eqref{AdS_localization}. Using the already mentioned fact that the boundary terms cancel completely, we find that the renormalized on-shell action
of Kerr-Newman black holes is given by
\begin{equation}
	\cI=\frac{\pi\,\beta^3}{24G}\,C_{IJK}\,\frac{\left(\iota_\xi\eta^I\right)\left(\iota_\xi\eta^J\right)\left(\iota_\xi\eta^K\right)}{\omega_1\omega_2}\:.
\end{equation}
The contractions of $\eta^I$ with $\xi$ are constants, using \eqref{iota_xi_eta} and \eqref{BH_single_gauge} we can write them as
\begin{equation}
	\iota_\xi\eta^I\,=\,\gauge^I\cdot\xi\,=%\,\xi_0\,\etacoho^I_1+\xi_1\,\etacoho^I_2+\xi_2\,\etacoho^I_0
		\frac1\beta\left(2\pi\ii\,\etacoho^I_1-\omega_1\,\etacoho^I_2-\omega_2\,\etacoho^I_0\right)\:.
\end{equation}
At last, if we fix the redundancy in the definition of the $\etacoho^I_a$ by setting $\etacoho^I_0=0=\etacoho^I_2$ we find
\begin{equation}
\label{BH_first_result}
	\cI=-\frac{\ii\pi^4}{3G}\,C_{IJK}\,\frac{\etacoho_1^I\,\etacoho_1^J\,\etacoho_1^K}{\omega_1\omega_2}\:,
\end{equation}
which perfectly matches the known results in the literature
\cite{Hosseini:2017mds,Hosseini:2018dob,Cabo-Bizet:2018ehj,Cassani:2019mms}
if $\etacoho^I_1$ is taken to be proportional to the (rescaled) electrostatic potential of the black hole;
in section \ref{alpha vs varphi} we will explain why this must be the case, and provide a more general argument for the identification of the $\etacoho^I_a$
with the thermodynamic potentials.

There is an important technical remark to make:
the polyform \eqref{BH_first_polyform}, is not the only possible choice of equivariantly closed polyfom,
for example we can also choose
\begin{equation}
\label{BH_second_polyform}
	\Phi^{JK}=\dd_X\wt\eta^J\wedge\dd_X\wt\eta^K=\dd\eta^J\wedge\dd\eta^K
		-\left(\iota_X\wt\eta^J\,\dd\eta^K+\iota_X\wt\eta^K\,\dd\eta^J\right)+\iota_X\wt\eta^J\,\iota_X\wt\eta^K\:,
\end{equation}
Where $\wt\eta^I$ was defined in \eqref{eta_tilde_explicit_def} and \eqref{eta_tilde_def}.
The above polyform also satisfies $0=\dd_X\Phi^{JK}=\cL_X\Phi^{JK}=\cL_\xi\Phi^{JK}$, as required.
This new $\Phi^{JK}$ is actually more convenient to work with, in the general setup with multiple patches.
However, since $\wt\eta^I$ is not regular everywhere, we cannot use the same trick as in \eqref{eta_restrictions} to compute the restriction of
$\iota_X\wt\eta^I$ at the localization loci $L_a$. Instead, we have to use the fact that the Chern classes
of the base can be extended to equivariantly closed basic polyforms $c_1^\bT(\cL_a)$ whose restrictions match the weights \eqref{weights}
times $-(2\pi)^{-1}$ \cite{Colombo:2025ihp}, that is
\begin{equation}
\label{weights_VS_Chern}
	c_1^\bT(\cL_{a-1})\,\big|_{L_a}=\frac1{2\pi}\frac{(\xi,X,V^a)}{(\xi,V^{a-1},V^a)}\:,\qquad
	c_1^\bT(\cL_a)\,\big|_{L_a}=-\frac1{2\pi}\frac{(\xi,X,V^{a-1})}{(\xi,V^{a-1},V^a)}\:,
\end{equation}
and $c_1^\bT(\cL_b)\,|_{L_a}=0$ for $b\ne a-1,a$. The restrictions of $\iota_X\wt\eta^I$ can be found from the above and
\eqref{eta_tilde_explicit_def} together with \eqref{Chern_class}. The localization formula \eqref{localization_formula} then gives
\begin{align}
\label{BH_localization_2}
	{}&\frac16C_{IJK}\bigg[\int_{M}\eta^I\wedge\Phi^{JK}-\int_{\text{AdS}}\eta^I\wedge\Phi^{JK}\bigg]=\\\nonumber
	{}&=-\frac{(2\pi)^3}6C_{IJK}\sum_{a=0}^2
		\frac{\left(\iota_\xi\eta^I\right)\big((\xi,X,V^{a-1})\,\etacoho^J_a-(\xi,X,V^a)\,\etacoho^J_{a-1}\big)\big(\text{same with }J\to K\big)}
		{(\xi,V^{a-1},V^a)(\xi,X,V^a)(\xi,V^{a-1},X)}+(\text{boundary})\:,
\end{align}
with the $a=0$ contribution coming from the AdS center and the $a=1,2$ terms coming from the North and South poles of the black hole horizon.
Again, the boundary terms cancel if the contribution of the form \eqref{alpha} is also taken into account.
When the sum over $a$ is performed explicitly the result that one obtains is the exact same as before, as expected.

In section \ref{example BH} we will revisit the localization computation of the on-shell action of Kerr-Newman black holes as a special case
of the general formula that we present in section \ref{The formula for the localized on-shell action}, keeping the gauges $\gauge^I_a$
as general as possible and showing that different gauge choices still lead to the same expression for $\cI$, modulo $2\pi\ii\,\bZ$.

\subsubsection{Simplification of the boundary integral}
\label{Simplification of the boundary integral}

In this section we show how to further simplify the boundary contribution in the localization formula \eqref{localization_formula} under the assumption that
$\Phi=\dd_X\Psi$ for a (not necessarily regular) basic polyform $\Psi$. This simplification will be useful when dealing with interface integrals
in section \ref{Localization in each patch}.
If we us assume that
\begin{equation}\label{eq:Phi_to_Psi}
	\Phi=\dd_X\Psi\:,\qquad\Psi\in\bigoplus_k\,\Omega^k\left(M\smallsetminus\bigcup_a\cD_a\right)\:,\qquad\iota_\xi\,\Psi=0\:,
\end{equation}
then we can write
\begin{equation}
	-\:\int_{\partial M}\eta\wedge\vartheta\wedge\big(\dd_X\vartheta\big)^{-1}\wedge\Phi=-\int_{\partial M}\eta\wedge\Psi
		-\sum_a\int_{\partial(\partial M\smallsetminus\cD_a)}\eta\wedge\vartheta\wedge\big(\dd_X\vartheta\big)^{-1}\wedge\Psi\:,
\end{equation}
where $\partial(\partial M\smallsetminus\cD_a)$ means that an infinitesimal tubular neighborhood of $\cD_a$ in $\partial M$ must be removed before
taking the boundary.
\footnote{Notice that if $\partial M\cap\cD_a=\emptyset$, then $\partial(\partial M\smallsetminus\cD_a)=\emptyset$ and the integral is zero.}
Let us focus on the case $D=5$, where $\Psi$ only has a three-form and a one-form components: $\Psi=\Psi_3+\Psi_1$.
Then using \eqref{Theta_def} we get
\begin{equation}
\label{dual_basis}
	-\:\int_{\partial M}\eta\wedge\vartheta\wedge\big(\dd_X\vartheta\big)^{-1}\wedge\Phi=-\int_{\partial M}\eta\wedge\Psi_3+
		\sum_a\int_{\partial(\partial M\smallsetminus\cD_a)}\eta\wedge\Theta\wedge\Psi_1\:.
\end{equation}
It is convenient to work in the basis of angular coordinates $\{\dd\varphi^a_\xi,\dd\varphi^a_X,\dd\varphi^a_V\}$ that is defined to be dual
to $\xi,X,V^a$,
\footnote{In other words, this basis is defined by requiring that
		\[\iota_A\,\dd\varphi^a_B=\begin{cases}1&\quad\text{if }A=B\\0&\quad\text{otherwise}\end{cases}\:,\qquad A,B\in\{\xi,X,V^a\}\:.\]}
for which we have
\begin{equation}
\label{dual_basis_applied}
	\dd\phi^0\wedge\dd\phi^1\wedge\dd\phi^2=\frac{\dd\varphi^a_\xi\wedge\dd\varphi^a_X\wedge\dd\varphi^a_V}{(\xi,X,V^a)}\:,\qquad
		\Theta\,\big|_{\partial(\partial M\smallsetminus\cD_a)}=\dd\varphi^a_X\:,
\end{equation}
where the value of $\Theta$ on $\partial(\partial M\smallsetminus\cD_a)$ is entirely fixed by the requirement that it is regular,
and thus $\iota_{V^a}\Theta$ vanishes on $\cD_a$, and also $\iota_\xi\Theta=0$, $\iota_X\Theta=1$. Then we can write
\begin{equation}
	\int_{\partial(\partial M\smallsetminus\cD_a)}\eta\wedge\Theta\wedge\Psi_1=
		\frac{\fks_a\,(2\pi)^3}{(\xi,X,V^a)}\:\iota_\xi\eta\cdot\iota_{V^a}\Psi_1\,\big|_{\partial M\cap\cD_a}\:,
\end{equation}
where $\fks_a$ is an orientation sign, positive if $\dd\phi^0\wedge\dd\phi^1\wedge\dd\phi^2$ has positive orientation
on $\partial(\partial M\smallsetminus\cD_a)$, and we have also used the fact that $\eta$ is a regular form an thus 
$\iota_{V^a}\eta|_{\partial(\partial M\smallsetminus\cD_a)}=0$.
We finally arrive at the following expression for the boundary integral:
\begin{equation}
\label{bdy_rewritten}
	-\:\int_{\partial M}\eta\wedge\vartheta\wedge\big(\dd_X\vartheta\big)^{-1}\wedge\Phi=-\int_{\partial M}\eta\wedge\Psi_3
		+\sum_a\frac{\fks_a\,(2\pi)^3}{(\xi,X,V^a)}\:\iota_\xi\eta\cdot\iota_{V^a}\Psi_1\,\big|_{\partial M\cap\cD_a}\:.
\end{equation}

\subsection{Localization in each patch}
\label{Localization in each patch}

In this section we apply the localization formula \eqref{localization_formula} to the bulk integrals over the patches $U_{(a)}$ in the on-shell action
\eqref{OSA_initial}. The one-form $\eta^I_{(a)}$ will take the place of $\eta$ in \eqref{localization_formula}, and we need to construct
an equivariantly-closed polyform $\Phi^{JK}$ with four-form component $\dd\eta^J\wedge\dd\eta^K$;
the choice $\Phi^{JK}=\dd_X\wt\eta^J\wedge\dd_X\wt\eta^K$ that we described in \eqref{BH_second_polyform} is the most convenient.
We will then apply the localization formula \eqref{localization_formula} with the identifications
\begin{equation}
	M\:\longrightarrow\:U_{(a)}\:,\qquad\eta\:\longrightarrow\:\eta^I_{(a)}\,\equiv\,\wt\eta^I+\gauge^I_a\cdot\dd\phi\:,
		\qquad\Phi\:\longrightarrow\:\Phi^{JK}=\dd_X\wt\eta^J\wedge\dd_X\wt\eta^K\:. 
\end{equation}
The patch $U_{(0)}$ covers the boundary $\partial M$ and it is thus special among the various patches; let us first focus on the patch $U_{(a)}$ with $a\ne0$,
for which we find
\begin{equation}
\begin{aligned}
\label{patch_integral}
	\frac16C_{IJK}\int_{U_{(a)}}\eta_{(a)}^I\wedge\Phi^{JK}\:=\:\,&
		\cI_a-\frac16C_{IJK}\int_{\partial U_{(a)}}\eta^I_{(a)}\wedge\wt\eta^J\wedge\dd\eta^K+\\
	&-\frac16C_{IJK}\sum_{b\,=\,a-1,\,a}\frac{\fks_b\,(2\pi)^3\,(\gauge^I_a\cdot\xi)}{(\xi,X,V^b)}\:
		\cdot\iota_{V^b}\wt\eta^J\cdot\iota_X\wt\eta^K\,\big|_{\partial U_{(a)}\cap\cD_b}\:,
\end{aligned}
\end{equation}
where we have also used \eqref{bdy_rewritten} with $\Psi=\wt\eta^J\wedge\dd_X\wt\eta^K$,
the orientations signs are $\fks_{a-1}=-\mathfrak s_a=+1$, $\fks_a=\mathfrak s_a=-1$, and $\cI_a$ is the contribution at the locus $L_a$:
\begin{equation}
\begin{aligned}
	\cI_a=\:&\frac16C_{IJK}\:\frac{(2\pi)^2\:\Phi_0^{JK}\big|_{L_a}\cdot\int_{L_a}\eta^I_{(a)}}
		{\,{\displaystyle\prod}_{\,i=1}^{\,\frac{D-1}2}\!\left(\frac{X_a^0}{\xi_a^0}\:\xi_a^i-X_a^i\right)}=\\[1mm]
	=\:&\frac{(2\pi)^3}6C_{IJK}\:(\gauge^I_a\cdot\xi)\:\frac{\Big((\xi,X,V^{a-1})\,\etacoho^J_a-(\xi,X,V^a)\,\etacoho^J_{a-1}\Big)
			\Big(\text{same with }J\to K\Big)}{(\xi,V^{a-1},V^a)(\xi,X,V^a)(\xi,X,V^{a-1})}\:.
\end{aligned}
\end{equation}
In the above expression we have used \eqref{weights}, \eqref{inteta}, and \eqref{ddeta_cohomology} together with \eqref{weights_VS_Chern}.

Let us simplify the second line of \eqref{patch_integral}; first, we have that
$\iota_{V^b}\wt\eta^J\,\big|_{\cD_b}=-\etacoho^J_b$ by \eqref{eta_tilde_def}. 
Then using \eqref{determinant_product_formula_origin}%
\footnote{Specifically, \eqref{determinant_product_formula_origin} implies that
		$X=(\xi,V^{a-1},V^a)^{-1}\big[(X,V^{a-1},V^a)\,\xi+(\xi,X,V^a)\,V^{a-1}-(\xi,X,V^{a-1})\,V^a\big]$.}
and \eqref{eta_tilde_def} we can write
\begin{equation}
	\iota_X\wt\eta^K=(\xi,V^{a-1},V^a)^{-1}\Big[(\xi,X,V^a)\,\iota_{V^{a-1}}\,\wt\eta^K-(\xi,X,V^{a-1})\,\iota_{V^a}\,\wt\eta^K\Big]\:,
\end{equation}
so that
\begin{equation}
\begin{aligned}
	\,&\frac16C_{IJK}\sum_{b\,=\,a-1,\,a}\frac{\fks_b\,(2\pi)^3\,(\gauge^I_a\cdot\xi)}{(\xi,X,V^b)}\:
		\etacoho^J_b\cdot\iota_X\wt\eta^K\,\big|_{\partial U_{(a)}\cap\cD_b}=\\
	\,&\qquad=-\frac{(2\pi)^3}6\,\frac{C_{IJK}\:(\gauge^I_a\cdot\xi)}{(\xi,V^{a-1},V^a)}\bigg[\etacoho_{a-1}^J
		\left(\iota_{V^a}\wt\eta^K\,\big|_{\partial U_{(a)}\cap\cD_{a-1}}+\frac{(\xi,X,V^a)}{(\xi,X,V^{a-1})}\etacoho^K_{a-1}\right)+\\
	\,&\qquad\qquad\qquad\qquad\qquad\qquad\quad+\etacoho_a^J
		\left(\iota_{V^{a-1}}\wt\eta^K\,\big|_{\partial U_{(a)}\cap\cD_a}+\frac{(\xi,X,V^{a-1})}{(\xi,X,V^a)}\etacoho^K_a\right)\bigg]\:.
\end{aligned}
\end{equation}
Combining this expression with \eqref{patch_integral}, the dependence on the arbitrary vector $X$ cancels out,%
\footnote{\label{orientation_signs}This cancellation must happen for consistency.
		Notice that this can be used to argue that all orientation signs $\mathfrak s_a$ are equal:
		this cancellation in the patch $U_{(a)}$ occurs because of $\fks_{a-1}=-\mathfrak s_a$ and $\fks_a=\mathfrak s_a$; in the patch
		$U_{(a+1)}$ one would need $\wt\fks_{a}=-\mathfrak s_{a+1}$ and $\wt\fks_{a+1}=\mathfrak s_{a+1}$, where we have denoted with
		$\wt\fks_b$ the orientation signs for $U_{(a+1)}$ analogous to the $\sigma_a$ for $U_{(a)}$. We must have $\sigma_a=-\wt\sigma_a$
		since $\partial U_{(a)}=U_{(a,a+1)}+\ldots$ and $\partial U_{(a+1)}=-U_{(a,a+1)}+\ldots$, which implies 
		$\mathfrak s_{a+1}=-\wt\fks_a=\fks_a=\mathfrak s_a$, and thus all the $\fks_a$ are equal and we can take them to be $-1$.}
and we are left with the following expression
for the bulk integral over the patch $U_{(a)}$ with $a\ne0$:
\begin{equation}
\begin{aligned}
\label{patch_integral_result}
	\,&\frac16C_{IJK}\int_{U_{(a)}}\eta_{(a)}^I\wedge\Phi^{JK}\:=\:-\frac{(2\pi)^3}3C_{IJK}\:\frac{(\gauge^I_a\cdot\xi)\,\etacoho^J_{a-1}\,\etacoho^K_a}
		{(\xi,V^{a-1},V^a)}-\frac16C_{IJK}\int_{\partial U_{(a)}}\eta^I_{(a)}\wedge\wt\eta^J\wedge\dd\eta^K-\\
	\,&\qquad\qquad-\frac{(2\pi)^3}6\,\frac{C_{IJK}\:(\gauge^I_a\cdot\xi)}{(\xi,V^{a-1},V^a)}\left(
		\etacoho_{a-1}^J\,\iota_{V^a}\wt\eta^K\,\big|_{\partial U_{(a)}\cap\cD_{a-1}}
		+\etacoho_a^J\,\iota_{V^{a-1}}\wt\eta^K\,\big|_{\partial U_{(a)}\cap\cD_a}\right)\:.
\end{aligned}
\end{equation}

\subsubsection{The boundary patch}

The boundary patch $U_{(0)}$ is special and must be treated separately. This patch does not contain any localization locus $L_a$,
so the contribution of the integral is purely made out of boundary terms.
The boundary of this patch has two connected components, one being the boundary of $M$
and the other is comprised of two patch interfaces: $\partial U_{(0)}=\partial M+U_{(0,1)}-U_{(d,0)}$.
For each one of these boundary contribution we can apply the same formulas that we used for the patch $U_{(a)}$ with $a\ne0$;
mutatis mutandis we have
\begin{equation}
\begin{aligned}
\label{bdy_patch_integral_result}
	\,&\frac16C_{IJK}\int_{U_{(0)}}\eta_{(0)}^I\wedge\Phi^{JK}\:=\:
		-\frac16C_{IJK}\int_{\partial M+U_{(0,1)}-U_{(d,0)}}\eta^I_{(0)}\wedge\wt\eta^J\wedge\dd\eta^K-\\
	\,&\qquad\qquad-\frac{(2\pi)^3}6\,\frac{C_{IJK}\:(\gauge^I_0\cdot\xi)}{(\xi,V^d,V^0)}\left(
		\etacoho_d^J\:\iota_{V^0}\,\wt\eta^K\,\big|_{U_{(d,0)}\cap\cD_d}
		+\etacoho_0^J\:\iota_{V^d}\,\wt\eta^K\,\big|_{U_{(0,1)}\cap\cD_0}\right)+\\
	\,&\qquad\qquad+\frac{(2\pi)^3}6\,\frac{C_{IJK}\:(\gauge^I_0\cdot\xi)}{(\xi,V^d,V^0)}\left(
		\etacoho_d^J\:\iota_{V^0}\,\wt\eta^K\,\big|_{\partial M\cap\cD_d}
		+\etacoho_0^J\:\iota_{V^d}\,\wt\eta^K\,\big|_{\partial M\cap\cD_0}\right)\:.
\end{aligned}
\end{equation}

\subsubsection{Simplifying the interface integrals}

When $a\ne0$ we have $\partial U_{(a)}=U_{(a,a+1)}-U_{(a-1,a)}$, and all the integrals over $\partial U_{(a)}$ from \eqref{patch_integral_result}
together with the analogous pieces in \eqref{bdy_patch_integral_result} combine in the following interface integrals:
\begin{equation}
	\frac16C_{IJK}\sum_{a=0}^d\int_{U_{(a,a+1)}}\!\!\!\!\!\!\!\!\!\!\dd\Lambda^I_{a,a+1}\wedge\wt\eta^J\wedge\dd\eta^K\:.
\end{equation}
We can combine the above with the interface integrals from \eqref{OSA_initial}:
\begin{align}
\label{interface_simplification}\nonumber
	\,&\frac16C_{IJK}\sum_{a=0}^d\int_{U_{(a,a+1)}}\!\!\!\!\!\!\!\!\!\!\dd\Lambda_{a,a+1}^I\wedge\Big(
		\scal^Je^0\wedge\dd\left(\scal^Ke^0-2\eta^K\right)-A_{(a)}^J\wedge F^K+\wt\eta^J\wedge\dd\eta^K\Big)=\\\nonumber
	&\qquad=\frac16C_{IJK}\sum_{a=0}^d\int_{U_{(a,a+1)}}\!\!\!\!\!\!\!\!\!\!\dd\Lambda_{a,a+1}^I\wedge
		\dd\left(\scal^Je^0\wedge\wt\eta^K-\gauge^J_{a}\cdot\dd\phi\wedge A^K_{(a+1)}\right)=\\\nonumber
	&\qquad=\frac16C_{IJK}\sum_{a=0}^d\Bigg[\int_{U_{(0,\ldots,d)}}\!\!\!\!\!\!\!\!\!\!\dd\Lambda_{a,a+1}^I\wedge
		(\gauge^J_{a}\cdot\dd\phi)\wedge A^K_{(a+1)}+\\
	&\qquad\qquad\qquad\qquad\:\:\:+\int_{\partial(U_{(a,a+1)}\smallsetminus\cD_a)}%\!\!\!\!\!\!\!\!\!\!\!\!\!\!\!\!\!\!\!\!\!\!\!\!\!
		\dd\Lambda_{a,a+1}^I\wedge\left(\scal^Je^0\wedge\wt\eta^K-\gauge^J_{a}\cdot\dd\phi\wedge A^K_{(a+1)}\right)\Bigg]\:,
\end{align}
where in the first step we have used $A^I_{(a)}=\scal^Ie^0-\wt\eta^I-\gauge^I_{a}\cdot\dd\phi$, while in the second step we used Stokes
(being careful that $\wt\eta^K$ and $\gauge^I_{a}\cdot\dd\phi$ are not regular at $U_{(a,a+1)}\cap\cD_a$) and $\sum_{a=0}^d\dd\Lambda_{a,a+1}^I=0$.

The integral over the intersection of interfaces $U_{(0,\ldots,d)}$ in \eqref{interface_simplification} combines with the analogous term in
\eqref{OSA_initial}, giving
\begin{align}\nonumber
	\frac16C_{IJK}\int_{U_{(0,\ldots,d)}}\!\!\!\!\!\!\!\!\!\!\dd\Lambda_{a,a+1}^I\wedge(\gauge^J_0\cdot\dd\phi)\wedge A^K_{(a+1)}\,&=
	-\frac16C_{IJK}\sum_{a=0}^d\int_{U_{(0,\ldots,d)}}\!\!\!\!\!\!\!\!\!\!(\gauge^I_0\cdot\dd\phi)\wedge(\gauge^J_a\cdot\dd\phi)
		\wedge(\gauge^K_{a+1}\cdot\dd\phi)=\\
	&=-\frac{(2\pi)^3}{6}C_{IJK}\sum_{a=0}^d\left(\gauge^I_0,\gauge^J_a,\gauge^K_{a+1}\right)
\end{align}
where in the first step we have split $A^K_{(a+1)}=-\gauge^K_{a+1}\cdot\dd\phi+(a\text{-independent})$
and used again that $\sum_{a=0}^d\dd\Lambda_{a,a+1}^I=0$, while in the second step we have simply performed the integral, the result
being a determinant that we denote with our usual short-hand notation.

In order to simplify the last line of \eqref{interface_simplification} we use a basis of angular coordinates dual to $\xi,V^a,V^{a+1}$,
following a logic similar to \eqref{dual_basis} and \eqref{dual_basis_applied}, obtaining%
\footnote{Also use $0=\iota_{V^a}e^0|_{\cD_a}=\iota_{V^a}A^K_{(a+1)}|_{\cD_a}=\iota_{V^a}\dd\Lambda_{a,a+1}^I|_{\cD_a}$ by regularity, 
		$\iota_{V^a}\wt\eta^K|_{\cD_a}=-\etacoho^K_a$ by \eqref{eta_tilde_def}, $\gauge^J_a\cdot V^a=\etacoho^J_a$ by \eqref{gauge_regularity},
		$\iota_{V^{a+1}}(X^Je^0)-\iota_{V^{a+1}}A^J_{(a+1)}=\iota_{V^{a+1}}\eta^J_{(a+1)}$ by \eqref{eta_def},
		$\iota_\xi(X^Je^0)-\iota_\xi A^J_{(a+1)}=\iota_\xi\eta^J_{(a+1)}=\xi\cdot\etacoho^J_{a+1}$ by \eqref{iota_xi_eta}, and lastly
		$\iota_\xi\dd\Lambda^I_{a,a+1}=\iota_\xi(\eta^I_{(a+1)}-\eta^I_{(a)})=\xi\cdot(\gauge^I_{a+1}-\gauge^I_a)$.}
\begin{align}
	\,&\frac16C_{IJK}\int_{\partial(U_{(a,a+1)}\smallsetminus\cD_a)}\dd\Lambda_{a,a+1}^I\wedge
		\left(\scal^Je^0\wedge\wt\eta^K-\gauge^J_{a}\cdot\dd\phi\wedge A^K_{(a+1)}\right)=\\\nonumber
	&=\frac16C_{IJK}\frac{(2\pi)^3}{(\xi,V^a,V^{a+1})}\Big(
		-(\gauge^J_{a+1}\cdot\xi)\:\etacoho_a^K\:\iota_{V^{a+1}}\,\dd\Lambda^I_{a,a+1}
		+\xi\cdot(\gauge^I_{a+1}-\gauge^I_a)\,\etacoho_a^K\:\iota_{V^{a+1}}\,\eta^J_{(a+1)}\Big)\Big|_{U_{(a,a+1)}\cap\cD_a}\:.
\end{align}
It is convenient to massage the terms involving a contraction with $V^{a+1}$ by first writing, using \eqref{gauge_constants_def}
and \eqref{gauge_regularity},
\begin{equation}
	\iota_{V^{a+1}}\,\eta^J_{(a+1)}=\iota_{V^{a+1}}\,\wt\eta^J+\etacoho^J_{a+1}\:,\qquad
		\iota_{V^{a+1}}\,\dd\Lambda^I_{a,a+1}=-\gauge^I_a\cdot V^{a+1}+\etacoho^I_{a+1}\:,
\end{equation}
and then applying the relation%
\footnote{This follows from \eqref{determinant_product_formula_origin}.}
\begin{equation}
	V^{a+1}=(\xi,V^{a-1},V^a)^{-1}\Big[(V^{a-1},V^a,V^{a+1})\,\xi+(\xi,V^{a-1},V^{a+1})\,{V^a}
		-(\xi,V^a,V^{a+1})\,{V^{a-1}}\Big]
\end{equation}
to the terms $\gauge^I_a\cdot V^{a+1}$ and $(\gauge^I_{a}\cdot\xi)\,\iota_{V^{a+1}}\,\wt\eta^J$, but leaving
the term $(\gauge^I_{a+1}\cdot\xi)\,\iota_{V^{a+1}}\,\wt\eta^J$ as it is. In particular,
\begin{equation}
\begin{aligned}
	\,&\gauge^I_a\cdot V^{a+1}=(\xi,V^{a-1},V^a)^{-1}\Big[(V^{a-1},V^a,V^{a+1})\,(\gauge^I_{a}\cdot\xi)
		+(\xi,V^{a-1},V^{a+1})\,\etacoho^I_a-(\xi,V^a,V^{a+1})\,\etacoho^I_{a-1}\Big]\:,\\
	&\iota_{V^{a+1}}\,\wt\eta^J\big|_{U_{(a,a+1)}\cap\cD_a}=(\xi,V^{a-1},V^a)^{-1}
		\Big[-(\xi,V^{a-1},V^{a+1})\,\etacoho_a^J
		-(\xi,V^a,V^{a+1})\,\iota_{V^{a-1}}\,\wt\eta^J\big|_{U_{(a,a+1)}\cap\cD_a}\Big]\:.
\end{aligned}
\end{equation}
We get
\begin{align}\nonumber
	\,&\frac16C_{IJK}\int_{\partial(U_{(a,a+1)}\smallsetminus\cD_a)}\dd\Lambda_{a,a+1}^I\wedge
		\left(\scal^Je^0\wedge\wt\eta^K-\gauge^J_{a}\cdot\dd\phi\wedge A^K_{(a+1)}\right)=\\\nonumber
	&\qquad=\frac{(2\pi)^3}6\,C_{IJK}\Bigg[\frac{\xi\cdot(\gauge_a^I+\gauge_{a+1}^I)\,\etacoho^J_a\:\etacoho^K_a\:(\xi,V^{a-1},V^{a+1})}
		{(\xi,V^{a-1},V^a)(\xi,V^a,V^{a+1})}-\frac{(\gauge^I_{a+1}\cdot\xi)\:\etacoho^J_{a-1}\:\etacoho^K_a}{(\xi,V^{a-1},V^a)}-\\\nonumber
	&\qquad\qquad\qquad\qquad\quad-\frac{(\gauge^I_{a}\cdot\xi)\:\etacoho^J_a\:\etacoho^K_{a+1}}{(\xi,V^a,V^{a+1})}
		+\frac{(\gauge^I_{a}\cdot\xi)\:(\gauge^J_{a+1}\cdot\xi)\:\etacoho^K_a\:(V^{a-1},V^a,V^{a+1})}{(\xi,V^{a-1},V^a)(\xi,V^a,V^{a+1})}\Bigg]+\\
	&\qquad\quad+\frac{(2\pi)^3}6\,C_{IJK}\left(
		\frac{(\gauge^I_{a+1}\cdot\xi)\,\etacoho_a^J\,\iota_{V^{a+1}}\wt\eta^K}{(\xi,V^a,V^{a+1})}
		+\frac{(\gauge^I_{a}\cdot\xi)\,\etacoho_a^J\,\iota_{V^{a-1}}\wt\eta^K}{(\xi,V^{a-1},V^a)}\right)\Big|_{U_{(a,a+1)}\cap\cD_a}\:.
\end{align}
Notice that the terms in the last line of this expression perfectly cancel with the analogous $U_{(a,a+1)}\cap\cD_a$ terms from
\eqref{patch_integral_result} and \eqref{bdy_patch_integral_result}.
With the exception of boundary terms, we have simplified all the terms that contribute to the on-shell action down to topological quantities.

\subsection{The main result for the localized on-shell action}
\label{The formula for the localized on-shell action}

Putting all the pieces together, the on-shell action \eqref{OSA_initial} can be written as
\begin{empheq}[box={\mymath[colback=white, colframe=black]}]{equation}
\begin{aligned}
\label{OSA_localized}
	\cI=&\:\,\widehat\cI\,+\,\cB\:,\\[1mm]
	\widehat\cI=&\:\frac{\ii\pi^2}{12G}\,C_{IJK}\sum_{a=0}^d\Bigg[\xi\cdot\big(\gauge_a^I+\gauge_{a+1}^I\big)\Bigg(
		\frac{\etacoho^J_{a-1}\:\etacoho^K_a}{(\xi,V^{a-1},V^a)}+\frac{\etacoho^J_a\:\etacoho^K_{a+1}}{(\xi,V^a,V^{a+1})}-\\
	&-\frac{\etacoho^J_a\:\etacoho^K_a\:(\xi,V^{a-1},V^{a+1})}{(\xi,V^{a-1},V^a)(\xi,V^a,V^{a+1})}\Bigg)-
		\frac{(\gauge^I_{a}\cdot\xi)\:(\gauge^J_{a+1}\cdot\xi)\:\etacoho^K_a\:(V^{a-1},V^a,V^{a+1})}
		{(\xi,V^{a-1},V^a)(\xi,V^a,V^{a+1})}+\\
	&+\left(\gauge^I_0,\gauge^J_a,\gauge^K_{a+1}\right)\Bigg]\:,
\end{aligned}
\end{empheq}
where we have denoted with $\widehat\cI$ the ``bulk contribution" to the on-shell action, 
coming from the localized contributions at the loci $L_a$ and from the interfaces, and $\cB$ is the boundary contribution which is explicitly given by
\begin{equation}
\begin{aligned}
\label{bdy_contribution}
	\cB\:=\:&\,
		\int_{\partial M}\text{GHY + counterterms}-\frac1{16\pi\ii G}\,\frac16C_{IJK}\int_{\partial M}\eta^I_{(0)}\wedge\wt\eta^J\wedge\dd\eta^K-\\
	&+\frac1{16\pi\ii G}\int_{\partial M}\scal_I\,e^0\wedge*_\gamma\dd\left(f\scal^I\right)+
		\frac16C_{IJK}\scal^Ie^0\wedge\eta_{(0)}^J\wedge\dd\left(\scal^Ke^0-2\eta^K\right)+\\
	&+\frac1{16\pi\ii G}\,\frac{(2\pi)^3}6\,\frac{C_{IJK}\:(\gauge^I_{0}\cdot\xi)}{(\xi,V^d,V^0)}\left(
		\etacoho_d^J\:\iota_{V^0}\,\wt\eta^K\,\big|_{\partial M\cap\cD_d}
		+\etacoho_0^J\:\iota_{V^d}\,\wt\eta^K\,\big|_{\partial M\cap\cD_0}\right)\:.
\end{aligned}
\end{equation}
As we will show in section \ref{Boundary analysis} using Feffermann-Graham expansion, the boundary term $\cB$ depends only on the leading order asymptotic boundary data. In particular, whenever one can apply background subtraction, $\cB$ will cancel between the solution and the subtraction manifold leaving $\widehat \cI$ as final result.   

The bulk contribution $\widehat \cI$ of the localization formula depends explicitly on the Killing vector $\xi$, the parameters $\alpha^I_a$ entering in the cohomological expansion \eqref{Chern_class} of the form $\dd \eta^I$ and the gauge shifts that connect the local expressions $\eta^I_{(a)} = \nu^I_a \cdot \dd \phi +\tilde \eta^I$ in the different patches. The $\nu^I_a$ contain topological information about the solutions and they can be expressed in terms of $\alpha^I_a$ using \eqref{gauge_regularity}, up to unphysical gauge shifts and possible flat connections.  As we will see in section \ref{Topology, thermodynamics and UV-IR relations}, the $\alpha^I_a$ have a clear physical meaning, being related to the thermodynamic potentials and magnetic fluxes of the solution.

\section{Topology, thermodynamics and UV-IR relations}
\label{Topology, thermodynamics and UV-IR relations}

In this section we explain how to rewrite the various constants that appear in our main result for the localized on-shell action \eqref{OSA_localized}
in terms of more physically relevant quantities, such as magnetic fluxes, flat connections and thermodynamic potentials.
It will also be important to establish certain topological properties of the solutions 
of interest, in order to properly discuss the gauge invariance (modulo $2\pi\ii\,\bZ$) of the on-shell action.
Then we will introduce the key thermodynamical quantities and relations that we will need in section \ref{sect:examples},
where we will consider various examples of solutions and their physical properties.

\subsection{Magnetic fluxes and quantization conditions}
\label{Topological constraints on the thermodynamic potentials}

\noindent
For any two-cycle $\cC$ in the geometry, the magnetic fluxes supported by $\cC$ are subject to a quantization condition that we can write as
\begin{equation}
\label{flux_quantization}
	\frac1{2\pi\kappa}\int_{\cC}F^I\,\in\,\bZ\:,\qquad\forall\:\cC\in H_2(M,\bZ)\:,
\end{equation}
where $\kappa$ is a constant related to the convention for the normalization of the gauge fields $A^I$,%
\footnote{Here we are implicitly assumed that the same constant $\kappa$ works for all the gauge fields $A^I$, which for generic choices of
		$V_I$ and $C_{IJK}$ might not be true. However, without loss of generality it is always possible to redefine the $V_I$ and $C_{IJK}$
		appropriately so that this is no longer an issue. In the following we will always assume that these parameters have been redefined in this manner.}
chosen so that $\kappa^{-1}A^I$ defines a connection on a U(1)-principal bundle.%
\footnote{When the manifold $M$ does not admit a spin structure, a spin$^c$ structure must be used instead, and the gauge field appearing
		in the covariant derivative of the spinors, that is $\frac32V_IA^I$, is ``one-half" of a connection on a U(1)-principal bundle.
		In particular for any two-cycle $\cC$, the quantization of the fluxes of $\frac32V_IA^I$ will be determined by the integral of
		the second Stiefel-Whitney class $w_2\in H^2(M,\bZ_2)$ as follows:
		$\frac1{2\pi}\int_\cC\frac32V_IF^I=\frac12\int_\cC w_2+\text{integer}$.
		This quantization condition is perfectly compatible with the quantization \eqref{flux_quantization}, and it is already implied by the linear
		constraint \eqref{etacoho_linear_relation}, after the relation \eqref{quant} between fluxes and the $\etacoho^I_a$
		is considered. We will comment this more in footnotes \ref{footnote_spinc2} and \ref{footnote_spinc3}.\label{footnote_spinc1}}
Since  the on-shell action is only defined modulo $2\pi\ii\,\bZ$, it will be important for us to distinguish which quantities are integers and which ones are not.
The gauge invariance of $\cI$ modulo $2\pi\ii\,\bZ$ is only possible when the appropriate Chern-Simons level quantization is imposed,
which for a generic manifold is given by%
\footnote{For some special classes of five-manifolds less restrictive quantization conditions may be imposed.
		For examples, the five-manifolds that appear in M-theory after compactification on a Calabi-Yau
		are spin-manifolds with vanishing first Pontryagin class, and for these manifolds
		the right hand side of \eqref{kappa_fixing} could be relaxed to $6^{-1}\bZ$ \cite{Witten:1996qb} (see also \cite{Intriligator:1997pq}).
		In this paper we will always impose the Chern-Simons level quantization appropriate for an arbitrary manifold.}
\begin{equation}
\label{kappa_fixing}
	\frac{\pi\kappa^3}{24G}\,C_{IJK}\in\bZ\:,
\end{equation}
where we have used that the Chern-Simons terms take the form $(96\pi\ii\,G)^{-1}C_{IJK}A^I\wedge F^J\wedge F^K$ and that the
gauge fields are $\kappa$ times a connection on a U(1)-principal bundle.

For supergravity models that are consistent truncations of type IIB supergravity in ten dimensions, the five-dimensional gauge fields $A^I$
are generated from the Kaluza-Klein ansatz, and thus \eqref{flux_quantization} and \eqref{kappa_fixing} should be equivalent to
conditions of regularity of the ten-dimensional metric,  with the value of $\kappa$ fixed by the specific choice of
uplift manifold. In this paper however we will keep a five-dimensional perspective and will not discuss the 10d uplift any further.

In the rest of this section we will first explore the quantization conditions on the gauge shifts $\dd\Lambda^I_{a,{a+1}}$, and argue that
they imply that certain integer linear combinations of the cohomological parameters $\etacoho^I_a$ defined in \eqref{ddeta_cohomology} are quantized.
Then we will show that these quantized integer linear combinations of the $\etacoho^I_a$ are exactly the magnetic fluxes
\eqref{flux_quantization} over the two-cycles of the geometry.

\subsubsection{Quantization of the gauge shifts and magnetic fluxes}
\label{Quantization of the gauge shifts and magnetic fluxes}

The possible gauge transformations of a U(1)-connection $A$ on a circle $S^1$ are parametrized by an integer:
\begin{equation}
	A'=A+n\,\dd\phi\:,\qquad n\in\bZ,\quad\phi\sim\phi+2\pi\:\text{ coordinate on }S^1.
\end{equation}
Given that the gauge shifts $\dd\Lambda^I_{a,{a+1}}=(\gauge^I_{a+1}\cdot\dd\phi)-(\gauge^I_a\cdot\dd\phi)$ are gauge transformations on a three-torus,
and the gauge fields $A^I$ differ from U(1)-connections by a factor of $\kappa$,
we have the following quantization conditions:
\begin{equation}
\label{gauge_shift_quantization}
	\left(\gauge^I_{a+1}\right)_i-\left(\gauge^I_{a}\right)_i\in\kappa\,\bZ\qquad\forall\:a=0,\ldots,d\:,\quad i=0,1,2\:.
\end{equation}
Let us keep $\kappa$ generic and explore the consequences of the above gauge shift quantization. Imposing \eqref{gauge_shift_quantization} leads to constraints among the cohomological parameters $\etacoho_a^I$, specifically there are linear combinations of the $\etacoho_a^I$ with integer coefficients
 that must belong in $\kappa\bZ$.
We will later argue that these linear combinations correspond to the magnetic fluxes supported by the two-cycles of the solution.

In order to derive the quantization conditions on the $\etacoho_a^I$,
let us define the ``GLSM charges" $Q^a_m\in\bZ$ such that the vectors $(Q^a_m)_{a=0}^d$
form a basis of the space of coefficients of vanishing integer linear combinations of the $V^a$.%
\footnote{Such a basis must exist because the space of coefficients of vanishing integer linear combinations of the $V^a$ is a torsion-less
finitely generated abelian group.}
In other words, the $Q^a_m\in\bZ$ satisfy
\begin{align}
	{}&\sum_{a=0}^dQ^a_m\,V^a=0\:,\\
\label{Qma_basis_condition}
	{}&\sum_{a=0}^d\gamma_a\,V^a=0\quad\text{for some }\gamma_a\in\bZ
		\quad\implies\quad\gamma_a=\sum_m\wt\gamma_m\,Q^a_m\quad\text{for some unique }\wt\gamma_m\in\bZ\:,
\end{align}
and the index $m$ either takes 
a total of $d-2$ possible values, or $d-1$ values;
the second case corresponds to solutions whose $V^a$ all lie in the same $\bZ^2$ sublattice, and thus the geometry contains a circle subgroup
of the U(1)$^3$ isometry that never shrinks.

We can thus derive the quantization conditions on $\etacoho_a^I$ from \eqref{gauge_shift_quantization} by observing that it is equivalent to
\begin {equation}
	\kappa\,\bZ\:\ni\:V^b\cdot\left(\gauge_b^I-\gauge_a^I\right)\:=\:-\gauge_a^I\cdot V^b+\etacoho^I_b\:,\qquad\forall\:a,b\:,
\end{equation}
where we have used \eqref{gauge_regularity}.
Contracting with $Q_m^b$ we find that the following linear combinations of the $\etacoho^I_a$ are quantized:
\begin{equation}\label{etacoho_quantization}
	\sum_{a=0}^dQ_m^a\,\etacoho_a^I\:\in\:\kappa\,\bZ\:.
\end{equation}
The above quantization conditions have the geometric interpretation of quantization of the magnetic fluxes of $F^I$ along the nontrivial two-cycles
of the topology of the solutions, as we will soon discuss in section \ref{Two-cycles and quantization of the fluxes}.

\subsubsection{Two-cycles and quantization of the magnetic fluxes}
\label{Two-cycles and quantization of the fluxes}

Let us describe the two-cycles of the geometry.
Given two pairs of consecutive fan vectors $V^{a-1}$, $V^a$ and $V^{b-1}$, $V^b$ there will always be a linear relation among them as long as they are
all distinct, since they are three-component vectors, and it is possible to pick the coefficients so that they are all integers:
\begin{equation}
\label{linear_combination_for_cycle}
	\gamma_{a-1}V^{a-1}+\gamma_aV^a+\gamma_{b-1}V^{b-1}+\gamma_bV^b=0\:,\qquad\gamma_{a-1},\gamma_a,\gamma_{b-1},\gamma_b\in\bZ\:.
\end{equation}
Notice that due to \eqref{Qma_basis_condition}, such linear combinations are encoded into the $Q_a^m$.
We can then construct a nontrivial two-cycle as follows: on the toric diagram of the base of $M$ the projection of this two-cycle is a line between $L_a$
and $L_b$. The two-cycle is constructed as a $S^1$ fibration over this line, with the $S^1$ degenerating at the two extrema $L_a$ and $L_b$.
The angular coordinate of the two-cycle is chosen so that the vector $\gamma_{a-1}V^{a-1}+\gamma_aV^a$ is tangent to the two-cycle,
or equivalently the $S^1$ fibers are orbits of the Killing vector $\gamma_{a-1}V^{a-1}+\gamma_aV^a$, which defines a U(1) subgroup
of the isometry of $M$, since $\gamma_{a-1},\gamma_a$ are integers. Now $V^{a-1}$ and $V^a$ vanish at $L_a$ and similarly $V^{b-1}$ and $V^b$ vanish at $L_b$.
Since $\gamma_{a-1}V^{a-1}+\gamma_aV^a= -\gamma_{b-1}V^{b-1}-\gamma_bV^b$,
this Killing vector by construction vanishes at both $L_a$ and $L_b$, which is the reason why its orbits degenerate to a point at the extrema $L_a$ and $L_b$.
Notice that this two-cycle is not invariant under the U(1)$^3$ isometry of $M$, and thus it is ``non-toric".
Nonetheless, the integral of the field strength $F^I$ on this two-cycle will be quantized according to \eqref{flux_quantization}.
Calling this two-cycle $\cC_{ab}$, we have
\begin{equation}\label{quant}
\begin{aligned}
	2\pi\kappa\,\bZ\:\ni\:\int_{\cC_{ab}}F^I\:&=-\int_{\cC_{ab}}\dd\eta^I=-2\pi\,\iota_{(\gamma_{a-1}V^{a-1}+\gamma_aV^a)}\left(
		\wt\eta^I\big|_{L_a}-\wt\eta^I\big|_{L_b}\right)=\\
	&=-2\pi\left(\iota_{(\gamma_{a-1}V^{a-1}+\gamma_aV^a)}
		\wt\eta^I\big|_{L_a}+\iota_{(\gamma_{b-1}V^{b-1}+\gamma_bV^b)}\wt\eta^I\big|_{L_b}\right)=\\[2mm]
	&=2\pi\Big(\gamma_{a-1}\etacoho^I_{a-1}+\gamma_a\etacoho^I_a+\gamma_{b-1}\etacoho^I_{b-1}+\gamma_b\etacoho^I_b\Big)\:,
\end{aligned}
\end{equation}
where we used \eqref{eta_def} and the regularity of $X^I e_0$, and \eqref{gauge_regularity}.
We observe that more generally if $\gamma_b=0$, we can move the endpoint of the two-cycle from $L_b$ to $\cD_{b-1}$,
and similarly if $\gamma_{a-1}=0$.
Generically each vanishing linear combination of $V^a$ can be written as a linear combination of linear combinations of the form
\eqref{linear_combination_for_cycle}, thus the $\kappa^{-1}\sum_{a=0}^dQ^m_a\,\etacoho_a^I\in\bZ$
are magnetic fluxes associated to some nontrivial two-cycle, giving a geometric interpretation to \eqref{etacoho_quantization}.

\subsection{Flat connections and gauge invariance}
\label{Flat connections and gauge invariance}

\noindent
In this section we provide a parametrization for the constants $\left(\gauge^I_a\right)_i$ that specify the gauge in each patch,
and discuss the gauge invariance of the on-shell action. It will be important for us that the on-shell action is  defined only modulo $2\pi\ii\,\mathbb{Z}$.
The $\gauge^I_a$  contain unphysical gauge-choice parameters, whose overall contribution to $\cI$ vanishes (modulo $2\pi\ii\,\bZ$),
as well as a physical parameter $\gaugereal^I$ associated to the possible flat connections.
We discuss and provide examples of geometries with a nontrivial fundamental group, which can support nonzero flat connections, either discrete or continuous.
These include solitonic solutions, geometries  whose boundary contains a lens space,
and general black saddles associated with Euclidean black holes.

The fundamental group of a (smooth) solution $M$ with U(1)$^3$ isometry and topology specified by the vectors $\{V^a\}_{a=0}^d$ 
can be found as
\begin{equation}
\label{fundamental_group}
	\pi_1(M)\,\cong\,\bZ^3/\text{Span}_{\,\bZ}\big(\{V^a\}_{a=0}^d\big)\:.
\end{equation}
Indeed, $M$ is a U(1)$^3$ fibration over a simply connected polytope, so each loop in $M$ is homotopic to a loop around the U(1)$^3$ fiber over a point $p$
of the polytope, and we are free to continuously move $p$ around the polytope.
Since $\pi_1($U(1)$^3)\cong\bZ^3$, each one of these loops can be specified by three integers $n_i$.
Each $V^a$ is a Killing vector vanishing at the locus $\cD_a$, where the fibers degenerate to U(1)$^2$ due to the presence of a U(1) isotropy subgroup
of U(1)$^3$ given by $\phi_i\,\longrightarrow\,\phi_i+V_i^a\,\varphi$, with $\varphi\in\bR/2\pi\bZ$.
If we move the base point $p$ of the loop inside $\cD_a$, we obtain the identification $n_i\cong n_i+V_i^a\,n$ amongst the integers $n_i$ that specify the loop,
for any $n\in\bZ$. Repeating this procedure for each $V^a$ leads to \eqref{fundamental_group}.

The space of possible flat connections on $G$-principal bundles over $M$ is classified by group homomorphisms
\begin{equation}
	\pi_1(M)\:\longrightarrow\:G 
\end{equation}
that specify the holonomy of the connection along each loop, since flat connections have homotopy-invariant parallel transport.
In particular, the possible flat connections of the gauge field $A^I$ correspond to homomorphism of the fundamental group \eqref{fundamental_group}
into $G=\,$U(1).

There are three possible cases to consider:
\begin{itemize}
\item $\pi_1(M)$ is trivial. In this case there cannot be any flat connections.
	We will discuss solutions with trivial fundamental group in section \ref{Solutions without flat connections}.
\item $\pi_1(M)$ is a non-trivial finite group (called the \emph{torsion} group). Then the geometry can support
	flat connections, which can be described by a parameter that takes a discrete number of possible values.
	We will not do a systematic analysis of these geometries with torsion,
	but we will consider a couple of examples of such solutions in section \ref{An example of a solution with torsion}.
\item $\pi_1(M)\,\cong\,\bZ$.\footnote{The case $\pi_1(M)\,\cong\,(\bZ\times\text{nontrivial finite group})$ is not possible:
	if $\pi_1(M)$ contains a $\bZ$ factor, then all $V^a$ can be put in the form $V^a=(0,v^a)$ for $v^a\in\bZ^2$ with an SL(3,$\bZ$) transformation.
	Then smoothness implies that $|\det(v^{a-1},v^a)|=1$ for all $a$, which leads to 
	$\pi_1(M)\,\cong\,\bZ^3/\text{Span}_{\,\bZ}\big(\{V^a\}_{a=0}^d\big)\,\cong\,\bZ$. }
	We call these geometries \emph{solitonic solutions}, and they are of the form $M\,\cong\,S^1\times M_4\:$, with a simply connected four-dimensional
	$M_4$. We will discuss solitonic solutions in section \ref{Solitonic solutions}.
\end{itemize}
Generically flat connections appear in the formula for the localized on-shell action \eqref{OSA_localized}
through the gauge constants $\left(\gauge^I_a\right)_i$. After the regularity condition \eqref{gauge_regularity} of the gauge field is imposed,
each triplet of constants $\gauge^I_a$ is determined up to a single arbitrary gauge-choice
constant; if the quantization of the gauge shifts \eqref{gauge_shift_quantization} is also imposed, then this arbitrary gauge-choice constant
can be written in terms of an unphysical arbitrary integer parameter $\gaugeint^I_a\in\bZ$, and a physical non-integer parameter $\gaugereal^I$
describing the flat connection component of the gauge field. 
Then the gauge invariance of the on-shell action $\widehat\cI$ (modulo $2\pi\ii\,\bZ$)
is equivalent to the statement that the dependence on the arbitrary gauge choice integers $\gaugeint^I_a$ can be factored out as follows:
\begin{equation}
\label{gauge_invariance_formula}
	\widehat\cI\:=\:\widehat\cI\big|_{\gaugeint^I_a=0}\,+\,\frac{\ii\pi^2\kappa^3}{12G}\,C_{IJK}\,\big(\text{integers}\big)\:,
\end{equation}
where we assume \eqref{kappa_fixing}.
In appendix \ref{Algebraic proof of gauge invariace} we provide a proof of the above statement, focusing on a large class of solutions whose boundary geometry is homeomorphic to $S^1\times S^3$. %, and also assuming that $\pi_1(M)$ is torsion-less.

\subsubsection{Solutions without flat connections}
\label{Solutions without flat connections}

Let us first consider solutions with trivial fundamental group, which cannot have any flat connections. 
Because of \eqref{fundamental_group}, a smooth solution with U(1)$^3$ isometry will have a trivial
fundamental group if and only if the vectors of the fan $V^a$ span the lattice $\bZ^3$. 
For simplicity, let us assume that there are three vectors of the fan $\{V^{a_1},V^{a_2},V^{a_3}\}$ that form a basis of the lattice $\bZ^3$,
which we order so that $(V^{a_1},V^{a_2},V^{a_3})=1$;
then for any $a$ there exists $\gamma^a_i\in\bZ$ such that
\begin{equation}
	\sum_{i=1,2,3}\gamma_i^a\,V^{a_i}=V^a\:. 
\end{equation}
Using \eqref{Qma_basis_condition} and \eqref{etacoho_quantization}, there must exist
$\flux^I_a\in\bZ$ such that 
\begin{equation}
\label{flux_def}
	\sum_{i=1,2,3}\gamma_i^a\,\etacoho^I_{a_i}=\etacoho^I_a+\kappa\,\flux^I_a\:.
\end{equation}
The $\flux^I_a$ have the interpretation of magnetic fluxes by \eqref{quant} and the discussion below.%
\footnote{In terms of the charges $Q^a_m\in\bZ$ defined in section \ref{Quantization of the gauge shifts and magnetic fluxes}, we have
		$\flux^I_m=\kappa^{-1}\sum_{a=0}^dQ^m_a\etacoho^I_a$, with $m\in\{0,\ldots,d\}\smallsetminus\{a_1,a_2,a_3\}$ 
		and $Q_m^a=\sum_{i=1}^3\gamma^a_i\,\delta^a_{a_i}-\delta^a_m$. Then by definition $\fn_0^I=\fn_d^I=\fn_{\wb a}^I=0$.}
Explicitly, the coefficients $\gamma_a^I$ are given by
\begin{equation}
	\gamma^a_i=(V^a,V^{a_{i+1}},V^{a_{i+2}})\:,
\end{equation}
where $a_{i+1}$ and $a_{i+2}$ for $i>1$ are defined cyclically.
We can use this to find the most general parametrization for the gauge choice $\gauge_a^I$ in each patch.
The regularity condition \eqref{gauge_regularity} for the guage fields requires that $\gauge_a^I\cdot V^a=\etacoho^I_a$ and
$\gauge_a^I\cdot V^{a-1}=\etacoho^I_{a-1}$. We can then solve this constraint in full generality by setting
\begin{equation}
\label{gauge_parametrization1}
	\left(\gauge^I_a\right)_i=\!\!\sum_{i=1,2,3}(e_i,V^{a_{i+1}},V^{a_{1+2}})\,\etacoho^I_{a_i}-\kappa\Big[(e_i,V^a,w^a)\,\flux^I_{a-1}-
		(e_i,V^{a-1},w^a)\,\flux^I_a+(e_i,V^{a-1},V^a)\,\gaugeint^I_a\Big]\:,
\end{equation}
where $(e_i)_j\equiv\delta_{ij}$ denote the standard basis of $\bZ$, $\gaugeint^I_a$ is an arbitrary gauge parameter and
$w^a$ is a choice of lattice vector that satisfies $(V^{a-1},V^a,w^a)=1$, which must exist for smooth geometries,
as discussed in section \ref{Geometry of solutions}.%
\footnote{In the case of asymptotically $S^1\times L(p,q)$ solutions the vector $w^0$ satisfying the require property $(V^{d},V^0,w^0)=1$
		does not exists, meaning that for these solutions a special parametrization choice for the boundary patch is required.}
The advantage of this seemingly overly-complicated parametrization is that now the quantization condition for the gauge shifts \eqref{gauge_shift_quantization}
is simply solved by
\begin{equation}
	\gaugeint^I_a\in\bZ\qquad\forall a\:.
\end{equation}
We observe that it is always possible to change the choice of lattice vector $w^a\to w^a+n_1V^{a-1}+n_2V^a$ where $n_{1,2}\in\bZ$ as long as one redefines $\gaugeint^I_a$ appropriately.

Let us focus our attention on solutions whose boundary topology is $S^1\times S^3$,  and whose fan contains a vector $V^{\wb a}$ such that $\{V^d,V^0,V^{\wb a }\}$ span the lattice $\bZ^3$.
It is then possible to set $a_1=d$, $a_2=0$, $a_3=\wb a$, % for some $\wb a$
and work in the gauge $\etacoho^I_0=0=\etacoho^I_d$. Then we have
\begin{equation}
\label{gauge_parametrization2}
	\left(\gauge^I_a\right)_i=(e_i,V^d,V^0)\,\etacoho^I_{\wb a}-\kappa\Big[(e_i,V^a,w^a)\,\flux^I_{a-1}-
		(e_i,V^{a-1},w^a)\,\flux^I_a+(e_i,V^{a-1},V^a)\,\gaugeint^I_a\Big]\:,
\end{equation}
and the general $\etacoho^I_a$ are found as
\begin{equation}
\label{special_quantization}
	\etacoho^I_a=(V^a,V^d,V^0)\,\etacoho_{\wb a}^I-\kappa\,\flux^I_a\:,
\end{equation}
with $\fn_0^I=\fn_d^I=\fn_{\wb a}^I=0$ by definition.
It is also useful to note that the linear relations \eqref{etacoho_linear_relation} for $a$ and $\bar a$, together with \eqref{special_quantization}
and the algebraic identity \eqref{determinant_product_formula}, implies that the fluxes $\flux^I_a$ are subject to the constraint%
\footnote{\label{footnote_spinc2}As already commented in footnote \ref{footnote_spinc1}, the spin$^c$ quantization condition implies that
		$3\kappa\,V_I\,\flux^I_a\mod2=\frac1{2\pi}\int_{\cC_a}3\kappa\,V_I\,F^I_a\mod2=\int_{\cC_a}w_2\mod2$,
		where we are denoting with $\cC_a$ the two-cycle that supports the flux $\flux^I_a$ and $w_2$ is the second Stiefel-Whitney class.
		Comparison with \eqref{flux_constraint_wba} suggests that
		$\int_{\cC_a}w_2=1-(V^a,V^0,V^{\wb a})-(V^a,V^{\wb a},V^d)-(V^a,V^d,V^0)\mod2$.
		We are able to verify that such a relation is indeed true for a simple class of geometries; see footnote \ref{footnote_spinc3}.}
\begin{equation}
\label{flux_constraint_wba}
	3\kappa\,V_I\,\flux^I_a\,=\,1-(V^a,V^0,V^{\wb a})-(V^a,V^{\wb a},V^d)-(V^a,V^d,V^0)\:.
\end{equation}
In appendix \ref{Algebraic proof of gauge invariace} we use the parametrization \eqref{gauge_parametrization2} to provide an algebraic proof that
the on-shell action of these solutions is gauge invariant, in the sense that setting the gauge choice parameters $\gaugeint^I_a$ to zero
only changes the value of the on-shell action by a physically irrelevant $2\pi\ii\,\bZ$ term.

\subsubsection{Solitonic solutions}
\label{Solitonic solutions}

Let us consider the case of (smooth) solutions with fundamental group $\pi_1(M)\cong\bZ$, which are of the form $M\,\cong\,S^1\times M_4$,
with a simply connected four-dimensional $M_4$.
In light of \eqref{fundamental_group}, there must exist a vector $W\in\bZ^3$ and two vectors of the fan $V^{a_1},V^{a_2}$
such that $(V^{a_1},V^{a_2},W)=1$ and  any other vector $V^a$ of the fan is a linear combination of $V^{a_1},V^{a_2}$.
Specifically,
\begin{equation}
	V^a=(V^a,V^{a_2},W)\,V^{a_1}+(V^a,W,V^{a_1})\,V^{a_2}\:.
\end{equation}
Similarly to \eqref{flux_def}, we can then define the magnetic fluxes $\flux^I_a\in\bZ$ such that
\begin{equation}
	(V^a,V^{a_2},W)\,\etacoho^I_{a_1}+(V^a,W,V^{a_1})\,\etacoho^I_{a_2}=\etacoho^I_a+\kappa\,\flux^I_a\:,
\end{equation}
and similarly to \eqref{gauge_parametrization1} we can generically write the gauges as 
\begin{equation}
\begin{aligned}
\label{gauge_parametrization_soliton1}
	\left(\gauge^I_a\right)_i=\,&(e_i,V^{a_2},W)\,\etacoho^I_{a_1}+(e_i,W,V^{a_1})\,\etacoho^I_{a_2}+\\
	&+h_a\,\kappa\Big[(e_i,V^a,W)\,\flux^I_{a-1}-(e_i,V^{a-1},W)\,\flux^I_a+(e_i,V^{a-1},V^a)\,\gaugereal^I_a\Big]\:.
\end{aligned}
\end{equation}
Indeed, imposing smoothness of the solution, for each pair of consecutive fan vectors $V^{a-1},V^{a}$
the determinant $(V^{a-1},V^{a},W)$ must be equal to a sign%
\footnote{Again, for asymptotically $S^1\times L(p,q)$ solutions the boundary patch requires special treatment since $|(V^d,V^0,W)|=p$.}
that we denote as $-h_a$.%
\footnote{Here we are being consistent with the typical conventions used in asymptotically-flat
five-dimensional ungauged supergravity papers such as \cite{Breunholder:2017ubu,Cassani:2025iix}.}
Then \eqref{gauge_parametrization_soliton1} corresponds to the most general choice of gauge that satisfies the required regularity conditions
$\gauge_a^I\cdot V^a=\etacoho^I_a$ and $\gauge_a^I\cdot V^{a-1}=\etacoho^I_{a-1}$.
Imposing the quantization condition \eqref{gauge_shift_quantization} is then equivalent to the requirement
\begin{equation}
	\gaugereal^I_{a+1}-\gaugereal^I_a\:\in\:\bZ\:,
\end{equation}
which is solved by setting
\begin{equation}
\label{flat_connection_parameter}
	\gaugereal^I_a\,\equiv\,\gaugereal^I-h_a\,\gaugeint^I_a\:,\qquad%\gaugereal^I\in\bR/\bZ\:,\quad
		\gaugeint^I_a\in\bZ\:.
\end{equation}
The continuous parameter $\gaugereal^I$ is defined up to integers and it has a physical meaning:
it is the component of the gauge field along the $S^1$ factor in the geometry.
In particular $\gaugereal^I$ parametrizes the flat connection component of the gauge field $A^I$.
This is consistent with the fact that flat connections are homomorphisms
\begin{equation}
	\pi_1(M)\,\cong\,\bZ\:\longrightarrow\:\text{U}(1)\:,
\end{equation}
since such homomorphisms are classified by $\gaugereal^I\in\bR/\bZ$ by $\bZ\,\ni\,n\mapsto e^{2\pi\ii\,n\gaugereal^I}$.%
\footnote{More in general, we consider solutions for which the fields, including $A^I$, can take complex values.
		Therefore, we may also allow for complex values of $\gaugereal^I$, which would correspond to  replacing U(1) with $\bC^*$. This would have consequences on the on-shell action of the solitonic solutions discussed in section \ref{sect:example top sol}, which have purely imaginary action if $M^I$ is real, and complex otherwise.}
The integer parameters $\gaugeint^I_a$ on the other hand are not physical, they parametrize an arbitrary gauge choice in each patch.
By gauge invariance, the on-shell action of solitonic solutions cannot depend on the $\gaugeint_a^I$, up to physically irrelevant $2\pi\ii\,\bZ$ terms.

It is useful to observe that for solutions whose boundary topology is $S^1\times S^3$, the quantity $-h_a\,(e_i,V^{a-1},V^a)$
is $a$-independent and it also equals $(e_i,V^{a_1},V^{a_2})$. Then the parametrization \eqref{gauge_parametrization_soliton1}
for the gauges of a solitonic solution, together with \eqref{flat_connection_parameter}, perfectly matches the parametrization \eqref{gauge_parametrization1}
for ther gauges of solutions without flat connections under the identifications
\begin{equation}
\label{soliton_mapping}
	V^{a_3}\:\longrightarrow\:W\:,\qquad w^a\:\longrightarrow\:-h_a\,W\:,\qquad\etacoho^I_{a_3}\:\longrightarrow\:-\kappa\,\gaugereal^I\:.
\end{equation}
In particular we do not need to provide a separate proof of the gauge invariance of the on-shell action of solitonic solutions,
we can simply observe that under the above identifications the argument of appendix \ref{Algebraic proof of gauge invariace}
trivially translates to the case of solitonic solutions.

\subsubsection{Examples of solutions with discrete flat connections}
\label{An example of a solution with torsion}

In this section we provide a couple of examples of solutions whose geometry has a finite and non-trivial fundamental group.
As we will see, often these geometries require two boundary patches in order to properly describe the flat connections that they can support.
The case of two boundary patches is not covered by our localized on-shell action formula \eqref{OSA_localized};
it is a technical difficulty that we postpone to future work.
However some geometries with torsion are fully describable without the need of multiple boundary patches: in this section we will provide one such example,
and in section \ref{Black saddles thermodynamics} we will also compute its on-shell action and discuss its physical properties.

\paragraph{Black holes with $L(p,q)$ horizon and $S^1\times L(p,q)$ boundary.}
A simple example of geometry whose fundamental group has torsion can be found by considering the known Euclidean Kerr-Newman black holes 
solutions in AdS$_5$ and taking the quotient of the solution by a $\bZ_p$ subgroup of the U(1)$^3$ isometry.
With the appropriate choice of $\bZ_p$ subgroup it is possible to obtain a geometry with an horizon with the topology of a lens space
$L(p,q)$, where $p,q$ are coprime integers, and boundary geometry $\partial M\,\cong\,S^1\times L(p,q)$.
While this is a somewhat trivial solution, we can use it to illustrate some of the key features of geometries with discrete flat connections.

The fan of the solution is%
\footnote{Starting from the black hole fan $\wb V^0=(0,0,1)$, $\wb V^1=(1,0,0)$, $\wb V^2=(0,1,0)$ with respect to a basis of $2\pi$-periodic angles
		$\wb\phi_0,\wb\phi_1,\wb\phi_2$, we consider the $\bZ_p$ action given by $\bZ_p\,\ni\,n\::\:(\wb\phi_0,\wb\phi_1,\wb\phi_2)\mapsto
		\big(\wb\phi_0,\wb\phi_1+\frac{2\pi}pn,\wb\phi_2+\frac{2\pi q}pn\big)$. In order to determine the fan of the solution obtained by taking the
		quotient with respect to this $\bZ_p$ action, it is convenient to first perform the SL(3,$\bZ$) transformation to the angles
		$(\wt\phi_0,\wt\phi_1,\wt\phi_2)=(\phi_0,\wb\phi_1,\wb\phi_2-q\,\wb\phi_1)$, which changes the vectors of the fan to
		$\wt V^0=(0,0,1)$, $\wt V^1=(1,0,0)$, $\wt V^2=(0,1,-q)$. In these new coordinates the $\bZ_p$ action simply shifts $\wt\phi_1$ 
		by $\frac{2\pi n}p$, $n\in\bZ_p$. Then taking the quotient with respect to this action is equivalent to demanding that the new coordinates
		$(\phi_0,\phi_1,\phi_2)=(\wt\phi_0,\,p\,\wt\phi_1,\,\wt\phi_2)$ should be $2\pi$-periodic; in these new coordinates the fan 
		becomes precisely the one in \eqref{quotiented_BH}.}
\begin{equation}
\label{quotiented_BH}
	V^0=(0,0,1)\:,\qquad V^1=(1,0,0)\:,\qquad V^2=(0,p,-q)\:,
\end{equation}
and its fundamental group is
\begin{equation}
	\pi_1(M)\,\cong\,\bZ^3/\text{Span}_{\,\bZ}\big(\{V^a\}_{a=0}^2\big)\,\cong\,\bZ_p\:,
\end{equation}
where we have used \eqref{fundamental_group};
the equivalence class of $(0,1,0)$ is the generator of $\bZ_p$. %\,\cong\,\bZ^3/\text{Span}_{\,\bZ}\big(\{V^a\}_{a=0}^2\big)$.
If we try to describe this geometry with a single boundary patch, we fail to find any presence of flat connections in the most generic possible
parametrization for the $\gauge^I_a$. Indeed, the flat connections on the lens space $L(p,q)$ require two patches to be properly described.
Without loss of generality, we may describe this solution with just two patches divided by an interface going from the boundary to $\cD_1$.
One patch contains $\cD_0$ and part of $\cD_1$, and its gauge is determined by $\gauge_1^I$ satisfying the regularity conditions
\begin{equation}
		\gauge^I_1\cdot V^0=\etacoho_0^I\,\equiv\,0\:,\qquad\gauge^I_1\cdot V^1=\etacoho_1^I\:.
\end{equation}
The other patch contains $\cD_2$ and part of $\cD_1$, and its gauge is determined by $\gauge_2^I$ satisfying the regularity conditions
\begin{equation}
		\gauge^I_2\cdot V^2=\etacoho_2^I\,\equiv\,0\:,\qquad\gauge^I_2\cdot V^1=\etacoho_1^I\:.
\end{equation}
The above conditions are generically solved by
\begin{equation}
	\gauge^I_1=(\etacoho^I_1,\,x^I,0)\:,\qquad
		\gauge^I_2=(\etacoho^I_1,\,q\,y^I,p\,y^I)\:.
\end{equation}
Imposing the quantization of the gauge shifts \eqref{gauge_shift_quantization} is equivalent to setting
\begin{equation}
	x^I=\kappa\,(q\,\gaugereal^I+\gaugeint_1^I)\:,\quad y^I=\kappa\,(\gaugereal^I+\gaugeint_2^I)\:,\quad\gaugeint^I_{1,2}\in\bZ\:,
		\quad p\cdot\gaugereal^I\in\{0,\ldots,p-1\}\:.
\end{equation}
The rational parameter $0\leq\gaugereal^I<1$ takes the value of a fraction with $p$ as denominator, and it parametrizes the possible flat connections that
the geometry can support. Indeed, the homomorphisms
\begin{equation}
	\bZ_p\,\cong\,\bZ^3/\text{Span}_{\,\bZ}\big(\{V^a\}_{a=0}^2\big)\:\longrightarrow\:\text{U(1)}
\end{equation}
that classify flat connections are realized by mapping the equivalence class of $(n_0,n_1,n_2)\in\bZ^3$ to $e^{2\pi\ii\,n_1q\gaugereal^I}$.
On the other hand, the integers $\gaugeint^I_{1,2}$ are unphysical gauge-choice parameters.

\paragraph{Black saddles}
\label{Black saddles geometry}

Let us consider the fan
\begin{equation}
	V^0=(0,0,1)\:,\qquad V^1=(n,x,y)\:,\qquad V^2=(0,1,0)\:,\qquad(\text{with }\:n>0)
\end{equation}
which corresponds to the Euclidean black saddles described in \cite{Aharony:2021zkr},
and whose local form of the solution is the same as the one of the Euclidean Kerr-Newman black hole up to redefinitions of the parameters and of the angles.
We will compute the on-shell action and discuss the physics of these solutions in section \ref{Black saddles thermodynamics},
for now let us focus on their topology. Using \eqref{fundamental_group} it is easy too see that
\begin{equation}
	\pi_1(M)\,\cong\,\bZ^3/\text{Span}_{\,\bZ}\big(\{V^a\}_{a=0}^2\big)\,\cong\,\bZ_{n}\:,
\end{equation}
and thus these solution can support discrete flat connections taking $n$ possible values.
We focus on smooth geometries, which must satisfy gcd$(n,x)=1=\:$gcd$(n,y)$.
Indeed, the equations
\begin{equation}
\label{black_saddle_smoothness}
	1=(V^0,V^1,w^1)=w^1_1\,n-w^1_0\,x\:,\qquad1=(V^1,V^2,w^2)=w^2_2\,n-w^2_0\,y
\end{equation}
are solvable for $w^1,w^2\in\bZ^3$ if and only if the pairs $n,x$ and $n,y$ are coprime, by B\'ezout's lemma.
For a smooth geometry we may then set $w^1_2=0=w^2_1$ and fix a choice of integers $w^1_0,w^1_1,w^2_0,w^2_2$
that solves equations \eqref{black_saddle_smoothness}.
We can use these values for the components of $w^1,w^2$ in formulas \eqref{p,q_fmla}, \eqref{cD_a_topology},
concluding that the topology of the Euclidean ``horizon" is a lens space, $\cD_1\,\cong\,L(n,w^1_0\,y)$.%
\footnote{We have also used that $L(p,q)\,\cong\,L(p,-q)$ to get rid of a minus sign.} 
For these geometries we can use the same disposition of patches that we have used throughout the paper and still capture the possible presence of the discrete
flat connections.
The following (over)parametrization of the gauge constants $\gauge^I_a$ in each patch satisfies the regularity conditions \eqref{gauge_regularity}
in the most general way:
\begin{equation}
\label{black_saddle_gauges}
\begin{aligned}
	{}&\gauge^I_0=\Big(\frac{\etacoho^I_1}n+\kappa\,\big(\gaugeint_0^I+x\,w^1_0\,\gaugereal^I\big),\,0,\,0\Big)\:,\\
	{}&\gauge^I_1=\Big(\frac{\etacoho^I_1}n+\kappa\,x\,\big(\gaugeint_1^I+w^1_0\,\gaugereal^I\big),\,
		-\kappa\,n\,\big(\gaugeint^I_1+w^1_0\,\gaugereal^I\big),\,0\Big)\:,\\
	{}&\gauge^I_2=\Big(\frac{\etacoho^I_1}n+\kappa\,y\,\big(\gaugeint_2^I+w^2_0\,\gaugereal^I\big),\,0,\,
		-\kappa\,n\,\big(\gaugeint^I_2+w^2_0\,\gaugereal^I\big)\Big)\:.\\
\end{aligned}
\end{equation}
The advantage of the above parametrization is that it is possible to generally satisfy the quantization condition \eqref{gauge_shift_quantization}
on the gauge shifts by demanding that $\gaugeint^I_0,\gaugeint^I_1,\gaugeint^I_2$ and also $n\cdot\gaugereal^I$ are all integers.
Furthermore, up to redefining $\gaugeint^I_1,\gaugeint^I_2$, we can restrict $\gaugereal^I$ to $0\leq\gaugereal^I<1$.
Then the parametrization \eqref{black_saddle_gauges} is no longer an over-parametrization, rather it is the most general parametrization
that satisfies both regularity and quantization of gauge shifts. The rational numbers $\gaugereal^I$ takes $n$ possible values, specifically
\begin{equation}
	\gaugereal^I\in\Big\{0,\,\frac1n\,,\,\ldots\,,\frac{n-1}n\Big\}\:,
\end{equation}
and it has the physical interpretation of parameterizing the possible discrete flat connections.
On the other hand, the integers $\gaugeint^I_0,\gaugeint^I_1,\gaugeint^I_2$ are arbitrary gauge-choice constants.

\subsection{Thermodynamic potentials}

In this section, following \cite{Kunduri:2013vka} (see also \cite{Cassani:2025iix}),
we  define thermodynamic potentials for our class of solutions with  U(1)$^3$ symmetry.
At the beginning, we will assume for simplicity that we have Dirichlet boundary conditions for the gauge fields $F^I|_{\text{boundary}}=0$.
However, even in the case of boundaries that are topologically $S^1\times S^3$, if the $S^3$ is squashed then Dirichlet boundary conditions are not possible,
as we discuss in section \ref{Boundary analysis}; at the end of this section we will comment which thermodynamic potential we can still defined in such cases.
In subsection \ref{Quantum statistical relation} we will briefly review the supersymmetric quantum statistical relation,
which relates the on-shell action $\cI$ and the thermodynamic potentials with the entropy and the charges  of the solution.

We first define the functions $\Pot_{V^a}^I$ by
\begin{equation}
	\iota_{V^a}F^I=\cpot\cdot\dd\Pot_{V^a}^I\:,\qquad\Pot_{V^a}^I\big|_{\text{boundary}}=0\:,
\end{equation}
where $\cpot$ is a normalization constant, which in our conventions is given by
\begin{equation}
\label{gamma_fixing}
	\gamma=-\frac{\beta\kappa}{2\pi\ii}\:.
\end{equation}
Notice that the Dirichlet boundary conditions imply that $\dd\Pot_{V^a}^I\big|_{\text{boundary}}=0$, and thus
$\Pot_{V^a}^I\big|_{\text{boundary}}$ is a constant that can be fixed to zero by definition.
The function $\Pot_{V^a}^I$ is also constant over $\cD_a$, since $F^I$ is a regular form and the Killing vector $V^a$ vanishes at $\cD_a$, leading to
\begin{equation}
	%V^a\big|_{\cD_a}=0\quad\implies
	\quad0=\iota_{V^a}F^I\big|_{\cD_a}=\cpot\cdot\dd\Pot_{V^a}^I\big|_{\cD_a}
		\quad\implies\quad\dd\left(\Pot_{V^a}^I\big|_{\cD_a}\right)=0\:.
\end{equation}
If $\cD_a$ intersects the boundary $\partial M$, which happens for $a=0$ and $a=d$,
then from $\Pot_{V^a}^I\big|_{\text{boundary}}=0$ and the above observation we conclude
that $\Pot_{V^0}^I\big|_{\cD_0}=0=\Pot_{V^d}^I\big|_{\cD_d}$. On the other hand, if $\cD_a$ does not intersect the boundary
we can define the thermodynamic potential $\Pot_{\cD_a}^I$ by
\begin{equation}
	\Pot_{\cD_a}^I=\Pot_{V^a}^I\big|_{\cD_a}\:,\qquad\Pot_{\cD_a}^I\text{ constant.}
\end{equation}
For Euclidean solutions that are the analytic continuation of Lorentzian ones, a locus $\cD_a$ that does not intersect the boundary is
either the analytic continuation
of a black hole horizon or the analytic continuation of a ``bubble". In the former case $\Pot_{\cD_a}^I$ is called \emph{electrostatic potential},
whereas in the latter case $\Pot_{\cD_a}^I$ is called \emph{magnetic potential}.

It is easy to see the connection between this definition of the electrostatic potential and the possibly more familiar definition
\begin{equation}
\label{usual_potential_def}
	\Pot^I_{\cD_a}=\cpot^{-1}\cdot\left(\iota_{V^a}A^I\big|_{\text{boundary}}-\iota_{V^a}A^I\big|_{\cD_a}\right)\:.
\end{equation}
To see this, choose a gauge $A^I$ that is not necessarily regular everywhere but satisfies $\cL_{V^a}A^I=0$. Then
\begin{equation}
	\dd\Pot_{V^a}^I=\cpot^{-1}\cdot\iota_{V^a}\dd\left(A^I\right)=\dd\left(-\cpot^{-1}\cdot\iota_{V^a}A^I\right)\:,
\end{equation}
from which we find
\begin{equation}
	\Pot_{V^a}^I=-\cpot^{-1}\cdot\iota_{V^a}A^I+\text{constant}\:.
\end{equation}
The value of the integration constant is fixed to be $\cpot^{-1}\cdot\iota_{V^a}A^I\big|_{\text{boundary}}$ by the requirement
$\Pot_{V^a}^I\big|_{\text{boundary}}=0$, leading to \eqref{usual_potential_def} when $\Pot_{V^a}^I$ is evaluated at $\cD_a$.

We also point out that the relation \eqref{usual_potential_def} can also be written as
\begin{equation}
\label{potential_integral_definition}
	\Pot^I_{\cD_a}=\frac1{2\pi\gamma}\int_{\cN_a}F^I\:,
\end{equation}
where $\cN_a$ is a a surface with the topology of a disk that is invariant under the orbits of $V^a$, touches the locus $\cD_a$ at one point, 
and $\partial\cN_a\subset\partial M$.

When the boundary condition for the gauge fields are not Dirichlet it is no longer generally true that 
$\dd\Pot_{V^a}^I|_{\text{boundary}}$ vanishes, and thus it is not always possible to set $\Pot_{V^a}^I$ to be zero over the boundary.
Generically one might define the potentials $\Pot^I_{\cD_a}$ as functions on the boundary, using the integral definition \eqref{potential_integral_definition}
which would now depend on the position of $\partial\cN_a$ within $\partial M$, similar to what has been done in \cite{BenettiGenolini:2024lbj}.
However, depending on what the boundary condition are it may still be possible to define some constant potentials; often one might demand that
$F^I$ at the boundary has no leg along the time circle $S^1$, that is $\iota_{\partial_{\phi_0}}F^I|_{\text{boundary}}=0$, which is the case
for the squashed boundaries that we discuss in section \ref{Boundary analysis} for example. Therefore for vectors of the form $V^a=(1,0,0)$
it would still be possible to define a constant $\Pot_{\cD_a}$ in the same manner as the beginning of this section, which is perfectly sufficient to
discuss the physics of a large class of solutions.

\subsubsection{Quantum statistical relation}
\label{Quantum statistical relation}

The supersymmetric quantum statistical relation \cite{Cabo-Bizet:2018ehj} makes it possible to extract physical information from the on-shell action
$\widehat\cI$, namely the entropy $S$ and the constraint among the charges, for a large class of supersymmetric solutions.
We will frequently use this relation in section \ref{sect:examples} when discussing examples.

Let us consider a real Lorentzian supersymmetric black hole solution with a single horizon; this solution could be topologically nontrivial, it might include
``bubbles" and also the horizon could be topologically something different from an $S^3$, namely a $L(p,q)$ or $S^1\times S^2$,
which is the case of the black lens and black ring solutions that we discuss in section \ref{sec:lensesandrings}.
For simplicity let us focus for the moment on solutions whose boundary at a fixed time is a round $S^3$ with the gauge fields $A^I$
having Dirichlet boundary conditions. We will briefly discuss the more general case of squashed boundaries and non-Dirichlet conditions afterwards.
The entropy $S$ is defined as the horizon area divided by $4G$, and the electric charges $Q_I$ and angular momenta $J_1$, $J_2$ are defined by the integrals
\begin{equation}
	Q_I=\frac\kappa{8\pi G}\int_{S^3\text{ in }\partial M}Q_{IJ}*_gF^J\:,\qquad J_i=\frac1{16\pi G}\int_{S^3\text{ in }\partial M}*_g\,\dd K_i\:,
\end{equation}
where $K_i$ are the one-forms dual to $\partial_{\phi_i}$. 
We can then analytically continue the real Lorentzian solution (which is extremal in the absence of closed time-like curves) to a complex non-extremal
Euclidean saddle, whose on-shell action can be computed by the formula \eqref{localization_formula} without explicit knowledge of the metric and fields.
The charges $Q_I$, $J_i$ then become complex, and so does the entropy which we will now denote as $\cS$.
We refer the reverse procedure of going back to the real Lorentzian solution as ``taking the BPS limit".

The quantum statistical relation \cite{Gibbons:1976ue,Papadimitriou:2005ii} expresses the on-shell action of the non-extremal Euclidean saddle
in terms of the Gibbs free energy:
\begin{equation}
\label{q_stat1}
	\cI=\beta\,G(\beta,\Pot^I,\Omega_i)=\beta E-\cS-\beta\,\Pot^IQ_I-\beta\Omega_1J_1-\beta\Omega_2J_2\:.
\end{equation}
Here $\Pot^I$ and $\Omega_i$ are chemical potentials, and $\Pot^I$ matches $\Pot_{\cD_{\wb a}}^I$ for the horizon $\cD_{\wb a}$,
with $V^{\wb a}=(1,0,0)$.
Regardless of extremality, supersymmetry imposes a constraint among the (generically complex) conserved charges and the chemical potentials
evaluated in the BPS limit,
\begin{equation}
\label{susy_charges}
	E-\Pot^I_{\text{BPS}}\,Q_I-\Omega^{\text{BPS}}_1J_1-\Omega^{\text{BPS}}_2J_2=E_{\text{Casimir}}\:,
\end{equation}
where unlike the derivation in \cite{Cabo-Bizet:2018ehj} we have set the right-hand-side equal to the Casimir energy,
see \cite{Ntokos:2021duk,Papadimitriou:2017kzw} for more details.
Since we are assuming that the gauge field have Dirichlet boundary conditions, the boundary contribution $\cB$ to the on-shell action is proportional
to the Casimir energy, or more precisely\footnote{Note that a priory $\beta E_{\text{Casimir}}$ is a scheme dependent object. In supersymmetric cases there is a preferred scheme that relates this object to the supersymmetric Casimir energy, however there are subtleties that we summarize in \cref{app:renormalization}. The important point is that $\widehat{\mathcal{I}}$ has the expected thermodynamic properties on all known solutions.}
\begin{equation}
\label{cI_vs_Casimir}
	\cI=\widehat\cI+\beta E_{\text{Casimir}}\:.
\end{equation}
Combining \eqref{q_stat1}, \eqref{susy_charges} and \eqref{cI_vs_Casimir} gives the supersymmetric quantum statistical relation:
\begin{equation}
\label{q_stat2}
	\widehat\cI=-\cS-\varphi^IQ_I-\omega_1J_1-\omega_2J_2\:,
\end{equation}
where the potentials $\varphi^I$ and $\omega_i$ are defined as
\begin{equation}
\label{rescaled_potentials}
	\varphi^I=\beta\,\big(\Pot^I-\Pot^I_{\text{BPS}}\big)\:,\qquad\omega_i=\beta\,\big(\Omega_i-\Omega^{\text{BPS}}_i\big)\:.
\end{equation}
The above $\omega_i$ appear in the $\partial_{\phi_i}$ components of the supersymmetric Killing vector $\xi$ as $\xi_i=-\beta^{-1}\omega_i$
\cite{Cabo-Bizet:2018ehj}. 
The entropy $S$ of the real BPS solution can be found as the extremal value of its complex analogue $\cS$ as long as a thermodynamic constraint between
$\varphi^I$ and $\omega_i$ is also imposed:
\begin{equation}
\begin{aligned}
\label{Legendre_transform_entropy}
	{}&S=\cS\,\big|_{\text{extremum}}\:,\qquad\text{where}\quad
		\partial_{\omega_i}\cS\big|_{\text{extremum}}=0=\partial_{\varphi^I}\cS\big|_{\text{extremum}}\quad\text{and}\\[2mm]
	{}&\cS=-\widehat\cI-\varphi^IQ_I-\omega_1J_1-\omega_2J_2+\Lambda\cdot\big(\omega_1+\omega_2-3\kappa\,\V_I\,\varphi^I-2\pi\ii\big)\:.
\end{aligned}
\end{equation}
In the above formula $\Lambda$ is a Lagrange multiplier imposing a constraint that we will explain in section \ref{alpha vs varphi},
and the extremization of $\cS$ is essentially a Legendre transform.
For Kerr-Newman black holes this Legendre transform matches the extremization principle first proposed in \cite{Hosseini:2017mds,Hosseini:2018dob}.
An important consequence of \eqref{Legendre_transform_entropy} is that it is possible to obtain the entropy $S$ of a single-horizon Lorentzian solution
as a function of the charges $Q_I$, $J_i$, together with a constraint among the charges coming from the condition Im$\,S=0$,
as long as the on-shell action $\widehat\cI$ is known as a function of the potentials $\varphi^I$, $\omega_i$.
For this reason in section \ref{alpha vs varphi} we will explain how to relate the cohomological parameters $\etacoho^I_a$ to $\varphi^I$;
the $\omega_i$ on the other hand are related to the components of $\xi$, as already mentioned.

Let us now discuss the case of squashed boundary and/or non-Dirichlet conditions.
The definition of the angular momenta must be generalized to
\begin{equation}
	J_i=\frac1{16\pi G}\int_\Sigma\left[*_g\,\dd K_i+2\,\iota_{\partial_{\phi_i}}A^I\left(Q_{IJ}*_gF^J+\frac16C_{IJK}A^J\wedge F^K\right)\right]\:,
\end{equation}
where $\Sigma$ is a three-dimensional Cauchy surface within the boundary, see \cite{Cassani:2018mlh}.
We must also introduce the Page electric charges:
\begin{equation}
	Q_I^{\text{Page}}=\frac\kappa{8\pi G}\int_\Sigma \left(Q_{IJ}*_gF^J+\frac14C_{IJK}A^J\wedge F^K\right)\:.
\end{equation}
Then the supersymmetric quantum statistical relation \eqref{q_stat2} is still valid provided that we make the substitution $Q_I\to Q_I^{\text{Page}}$
\cite{Ntokos:2021duk}.%
\footnote{There are subtleties: if we write $\cI=\cI_0+\beta E_{\text{Casimir}}$ in a similar manner as \eqref{cI_vs_Casimir},
		then $\cI_0\ne\widehat\cI$, rather the difference $\cI_0-\widehat\cI$ is a boundary term. In \cite{Ntokos:2021duk} the
		supersymmetric quantum statistical relation is then written in terms of $\cI_0$, $Q_I^{\text{Page}}$ and $J_i^{\text{Page}}$,
		where $J_i^{\text{Page}}$ are the Page angular momenta. It turns out that the difference $\cI_0-\widehat\cI$ precisely compensates
		the difference between $J_i^{\text{Page}}$ and $J_i$, so the supersymmetric quantum statistical relation can be written in terms of
		$\widehat\cI$, $Q_I^{\text{Page}}$ and $J_i$. We find the latter rewriting much more convenient due to the topological nature of $\widehat\cI$.}
With this knowledge we can then reproduce the entropy of the known AdS black hole solutions with squashed boundary \cite{Cassani:2018mlh,Bombini:2019jhp},
see section \ref{example BH}. 

\subsection{UV-IR relations}
\label{alpha vs varphi}

 The thermodynamic potentials $\Pot^I_{\cD_a}$  can be expressed in terms of the $\etacoho^I_a$ defined in section \ref{Geometry of solutions}.
This can be done by writing
\begin{equation}
	F^I=\dd\left(\scal^Ie^0\right)-\dd\eta^I=\dd\left(\scal^Ie^0-\wt\eta^I\right)\
\end{equation}
and then using \eqref{usual_potential_def} to obtain
\begin{equation}
\label{potentials_almost_there}
	\Pot^I_{\cD_a}=\cpot^{-1}\cdot\left(\iota_{V^a}\left(\scal^Ie^0-\wt\eta^I\right)\big|_{\text{boundary}}
		-\iota_{V^a}\left(\scal^Ie^0-\wt\eta^I\right)\big|_{\cD_a}\right)\:.
\end{equation}
Since $X^Ie^0$ is regular everywhere we have, using \eqref{eta_tilde_def},
\begin{equation}
	\iota_{V^a}\left(\scal^Ie^0-\wt\eta^I\right)\big|_{\cD_a}=\etacoho^I_a\:.
\end{equation}
In order to simplify the other term, let us recall that by convention the $\cD_a$ that intersect the boundary $\partial M$ are $\cD_0$ and $\cD_d$.
We notice that when $V^0$ and $V^d$ are not collinear, from the relation
\begin{equation}
	(V^0,V^a,V^d)\,\xi-(\xi,V^a,V^d)\,V^0+(\xi,V^0,V^d)\,V^a-(\xi,V^0,V^a)\,V^d=0
\end{equation}
we can derive
\begin{equation}
\label{contraction_expansion}
	\iota_{V^a}=(\xi,V^0,V^d)^{-1}\Big[(\xi,V^a,V^d)\,\iota_{V^0}+(\xi,V^0,V^a)\,\iota_{V^d}-(V^0,V^a,V^d)\,\iota_\xi\Big]\:.
\end{equation}
This is not possible if $V^0$ and $V^d$ are collinear, for example as in the case of asymptotically $T^2\times S^2$ solutions;
however we will not discuss this case in the paper. It is now easy to compute the contractions $\iota_{V^0}$, $\iota_{V^d}$, $\iota_\xi$
of the form $X^Ie^0-\wt\eta^I$ at the boundary. Using the fact that the following contractions must be constant
(or equivalently using \eqref{potentials_almost_there} and $\Pot^I_{\cD_0}=0=\Pot^I_{\cD_d}$), we have
\begin{equation}
	\iota_{V^0}\left(\scal^Ie^0-\wt\eta^I\right)\big|_{\text{boundary}}=\iota_{V^0}\left(\scal^Ie^0-\wt\eta^I\right)\big|_{\cD_0}=\etacoho^I_0\:,
\end{equation}
and similarly for $\iota_{V^d}$. Since $\iota_\xi\wt\eta^I=0$, $\iota_\xi e^0=f$ and 
$f$, $X^I$ take constant values $\bar f$, $\bar X^I$ at the boundary, we can write
\begin{equation}
	\iota_\xi\left(\scal^Ie^0-\wt\eta^I\right)\big|_{\text{boundary}}=\bar f\bar X^I\:.
\end{equation}
Putting everything together we obtain that the thermodynamic potentials can be written as
\begin{equation}
\label{UV_IR}
	\Pot^I_{\cD_a}=\frac{(\xi,V^a,V^d)\,\etacoho^I_0+(\xi,V^0,V^a)\,\etacoho^I_d
		-(\xi,V^0,V^d)\,\etacoho^I_a-(V^0,V^a,V^d)\,\bar f\bar X^I}{\gamma\cdot(\xi,V^0,V^d)}\:.
\end{equation}
This equation is a type of UV-IR relation, connecting the quantities $\etacoho^I_a$ that appear in the localization contributions at $L_a$ in the IR
with the thermodynamic potentials $\Pot^I_{\cD_a}$ and the asymptotic moduli $\bar X^I$ in the UV.

Let us fix the ambiguity in the definition of the cohomological coefficients $\etacoho^I_a$ by setting $\etacoho_0^I=0=\etacoho_d^I$.
With this choice it is then easy to express the $\etacoho^I_a$ in terms of the potentials $\Pot^I_{\cD_a}$ and the
asymptotic moduli $X^I_{\text{bdy}}$ as
\begin{equation}
	\etacoho^I_a=-\,\gamma\cdot\Pot^I_{\cD_a}-\frac{(V^0,V^a,V^d)}{(\xi,V^0,V^d)}\,\bar f\bar X^I\:.
\end{equation}
We can define
\begin{equation}
\label{varphi_DEF}
	\varphi^I_a\equiv\beta\left(\Pot_{\cD_a}^I-\frac{2\pi\ii}{\beta\kappa}\,\frac{(V^0,V^a,V^d)}{(\xi,V^0,V^d)}\,\bar f\bar X^I\right)\:,
\end{equation}
which matches the definition \eqref{rescaled_potentials} of $\varphi^I$ when the Euclidean solution is the analytic continuation
of a single-horizon Lorentzian solution and $a=\wb a$, that is $\varphi^I\equiv\varphi^I_{\wb a}$ for $\wb a$ such that $V^{\wb a}=(1,0,0)$.%
\footnote{Indeed, in the BPS limit $\xi\sim\frac{2\pi\ii}\beta V^{\wb a}$, so the condition $\iota_\xi\wt\eta^I=0$ implies that
		$\frac1\beta\,\etacoho^I_{\wb a}=-\frac1\beta\,\iota_{V^{\wb a}}\,\wt\eta^I\,|_{\cD_a}$ becomes zero in the BPS limit, whereas the $\bar f\bar X^I$ term is unchanged.
		Therefore if we multiply both sides of \eqref{varphi_DEF} by $\beta^{-1}$, the left hand side vanishes in the BPS limit, and we can identify the piece proportional to $\bar f\bar X^I$
		with the BPS limit of $\Phi^I_{\cD_{\wb a}}$, giving us an expression identical to \eqref{rescaled_potentials}.}
Then the UV-IR relation \eqref{UV_IR} can be equivalently restated as
\begin{empheq}[box={\mymath[colback=white, colframe=black]}]{equation}
\label{etacoho_vs_varphi}
	\etacoho^I_a=\frac{\kappa}{2\pi\ii}\,\varphi^I_a\:,\qquad\text{with }\:\etacoho_0^I=0=\etacoho_d^I\:.
\end{empheq}
where we have substituted the value \eqref{gamma_fixing} for the constant $\gamma$.

Combining the linear relation \eqref{etacoho_linear_relation} among the $\etacoho^I_a$
 and the proportionality \eqref{etacoho_vs_varphi} between $\etacoho^I_a$ and $\varphi_a^I$,
we find the following constraint on the thermodynamic potentials:
\begin{empheq}[box={\mymath[colback=white, colframe=black]}]{equation}
\label{etacoho_vs_varphi2}
	\V_I\,\varphi_a^I=\frac{2\pi\ii}{3\kappa}\,\frac{(\xi,V^0,V^a)+(\xi,V^a,V^d)-(\xi,V^0,V^d)}{(\xi,V^0,V^d)}\:.
\end{empheq}
When we consider Euclidean saddles that are the analytic continuation of Lorentzian single-horizon solutions,
the above relation for $a=\wb a$ with $V^{\wb a}=(1,0,0)$ gives the thermodynamic constraint that must be imposed in the Legendre transform
that computes the entropy $S$ of the black hole solution from the on-shell action $\widehat\cI$. Specifically, for saddles with boundary
$S^1\times S^3$ the constraint in \eqref{Legendre_transform_entropy} is reproduced provided that the supersymmetric Killing vector takes the form
\begin{equation}
	\xi=\frac1\beta\left(2\pi\ii\,\partial_{\phi_0}-\omega_1\partial_{\phi_1}-\omega_2\partial_{\phi_2}\right)
\end{equation}
and we use $V^0=(0,0,1)$, $V^d=(0,1,0)$ as in \eqref{V^0,V^d} and \eqref{boundary_topology}.

\section{Examples}
\label{sect:examples}

In this section we describe  the examples of Kerr-Newman black holes, black rings and black lenses and topological solitons. For black rings and black lenses
we reproduce known results in the asymptotically flat limit and we provide a formula for the on-shell action and the entropy of AdS solutions still to be found. We conclude by discussing a fairly general class of fans where we can simplify our general formula  \eqref{OSA_localized} and rewrite the result in terms of the equivariant volume of an associated geometry in the spirit of \cite{Martelli:2023oqk}. 

\subsection{Kerr-Newman black holes and black saddles}
\label{example BH}

We have already discussed the localization of the on-shell action of Kerr-Newman black holes in section \ref{BH single patch},
where we used some ad-hoc simplifications that are not applicable to general solutions in order to provide a simple example of localization.
We will now revisit the Kerr-Newman black holes, 
with the aim to illustrate how the general formulas of sections \ref{The formula for the localized on-shell action} and 
\ref{Topology, thermodynamics and UV-IR relations} work in a concrete example.
We will also discuss the black saddles of \cite{Aharony:2021zkr}, which locally have the same metric as the
(analytically continued \cite{Cabo-Bizet:2018ehj}) Kerr-Newman black holes
of \cite{Gutowski:2004ez,Gutowski:2004yv,Chong:2005da,Chong:2005hr,Kunduri:2006ek}, but generically have different topology.

Let us start from the general formula \eqref{OSA_localized}, and specialize it to the case of
black holes solutions whose geometry is described by the fan
\begin{equation}
\label{BH_fan2}
	V^0=(0,0,1)\:,\qquad V^1=(1,0,0)\:,\qquad V^2=(0,1,0)\:.
\end{equation}
For an explanation of the 
above fan we refer to the discussion of section \ref{BH single patch}.
For the gauge constants $\gauge^I_a$ we can use the gauge  $\alpha_0^I=\alpha_d^I=0$ and the parametrization \eqref{gauge_parametrization2} with $w^a=V^{a+1}$, which reads
\begin{equation}
\label{BH_multigauge}
	\gauge^I_0=\left(\etacoho^I_1-\kappa\,\gaugeint^I_0,\,0,\,0\right)\:,\qquad\gauge^I_1=\left(\etacoho^I_1,-\kappa\,\gaugeint^I_1,\,0\right)\:,\qquad
		\gauge^I_2=\left(\etacoho^I_1,\,0,-\kappa\,\gaugeint^I_2\right)\:,
\end{equation}
where  for consistency $\fn_0^I=\fn_d^I=\fn_{1}^I=0$.
If the gauge-choice integers $\gaugeint^I_a$ are also set to zero, the three patches effectively become a single patch and we go back to the parametrization
\eqref{BH_single_gauge} that we used in section \ref{BH single patch}. Here we will keep the $\gaugeint^I_a$ general in order to showcase
the gauge invariance of the on-shell action.
Just like in section \ref{BH single patch} we set the supersymmetric Killing vector to
\begin{equation}
\label{BH_susy_xi}
	\xi=\frac1\beta\left(2\pi\ii\,\partial_{\phi_0}-\omega_1\partial_{\phi_1}-\omega_2\partial_{\phi_2}\right)\:.
\end{equation}
Plugging \eqref{BH_fan2}, \eqref{BH_multigauge} and \eqref{BH_susy_xi} (together with $\etacoho_0^I=0=\etacoho^I_2$)
in the general formula \eqref{OSA_localized} for the localized on-shell action, we obtain
\begin{equation}
	\widehat\cI=-\,\frac{\ii\pi^2\kappa^3}{12G}\,C_{IJK}\left(\frac{(2\pi)^2}{\kappa^3}\,
		\frac{\etacoho^I_1\etacoho^J_1\etacoho^K_2}{\omega_1\omega_2}+\gaugeint^I_0\,\gaugeint^J_1\,\gaugeint^K_2\right)\:,
\end{equation}
which matches perfectly with \eqref{BH_first_result} when the $\gaugeint^I_a$ are set to zero, as expected.
When the $\gaugeint^I_a$ are nonzero, their dependence neatly separates from the rest in a term that is of the form $2\pi\ii\,$(integer),
provided that the normalization $\kappa$ of the gauge field is taken as in \eqref{kappa_fixing}.
We can now use the relation \eqref{etacoho_vs_varphi} between $\etacoho^I_1$ and the thermodynamic potential $\varphi^I_1$ 
(which for simplicity we will denote as $\varphi^I$) in order to write the on-shell action as
\begin{equation}
\label{OSA_BH}
	\widehat\cI=\frac{\pi\kappa^3}{24G}\,C_{IJK}\,\frac{\varphi^I\varphi^J\varphi^K}{\omega_1\omega_2}\: , 
\end{equation}
matching various known results in the literature
\cite{Hosseini:2017mds,Hosseini:2018dob,Cabo-Bizet:2018ehj,Cassani:2019mms}.
The thermodynamic potentials are subject to the constraint \eqref{etacoho_vs_varphi2}, which can be written as
\begin{equation}
\label{BH_AdS_constraint}
	3\kappa\,\V_I\,\varphi^I=\omega_1+\omega_2-2\pi\ii\:.
\end{equation}
For the STU model \eqref{STU}, we have $\V_I=(3\,\ell)^{-1}$, where $I=1,2,3$ and
$\ell$ is the AdS radius, leading to
\begin{equation}
\label{BH_AdS_constraint_STU}
	\frac\kappa\ell\left(\varphi^1+\varphi^2+\varphi^3\right)=\omega_1+\omega_2-2\pi\ii\qquad(\text{STU model})\:.
\end{equation}
In order to better match the literature for the STU model  in the following we will find convenient to rescale $\varphi^I\to(\ell/\kappa)\,\varphi^I$ and
$Q_I\to(\kappa/\ell)\,Q_I$.

If we take the limit to ungauged supergravity, the AdS black holes become the asymptotically\texttt{-}flat black hole solutions of  \cite{Breckenridge:1996is,Breunholder:2017ubu}.
The on-shell action \eqref{OSA_BH} remains unaltered in this limit, but in the constraint \eqref{BH_AdS_constraint} we need to set $\V_I=0$, finding
\begin{equation}
	\omega_1+\omega_2-2\pi\ii=0\qquad(\text{asymptotically-flat limit})\:,
\end{equation}
which is the constraint \eqref{AF_constraint}.

\paragraph{Thermodynamics of asymptotically-AdS black holes.}

The entropy of  AdS$_5$ black holes as a function of the electric charges $Q_I$ and angular momenta $J_1$, $J_2$
can be found by taking the Legendre transform of the on-shell action \eqref{OSA_BH}.
In other words, the black hole entropy is the extremal value of the functional%
\footnote{Notice that the well definiteness of $\widehat\cI$ modulo $2\pi\ii\,\bZ$ can make this statement ambiguous. Lifting this ambiguity by imposing that $\widehat\cI$
		is a homogeneous function of the potentials leads to the correct value of entropy and constraint among the charges for the AdS black holes and all the asymptotically-flat solutions that we
		compare to, and we will follow this prescription hereafter.}
\begin{equation}
\label{BH_functional}
	\cS=-\widehat\cI-\varphi^IQ_I-\omega_1J_1-\omega_2J_2+\Lambda\left(\omega_1+\omega_2-3\kappa\,\V_I\,\varphi^I-2\pi\ii\right)\:,
\end{equation}
where $\Lambda$ is a Legendre multiplier implementing the constraint \eqref{BH_AdS_constraint}.
The functional \eqref{BH_functional} can be derived using the supersymmetric quantum statistical relation that we have reviewed in section
\ref{Quantum statistical relation}.
Here we will briefly review the extremization of \eqref{BH_functional}, which will be helpful for the discussion of sections \ref{sec:lensesandrings}
and \ref{sect:example top sol}.

For simplicity, let us focus on the STU model first, and discuss more general supergravities later.
For the STU model, the functional \eqref{BH_functional} can be written as
\begin{equation}
	\cS=-\frac{\pi\,\ell^3}{4G}\,
		\frac{\varphi^1\varphi^2\varphi^3}{\omega_1\omega_2}
		-\varphi^IQ_I-\omega_1J_1-\omega_2J_2+\Lambda\left(\omega_1+\omega_2-\varphi^1-\varphi^2-\varphi^3-2\pi\ii\right)\:,
\end{equation}
where we have rescaled $\varphi^I\to(\ell/\kappa)\,\varphi^I$ and $Q_I\to(\kappa/\ell)\,Q_I$ for convenience.
An argument based on homogeneity or a  short computation reveals that the extremal value of $\cS$, which matches the black hole entropy $S$, is proportional to the value of the Lagrange
multiplier $\Lambda$ at the extremum:
\begin{equation}
\label{entropy_vs_Lagrange}
	S=\cS\big|_{\text{extremum}}=-2\pi\ii\,\Lambda\:,\qquad\text{where}\quad
		\partial_{\omega_i}\cS\big|_{\text{extremum}}=0=\partial_{\varphi^I}\cS\big|_{\text{extremum}}\:.
\end{equation}
One can also see that 
at the extremum the following polynomial in the charges and Lagrange multiplier vanishes:
\begin{equation}
\label{BH_polynomial}
	0=\frac{4G}{\pi\ell^3}\big(Q_1+\Lambda\big)\big(Q_2+\Lambda\big)\big(Q_3+\Lambda\big)
		+\big(J_1-\Lambda\big)\big(J_2-\Lambda\big)\,\equiv\,p_3\Lambda^3+p_2\Lambda^2+p_1\Lambda+p_0\:,
\end{equation}
where we have defined the coefficients of the power of $\Lambda$ as $p_i\equiv p_i(Q_I,\,J_i)$.
Both \eqref{entropy_vs_Lagrange} and \eqref{BH_polynomial} can easily be checked by substituting the value of the charges $Q_I,\,J_i$
as a function of the chemical potentials $\varphi^I,\,\omega_i$ and $\Lambda$ that solves the extremization equations
$\partial_{\omega_i}\cS=0=\partial_{\varphi^I}\cS$.
The value of $\Lambda$ that solves the polynomial equation \eqref{BH_polynomial} can then be used to determine the black hole entropy as per
\eqref{entropy_vs_Lagrange}. Further demanding that the entropy $S$ is real, leads to
\begin{equation}
\label{cubic_real_solutions}
	p_0\,p_3\,=\,p_1\,p_2\:,\qquad \frac{p_1}{p_3}>0\:,\qquad S=2\pi\sqrt{\frac{p_1}{p_3}}\:,
\end{equation}
which translate into the following value the the entropy
\begin{equation}
	S(Q,J)=2\pi\sqrt{Q_1Q_2+Q_2Q_3+Q_3Q_1-\frac{\pi\ell^3}{4G}\,(J_1+J_2)}\:,
\end{equation}
together with a non-linear constraint among the charges of a supersymmetric black hole,
\begin{equation}
\label{BH_nonlinear_constraint}
	\frac{4G}{\pi\ell^3}(Q_1+Q_2)(Q_2+Q_3)(Q_3+Q_1)-(J_1+J_2)(Q_1+Q_2+Q_3)-J_1J_2+\left(\frac{S(Q,J)}{2\pi}\right)^2=0\:,
\end{equation}
in perfect agreement with the values computed from the explicit black hole solution \cite{Kim:2006he}.

For a more general supergravity, we can follow the approach of \cite{Cassani:2019mms,Cassani:2024tvk}.
Let us assume that there exists a symmetric tensor $C^{IJK}$ that satisfies the relation \eqref{CIJK_inverse},
then we can generalize the vanishing polynomial in the charges \eqref{BH_polynomial} as follows:
\begin{equation}
\label{BH_polynomial2}
	0=\frac{2G}{3\pi\kappa^3}\,C^{IJK}\big(Q_I+3\kappa\V_I\Lambda\big)
		\big(Q_J+3\kappa\V_J\Lambda\big)\big(Q_K+3\kappa\V_K\Lambda\big)
		+\big(J_1-\Lambda\big)\big(J_2-\Lambda\big)\:.
\end{equation}
Following similar steps as before, we find that the black hole entropy is given by
\begin{equation}
\label{entropy_AdS_BH_general_sugra}
	S=2\pi\sqrt{\frac{3\ell^3}{2\kappa^2}\:C^{IJK}\,V_I\,Q_J\,Q_K-\frac{\pi\ell^3}{4G}\,(J_1+J_2)}\:,
\end{equation}
where we have used that $C^{IJK}\V_I\,\V_J\,\V_K$ equals $\frac29\,\ell^{-3}$ as per \eqref{AdS_radius_symmetric}.
There is also a non-linear constraint among the charges that generalizes \eqref{BH_nonlinear_constraint}, which we will not transcribe here
but can be easily derived from \eqref{BH_polynomial2} by imposing $p_0\,p_3\,=\,p_1\,p_2$.

\paragraph{Thermodynamics of black holes with squashed boundary.}

Let us now consider asymptotically-locally-AdS black holes with a squashed $S^3$ at the boundary, whose known extremal solutions have been found, numerically, in \cite{Cassani:2018mlh,Bombini:2019jhp}. 
 As discussed at the end of section \ref{Quantum statistical relation}, the extremization principle
that computes the entropy and constraint among the charges of solutions with squashed boundary takes the same form, up to replacing the electric charges
$Q_I$ with the Page electric charges $Q_I^{\text{Page}}$. The functional is then
\begin{equation}
\label{BH_functional_squashed}
	\cS=-\widehat\cI-\varphi^IQ_I^{\text{Page}}-\omega_1J_1-\omega_2J_2+\Lambda\left(\omega_1+\omega_2-3\kappa\,\V_I\,\varphi^I-2\pi\ii\right)
\end{equation}
and its extremization leads to the entropy
\begin{equation}
	S=2\pi\sqrt{\frac{3\ell^3}{2\kappa^2}\:C^{IJK}\,V_I\,Q_J^{\text{Page}}\,Q_K^{\text{Page}}-\frac{\pi\ell^3}{4G}\,(J_1+J_2)}\:,
\end{equation}
which is the same as \eqref{entropy_AdS_BH_general_sugra} up to $Q_I\to Q_I^{\text{Page}}$. The above formula is precisely the same result as
\cite{Cassani:2018mlh,Bombini:2019jhp}, with $(Q_I^{\text{Page}})_{\text{our}}=\kappa(P_I)_{\text{their}}$ and
$(J_1+J_2)_{\text{our}}=2(J)_{\text{their}}$. See also \cite{Park:2025fon} (3.53)-(3.55) for a recent discussion of the
on-shell action of $\text{AdS}_5$ black holes with $S^1 \times S^3$ boundary where the $S^3$ is taken: round, biaxially squashed or elliptically squashed.

\paragraph{Thermodynamics of asymptotically-flat black holes.}

We will now briefly review how the thermodynamic analysis changes for asymptotically-flat black holes; for more details see \cite{Cassani:2024tvk}.
We can write a functional whose extremum gives the entropy of the supersymmetric black holes in ungauged supergravities by taking the functional 
\eqref{BH_functional} and setting the Fayet-Iliopoulos gauging parameters $\V_I$ to zero:
\begin{equation}
\label{BH_functional_flat}
	\cS=-\widehat\cI-\varphi^IQ_I-\omega_1J_1-\omega_2J_2+\Lambda\left(\omega_1+\omega_2-2\pi\ii\right)\:,
\end{equation}
The relation \eqref{BH_polynomial} between the charges $J_i$, $Q_I$ and the critical value of the Lagrange multiplier $\Lambda$ is also easily adapted to
the ungauged case by setting $\V_I=0$, finding
\begin{equation}
\label{BH_polynomial_flat}
	0=\frac{2G}{3\pi\kappa^3}\,C^{IJK}Q_IQ_JQ_K
		+\big(J_1-\Lambda\big)\big(J_2-\Lambda\big)\:.
\end{equation}
Unlike the symptotically-AdS case, the above equation is quadratic in $\Lambda$ rather than cubic.
Nonetheless, by solving it and imposing that the solution is purely imaginary, we find the following real value for the entropy of asymptotically-flat black holes,
\begin{equation}
	S=\cS\big|_{\text{extremum}}=-2\pi\ii\,\Lambda=2\pi\sqrt{\frac{8G}{3\pi\kappa^3}\,C^{IJK}Q_IQ_JQ_K-(J_1-J_2)^2}\:,
\end{equation}
together with the constraint $J_1+J_2=0$, matching the entropy derived from the explicit solution of \cite{Cvetic:1996xz}.

\subsubsection{Black saddles}
\label{Black saddles thermodynamics}

There exists a family of Euclidean black saddles that locally have the same metric as the Kerr-Newman black holes,
which were first described in \cite{Aharony:2021zkr}
as asymptotically AdS$_5\times S^5$ solutions of type IIB supergravity, and identified as the dual of known contributions \cite{Hong:2018viz}
to the Bethe Ansatz expansion \cite{Closset:2017bse,Benini:2018ywd} of the superconformal index of $\cN=4$ super-Yang-Mills
\cite{Romelsberger:2005eg,Kinney:2005ej}.
Recently, the asymptotically-flat analogue of these black saddles has been discussed in \cite{Cassani:2025iix} from the perspective of five-dimensional
ungauged supergravity.

When the ten-dimensional saddles of \cite{Aharony:2021zkr} are reduced to five dimensions, their geometry can be described by the fan
\begin{equation}
\label{black_saddle_fan2}
	V^0=(0,0,1)\:,\qquad V^1=(n,x,y)\:,\qquad V^2=(0,1,0)\:,\qquad(\text{with }\:n>0)
\end{equation}
with the supersymmetric Killing vector taking the usual form:
\begin{equation}
	\xi=\frac1\beta\left(2\pi\ii\,\partial_{\phi_0}-\omega_1\partial_{\phi_1}-\omega_2\partial_{\phi_2}\right)\:.
\end{equation}
Notice that when $n=1$ the fan \eqref{black_saddle_fan2} differs from the Kerr-Newman black hole fan \eqref{BH_fan2}
by an SL($3,\bZ$) transformation; this same transformation when applied to $\xi$ has the effect of shifting $\omega_1$ and $\omega_2$
by integer multiples of $2\pi\ii$. Therefore the $n=1$ subfamily of saddles essentially differs from the Kerr-Newman black holes just by the value of
$\omega_{1,2}$. Indeed, this subfamily of saddles was argued to exist in \cite{Aharony:2021zkr} based on the observation that these
shifts of $\omega_{1,2}$ were allowed. Then the saddles with $n>1$ can be found by orbifolding the $n=1$ saddles with respect to a $\bZ_n$ action
that rotates the asymptotic time-circle $S^1$ but not the asymptotic $S^3$.
To be precise, the orbifolding $\bZ_n$ action described in \cite{Aharony:2021zkr} also involves the three $2\pi$-periodic angles of the internal manifold
$S^5$, with three integers $s^I$ ($I=1,2,3$) determining the $\bZ_n$ action on $S^5$.
The effect that these integers $s^I$ have on the $S^1\times S^3$ boundary is that the holonomy of the gauge fields $A^I$ along the $S^1$ is further shifted
by a term proportional to $s^I$.
From a purely five-dimensional perspective, such shifts correspond to turning on a discrete flat connection parameter $\gaugereal^I$
(see \eqref{black_saddle_holonomy} below).

We have already discussed the geometry described by the fan \eqref{black_saddle_fan2}
in section \ref{An example of a solution with torsion}, \ref{Black saddles geometry}.
The fundamental group of these solutions is $\pi_1(M)\,\cong\,\bZ_n$, so they can support a discrete flat connection parameter $\gaugereal^I$,
which is incompatible with a single globally defined gauge when $\gaugereal^I\ne0$, thus requiring patches.
The gauge constants $\gauge^I_a$ in each patch can be fixed as in equation \eqref{black_saddle_gauges}.
We now have all the ingredients needed for applying the localized on-shell action formula \eqref{OSA_localized}, which gives
\begin{equation}
\begin{aligned}
	\widehat\cI=\frac{\ii\pi^2\kappa^3}{12G}\,C_{IJK}\Bigg({}&\frac{\left(\frac{2\pi\ii}\kappa\right)^3\etacoho^I_1\etacoho^J_1\etacoho^K_1}
		{2\pi\ii\,n^3\,(\omega_1+2\pi\ii\,\frac xn)(\omega_2+2\pi\ii\,\frac yn)}\:+\\
	{}&+n^2\big(w^1_0\,\gaugereal^I+\gaugeint^I_1\big)
		\big(w^2_0\,\gaugereal^J+\gaugeint^J_2\big)\big(x\,w^1_0\,\gaugereal^K+\gaugeint^K_0\big)\Bigg)\:,
\end{aligned}
\end{equation}
where the integers $w^1_0,w^2_0$ are defined by \eqref{black_saddle_smoothness}, the rational number $\gaugereal^I$
is a flat connection parameter such that $n\cdot\gaugereal^I\in\{0,\,\ldots\,,n-1\}$, and the integers $\gaugeint^I_a$ are arbitrary gauge choices in each patch.
We can now use \eqref{etacoho_vs_varphi} to relate the parameter $\etacoho^I_1$ with the rescaled electrostatic potential $\varphi^I_1$
associated to the Euclidean $L(n,w^1_0\,y)$ horizon, and we can also discard any $2\pi\ii\,\bZ$ terms from $\widehat\cI$ using the Chern-Simons level
quantization \eqref{kappa_fixing}, finding
\begin{equation}
	\widehat\cI=\frac{\pi\kappa^3}{24G}\,C_{IJK}\:\frac{\varphi^I_1\,\varphi^J_1\,\varphi^K_1}
		{n^3\,(\omega_1+2\pi\ii\,\frac xn)(\omega_2+2\pi\ii\,\frac yn)}\:-
		\underbrace{n^2\,w^1_0\,w^2_0\:\frac{\ii\pi^2\kappa^3}{12G}\,C_{IJK}\,\gaugereal^I\gaugereal^J\gaugereal^K}_{\in\:n^{-1}\,2\pi\ii\,\bZ}\:.
\end{equation}
When considering $e^{-\cI}$, the second term in the above expression is a phase term corresponding to a $n$-th root of unity.
Even if in holography the ratio $\kappa^3/G$ is proportional to $N^2$, such a phase is actually a $\cO(N^0)$ term.
Ignoring this phase term, the flat connection $\gaugereal^I$ has seemingly vanished from the main part of the on-shell action.
However, the dependence on $\gaugereal^I$ comes back if we express $\widehat\cI$ in terms of the holonomy $\varphi^I$ around the asymptotic $S^1$,
\begin{equation}
\label{black_saddle_holonomy}
	\holonomy^I\:\equiv\:\frac{2\pi\ii}{\kappa}\,\left(\gauge^I_a\right)_0\:\text{ mod }2\pi\ii\:=\:
		\frac{\varphi^I_1}n-2\pi\ii\,\gaugereal^I\:\text{ mod }2\pi\ii\:,
\end{equation}
finding a good match with \cite{Aharony:2021zkr} and also \cite{Cassani:2025iix}.%
\footnote{As for \cite{Cassani:2025iix}, in asymptotically-flat space-times must be of the form \eqref{flat_fan}, which in this case imposes $n+x+y=1$.}

\subsection{Black holes with non-trivial horizon topology: black lens and black ring}\label{sec:lensesandrings}

In five dimensions, the horizon of a stationary asymptotically-flat Lorentzian black hole can have the topology of $S^3$, $S^3/\Gamma$, $S^1\times S^2$,
or a connected sum of these (assuming a single connected horizon) \cite{Galloway:2005mf},
showcasing a much richer structure compared to their four-dimensional
counterparts, which may only have $S^2$ horizon topology \cite{Hawking:1971vc}.
If U(1)$^2$ rotational symmetry is also assumed, the possibilities for horizon topologies are further restricted to $S^3$, $S^1\times S^2$ and $L(p,q)$
\cite{Hollands:2007aj}; black holes with $S^1\times S^2$ horizons are called black rings, whereas black holes with $L(p,q)$ horizon are called black lenses.
The first explicit solution with horizon topology different from $S^3$ was the black ring of \cite{Emparan:2001wn};
a black lens solution was first constructed in \cite{Kunduri:2014kja}.
There are various examples of asymptotically-flat, supersymmetric black ring and black lens solutions in the literature, recently extended to 
(non-extremal, complex) Euclidean saddles in  \cite{Cassani:2025iix, Boruch:2025sie},%
\footnote{In this section, whenever we describe a solution as Euclidean, it will be implicit that that solution is complex and non-extremal.}
although none have been successfully constructed in asymptotically-AdS space-time.
In this section we derive the on-shell action of supersymmetric black ring and black lens solutions from the general formula \eqref{OSA_localized},
finding a perfect match with various results in the asymptotically-flat literature,
and also providing novel predictions for the thermodynamics of AdS black rings and lenses, assuming that they do exist.

\subsubsection{Black rings}

The first Lorentzian supersymmetric black hole solution with $S^1\times S^2$ horizon was constructed in
\cite{Elvang:2004rt} for minimal ungauged supergravity, and it has been extended to an Euclidean non-extremal saddle in \cite{Cassani:2025iix}.
The solution of \cite{Elvang:2004rt} has been generalized to Lorentzian multi-charge black rings in \cite{Bena:2004de,Elvang:2004ds} 
for the ungauged STU model, and in \cite{Gauntlett:2004qy} for an arbitrary number of vector multiplets (and also an arbitrary number of  concentric
ring horizons). No asymptotically-AdS supersymmetric black rings have been constructed so far;
the near-horizon analysis of \cite{Kunduri:2006uh,Grover:2013hja} excludes the possibility of the existence of black rings in minimal gauged supergravity,
however there is no such obstruction to the existence of multi-charge AdS black rings, whose near-horizon geometry has been constructed in
\cite{Kunduri:2007qy} for the STU model.

The geometry of an Euclidean black ring solution %is associated to the following fan \cite{Cassani:2025iix}:
with a single bubble is associated to a fan of the following form:
\begin{equation}
\label{black_ring_fan}
	V^0=(0,0,1)\:,\qquad V^1=(1,0,0)\:,\qquad V^2=(0,0,\pm1)\:,\qquad V^3=(0,1,0)\:.
\end{equation}
Indeed, using \eqref{cD_a_topology} we find that $\cD_1\,\cong\,S^1\times S^2$ and $\cD_2\,\cong\,S^3$, where $\cD_1$ is the (Euclidean) horizon
and $\cD_2$ is the bubble.
%The vectors of the above fan are of the form \eqref{flat_fan}, as required for topologies that can support supersymmetric asymptotically-flat solutions.
The asymptotically-flat black ring solutions correspond to the $+$ choice of sign in $V^2$ of the above fan, since this is the only choice compatible with the condition \eqref{flat_fan},
which is required for topologies that can support supersymmetric asymptotically-flat solutions, and is in agreement with \cite{Cassani:2025iix}.
It is not known whether the topology described by the above fan can support asymptotically-AdS solutions; if they can, comparison with the near horizon solutions of \cite{Kunduri:2007qy} suggests
that they would correspond with the $-$ choice of sign in $V^2$, as we will explain later.
If AdS black rings do exist, our localization computation can predict their on-shell action and their thermodynamical properties, including their entropy.

In order to use the formula \eqref{OSA_localized} for the localized on-shell action, we parametrize the cohomological constants $\etacoho^I_a$
using \eqref{special_quantization} with $\wb a=1$ and the gauge constants $\gauge^I_a$ using \eqref{gauge_parametrization2} with
$w^0=(1,0,0)=\mp w^3$ and $w^1=(0,1,0)=\mp w^2$, giving us
\begin{equation}
\begin{aligned}
	\,&\gauge^I_0=\left(\etacoho^I_1-\kappa\,\gaugeint^I_0,\,0,\,0\right)\:,\qquad
		&&\gauge^I_1=\left(\etacoho^I_1,-\kappa\,\gaugeint^I_1,\,0\right)\:,\\[1mm]
	\,&\gauge^I_2=\left(\etacoho^I_1,\pm\kappa\,\gaugeint^I_2,\mp\kappa\,\flux^I_2\,\right)\:,
		\qquad&&\gauge^I_3=\left(\etacoho^I_1\pm\kappa\,\gaugeint^I_3,\,0,\mp\kappa\,\flux^I_2\,\right)\:,
\end{aligned}
\end{equation}
together with $\etacoho^I_0=0=\etacoho^I_3$ and $\etacoho^I_2=-\kappa\,\flux^I_2$, where $\flux^I_2\in\bZ$ has the interpretation of magnetic
flux across the $S^2$ component of the $S^1\times S^2$ horizon, while the $\gaugeint^I_a$ are arbitrary gauge-choice integers.
As usual we take
\begin{equation}
	\xi=\frac1\beta\left(2\pi\ii\,\partial_{\phi_0}-\omega_1\partial_{\phi_1}-\omega_2\partial_{\phi_2}\right)\:,
\end{equation}
and when we substitute all these ingredients in the formula \eqref{OSA_localized} we obtain the following expression for the on-shell action:
\begin{equation}
\begin{aligned}
	\widehat\cI=\frac{\ii\pi^2\kappa^3}{12G}\,C_{IJK}\Bigg(&\frac{\ii\,\flux^I_2\,\big(\pm3(\frac{2\pi\ii}\kappa)^2\alpha^J_1\alpha^K_1+
		3\,\omega_2\,(\frac{2\pi\ii}\kappa)\,\alpha^J_1\,\flux^K_2\pm\omega_2^2\,\flux^J_2\,\flux^K_2\big)}{2\pi\,\omega_1}\:+\\
	&+\flux^I_2\big(\gaugeint^J_0\gaugeint^K_2\mp\gaugeint^J_0\gaugeint^K_1\mp\gaugeint^J_2\gaugeint^K_3\big)\Bigg)\:.
\end{aligned}
\end{equation}
When the Chern-Simons level quantization \eqref{kappa_fixing} is imposed, the dependence on the arbitrary $\gaugeint^I_a$ disappears modulo $2\pi\ii\,\bZ$,
and we are left with
\begin{equation}
\label{OSA_ring}
	\widehat\cI=\mp\frac{\pi\kappa^3}{24G}\,C_{IJK}\,\frac{\flux^I}{\omega_1}\Big(3\,\varphi^J\varphi^K\pm
		3\,\omega_2\,\varphi^J\,\flux^K+\omega_2^2\,\flux^J\flux^K\Big)\:,
\end{equation}
where we have used the relation\eqref{etacoho_vs_varphi} to rewrite $\etacoho^I_1$ in terms of the thermodynamic potential
$\varphi^I_1\equiv\varphi^I$, and simplified the notation for $\flux^I_2\equiv\flux^I$.
The on-shell action \eqref{OSA_ring}, valid for supergravities with vector multiplets both gauged and ungauged, generalizes the on-shell action of black rings
in minimal ungauged supergravity computed in \cite{Cassani:2025iix}.

The thermodynamic potentials $\varphi^I$ are subject to the constraint \eqref{etacoho_vs_varphi2}, and the magnetic fluxes 
$\flux^I$ must satisfy the linear relation \eqref{flux_constraint_wba}. These constraints can be written as
\begin{equation}
\label{ring_constraints}
	3\kappa\,\V_I\,\varphi^I=\omega_1+\omega_2-2\pi\ii\:,\qquad 3\kappa\,\V_I\,\flux^I=1\mp1\:.
\end{equation}
When we consider ungauged supergravities the expression \eqref{OSA_ring} remains identical compared to the gauged case, whereas in the above constraints
we need to set $\V_I=0$; then the first constraint becomes the same as \eqref{AF_constraint}, while the second becomes trivial  for $\mp\to-$.
It is easy to see that the choice $\mp\to+$ is not possible in asymptotically-flat.

If asymptotically-AdS black ring solutions that are smooth and have a single bubble exist in the STU model, we can compare the constraint \eqref{ring_constraints} on the fluxes with
equations (144-145) of \cite{Kunduri:2007qy}, obtained from the most general near horizon solution of a black ring with U(1)$^2$ rotational symmetry. We find a match only for $\mp \to +$,
suggesting that if such black ring solutions in AdS exist, they must have $V^2=(0,0,-1)$ in the fan, the opposite sign compared to the asymptotically-flat ones.

\paragraph{Thermodynamics of the asymptotically-flat black ring.}

We can follow a procedure similar to the one that we outlined in section \ref{example BH} for black holes with $S^3$ horizon,
and find the entropy of asymptotically flat black rings from the on-shell action \eqref{OSA_ring} by extremizing the functional
\begin{equation}
\label{ring_extremand}
	\cS=-\widehat\cI-\varphi^IQ_I-\omega_1J_1-\omega_2J_2+\Lambda\,(\omega_1+\omega_2-2\pi\ii)\:,
\end{equation}
where $\widehat\cI$ is given by \eqref{OSA_ring} with $\pm\to+$.
Again, the entropy of the solution can be derived from the critical value of the Lagrange multiplier $\Lambda$:
\begin{equation}
	S=\cS\big|_{\text{extremum}}=-2\pi\ii\,\Lambda\:,\qquad\text{where}\quad
		\partial_{\omega_i}\cS\big|_{\text{extremum}}=0=\partial_{\phi^I}\cS\big|_{\text{extremum}}\:.
\end{equation}
For concreteness, let us first consider the STU model, for which the on-shell action \eqref{OSA_ring} can be expressed as
\begin{equation}
\label{ring_extremand_STU}
	\widehat\cI=-\frac{\pi\kappa^3}{4G}\,\frac1{\omega_1}\big(\flux^1\varphi^2\varphi^3+\varphi^2\varphi^3\flux^1+\flux^3\varphi^1\varphi^2
		+\omega_2\,(\flux^1\flux^2\varphi^3+\flux^2\flux^3\varphi^1+\flux^3\flux^1\varphi^2)+\omega_2^2\,\flux^1\flux^2\flux^3\big)\:,
\end{equation}
and it is easy to verify that the following expression of the charges $J_i,Q^I$
and $\Lambda$ vanishes when the extremization equations $\partial_{\omega_i}\cS=0=\partial_{\varphi^I}\cS$ are imposed:%
\footnote{Notice that in one of the terms of this expression we have not used Einstein summation over the index $I$, as evident
		by the presence of an explicit summation in $I$; we have added extra superfluous parenthesis around quantities with an index $I$
		that is not summed with Einstein convention to make it more explicit. The same comment will also apply to formula
		 \eqref{asymptotically_flat_ring_entropy}.}
\begin{align}
\label{black_ring_polynomial_STU}
	0=&\,\frac{\pi\kappa^3}{4G}\,\flux^1\flux^2\flux^3\,(J_1-\Lambda)
		+\frac13\sum_{I=1}^3\left(\frac{\flux^1\flux^2\flux^3Q_1Q_2Q_3}{(\flux^I)(Q_I)}-(\flux^I)^2(Q_I)^2\right)+\\\nonumber
	&+4\Big(J_2-\Lambda-\frac12\flux^IQ_I\Big)^2+3\Big(J_2-\Lambda-\frac23\flux^IQ_I\Big)^2
		-6\Big(J_2-\Lambda-\frac12\flux^IQ_I\Big)\Big(J_2-\Lambda-\frac23\flux^IQ_I\Big)\:.
\end{align}
We can then solve the above quadratic equation in $\Lambda$; imposing that the solution is purely imaginary so that the entropy $S=-2\pi\ii\,\Lambda$
is real give the constraint
\begin{equation}
	2J_2+\frac{\pi\kappa^2}{4G}\,\flux^1\flux^2\flux^3-\flux^IQ_I=0\:,
\end{equation}
and the following value for the entropy of asymptotically flat black rings:
\begin{equation}
\label{asymptotically_flat_ring_entropy}
	S=\pi\sqrt{\frac{\pi\kappa^3}{2G}\,\flux^1\flux^2\flux^3\left(2J_1-2J_2+\flux^IQ_I-\frac{\pi\kappa^3}{8G}\,\flux^1\flux^2\flux^3\right)+
		\sum_{I=1}^3\left(2\,\frac{\flux^1\flux^2\flux^3Q_1Q_2Q_3}{(\flux^I)(Q_I)}-(\flux^I)^2(Q_I)^2\right)}\:.
\end{equation}
The above constraint and entropy match perfectly with the ones derived from the explicit solution of \cite{Bena:2004de}.%
\footnote{In order to match our conventions with the ones of \cite{Bena:2004de},
		set $G=1$, $(J_i)_{\text{our}}=-(J_i)_{\text{their}}$, $(Q_I)_{\text{our}}=(N_i)_{\text{their}}$,
		$(\flux^I)_{\text{our}}=(n_i)_{\text{their}}$, $\kappa^3=4/\pi$. With this identifications, the constraint among the charges that we obtain
		immediately matches their equation (68). Then their equations (70) and (71) give a value for the black ring entropy that matches ours
		(using the constraint among the charges to see that the difference vanishes).}

For a more general supergravity model, assuming there exist a symmetric tensor satisfying the relation \eqref{CIJK_inverse},
then we can generalize vanishing expression \eqref{black_ring_polynomial_STU} as follows:
\begin{align}
\label{black_ring_polynomial_flat}
	0=&\,\frac{\pi\kappa^3}{24G}\,C_{IJK}\flux^I\flux^J\flux^K\,(J_1-\Lambda)-\frac13(\flux^IQ_I)^2
		+\frac14C^{IJK}C_{KLM}Q_IQ_J\flux^L\flux^M+\\\nonumber
	&+4\Big(J_2-\Lambda-\frac12\flux^IQ_I\Big)^2+3\Big(J_2-\Lambda-\frac23\flux^IQ_I\Big)^2
		-6\Big(J_2-\Lambda-\frac12\flux^IQ_I\Big)\Big(J_2-\Lambda-\frac23\flux^IQ_I\Big)\:.
\end{align}
Using the extremization equations $\partial_{\omega_i}\cS=0=\partial_{\varphi^I}\cS$ and \eqref{CIJK_inverse}
one can verify that the above identity must hold for the critical value of $\Lambda$.
It is then easy to derive the black ring entropy $S=-2\pi\ii\Lambda$
and the supersymmetric constraint among the charges corresponding to Im$\,S=0$
by solving the quadratic equation \eqref{black_ring_polynomial_flat} for $\Lambda$.

\paragraph{Thermodynamics of the asymptotically-AdS black ring.}

If supersymmetric asymptotically\texttt{-}AdS black ring solutions with a single bubble and no singularities exist,
the on-shell action \eqref{OSA_ring} can give a prediction for their entropy and constraint among the charges.
%The same expression for the on-shell action \eqref{OSA_ring} is also valid for gauged supergravities, with the only difference being in the constraints
%\eqref{ring_constraints}, where now the Fayet-Iliopoulos parameters $V_I$ are no longer zero.
%In particular this means that the extremand now reads
The extremand now reads
\begin{equation}
\label{ring_extremand_AdS}
	\cS=-\widehat\cI-\varphi^IQ_I-\omega_1J_1-\omega_2J_2+\Lambda\left(\omega_1+\omega_2-3\kappa\,\V_I\,\varphi^I-2\pi\ii\right)\:,
\end{equation}
and the fluxes must satisfy 
\iffalse
$\V_I\flux^I=0$. Notice how the extremand \eqref{ring_extremand} valid for ungauged supergravities can be formally
matched with \eqref{ring_extremand_AdS} if one shifts $Q_I\to Q_I+3\kappa\V_I\Lambda$.
It is then easy to show that the identity \eqref{black_ring_polynomial_flat} generalizes easily to the case of gauged supergravities by simply shifting
$Q_I\to Q_I+3\kappa\V_I\Lambda$. The resulting equation is still quadratic in $\Lambda$, in contrast to the case of Kerr-Newman black holes, for which
the equation \eqref{BH_polynomial2} was cubic in $\Lambda$. Solving this quadratic equation, we find that the entropy $S=-2\pi\ii\Lambda$
of an asymptotically-AdS black ring would be
\begin{equation}
	S=2\pi\sqrt{\frac{J_2^2-J_2\,\flux^IQ_I+\frac14C^{IJK}C_{KLM}Q_IQ_J\flux^L\flux^M+\frac{\pi\kappa^3}{24G}\,J_1\,C_{IJK}\flux^I\flux^J\flux^K}
		{1+\frac{9\kappa^2}{4}C^{IJK}C_{KLM}V_IV_J\flux^L\flux^M}}\:,
\end{equation}
and the charges would be subject to the constraint (equivalent to Im$\,S=0$)
\begin{equation}
	2J_2+\frac{\pi\kappa^3}{24G}\,C_{IJK}\flux^I\flux^J\flux^K-\flux^IQ_I-\frac{3\kappa}2C^{IJK}C_{KLM}V_IQ_J\flux^L\flux^M=0\:.
\end{equation}
\fi
$3\kappa\,\V_I\,\flux^I=1\mp1$; the comparison with the near horizon solutions of \cite{Kunduri:2007qy} leads to setting this sign to $\mp \to +$.
Notice how the extremand \eqref{ring_extremand} valid for ungauged supergravities can be formally
matched with \eqref{ring_extremand_AdS} if one shifts $Q_I\to Q_I+3\kappa\,\V_I\Lambda$ and also flips the sign $\flux^I\to-\flux^I$ to compensate for the
different sign in $V^2$. It is then easy to show that the identity \eqref{black_ring_polynomial_flat} generalizes to the case of gauged supergravities by simply
applying $Q_I\to Q_I+3\kappa\,\V_I\Lambda$ and $\flux^I\to-\flux^I$.
The resulting equation is still quadratic in $\Lambda$, in contrast to the case of Kerr-Newman black holes, for which
the equation \eqref{BH_polynomial2} was cubic in $\Lambda$. Solving this quadratic equation, we find that the entropy $S=-2\pi\ii\Lambda$
of an asymptotically-AdS black ring would be
\begin{equation}
	S=2\pi\sqrt{\frac{J_2^2+(J_2-J_1)\flux^IQ_I+\frac14C^{IJK}C_{KLM}Q_I(Q_J+6\kappa J_1V_J)\flux^L\flux^M}{-1+\frac{9\kappa^2}4C^{IJK}C_{KLM}V_IV_J\flux^L\flux^M}}\:,
\end{equation}
and imposing Im$\,S=0$ would lead to the constraint
\begin{equation}
	\frac{\pi\kappa^3}{24G}\,C_{IJK}\flux^I\flux^J\flux^K-\flux^IQ_I+\frac{3\kappa}2C^{IJK}C_{KLM}V_IQ_J\flux^L\flux^M=0\:.
\end{equation}
From our analysis we can only derive a single constraint and an inequality (the positivity of the argument of the square root that gives the entropy).
However the smooth near horizon solutions of \cite{Kunduri:2007qy} are much more constrained: for STU, they depend on three parameters, which in turn must obey an inequality, which, in particular, forbids solutions in minimal gauged supergravity.
At present, we don't know if there is a method to derive such strong constraints without explicitly solving the equations of motion.

\subsubsection{Black lens}
\label{black lens section}

Smooth asymptotically-flat Lorentzian solutions featuring an event horizon with topology $L(2,1)$ have been constructed in \cite{Kunduri:2014kja} 
for minimal ungauged supergravity, and generalized to the STU model in \cite{Kunduri:2016xbo}.
The $L(2,1)$ black lens of \cite{Kunduri:2014kja} has a single bubble outside the horizon, and it is a special case of a wider class of $L(p,1)$ black lenses
with a single bubble that have orbifold singularities for $p\ne2$. These solutions have been analytically continued to Euclidean saddles in \cite{Cassani:2025iix},
where their on-shell action has been computed via direct integration.
In asymptotically-flat space-time, single-horizon smooth $L(p,1)$ black lenses require multiple bubbles,
such as the solutions of \cite{Tomizawa:2016kjh,Tomizawa:2019yzb};
multi-black lens solutions have also been constructed \cite{Tomizawa:2017suc}.
Despite the numerous explicit examples of asymptotically-flat black lenses, no supersymmetric black lens solutions has been constructed in asymptotically-AdS
space-time so far. In this section we will consider black lens solutions with a single $L(2,1)$ horizon and a single bubble, both in asymptotically-flat and
asymptotically-AdS space-times, reproducing known results in the former case and providing novel predictions in the latter.

Smooth asymptotically-flat Euclidean saddles with a single lens space horizon and a single bubble that can be analytically continued to extremal Lorentzian
solutions have the following fan \cite{Cassani:2025iix}:
\begin{equation}
\label{black_lens_fan}
	V^0=(0,0,1)\:,\qquad V^1=(1,0,0)\:,\qquad V^2=(0,2,-1)\:,\qquad V^3=(0,1,0)\:.
\end{equation}
Indeed, the above vectors are of the form \eqref{flat_fan}, which is required for asymptotically-flat solutions that are supersymmetric.
Furthermore, the fan \eqref{black_lens_fan} has no orbifold singularities: it is possible to choose the lattice vectors $w^a$ satisfying \eqref{wa_def} as
$w^0=w^3=(1,0,0)$ and $w^1=w^2=(0,1,0)$.
If $V^2$ was replaced with a generic bubble vector of the form \eqref{flat_fan}, that is $V^2=(0,p_2,1-p_2)$, a vector $w^3\in\bZ^3$ satisfying
$(V^2,V^3,w^3)=1$ would exist only if $p_2=0$ or $p_2=2$; the choice $p_2=0$ would give the black ring fan \eqref{black_ring_fan} that we have already
discussed, whereas $p_2=2$ leads to the fan \eqref{black_lens_fan} that we are considering now.
If we apply \eqref{cD_a_topology} and \eqref{p,q_fmla} to the fan \eqref{black_lens_fan}, we can see that $\cD_1\,\cong\,L(2,1)$ and $\cD_2\,\cong\,S^3$,
where $\cD_1$ is the horizon while $\cD_2$ is the bubble.

Asymptotically-AdS black lenses, if any exist, may have horizon topology $L(p,1)$ for any $p$ even in the presence of a single bubble,
since $V^2$ would no longer be restricted to be of the form \eqref{flat_fan}. Focusing on the case of $L(2,1)$ horizon, there are a priori
%two possibilities for the fan: either the same as \eqref{black_lens_fan} or a similar fan with $V^2$ replaced with $(0,2,1)$.
multiple possibilities for the fan, differing from \eqref{black_lens_fan} by a change of sign in each component of $V^2$.
For the moment, let us focus on the fan \eqref{black_lens_fan}, we will discuss % the  fan with $(0,2,1)$ later.
other possibilities later.

We can use the parametrization \eqref{gauge_parametrization2} with $\wb a=1$ for the constants $\gauge^I_a$ that determine the gauge in each patch,
which gives
\begin{equation}
\begin{aligned}
	\,&\gauge^I_0=\left(\etacoho^I_1-\kappa\,\gaugeint^I_0,\,0,\,0\right)\:,\qquad
		&&\gauge^I_1=\left(\etacoho^I_1,-\kappa\,\gaugeint^I_1,\,0\right)\:,\\[1mm]
	\,&\gauge^I_2=\left(\etacoho^I_1,-\kappa\,\gaugeint^I_2,\,\kappa\,(\flux^I_2-2\,\gaugeint^I_2)\,\right)\:,
		\qquad&&\gauge^I_3=\left(\etacoho^I_1-\kappa\,\gaugeint^I_3,\,0,\,\kappa\,\flux^I_2\,\right)\:,
\end{aligned}
\end{equation}
where $\flux^I_2\in\bZ$ is a magnetic flux whereas the $\gaugeint^I_a$ are arbitrary integers.
The $\etacoho^I_a$ that appear in the localized on-shell action \eqref{OSA_localized} are related to $\etacoho^I_1$ and $\flux^I_2$ by 
\eqref{special_quantization}. As usual, we take the supersymmetric Killing vector to be of the form
\begin{equation}
	\xi=\frac1\beta\left(2\pi\ii\,\partial_{\phi_0}-\omega_1\partial_{\phi_1}-\omega_2\partial_{\phi_2}\right)\:.
\end{equation}
We now have all the ingredients needed to apply the formula \eqref{OSA_localized}, which gives the following value for the localized on-shell action:
\begin{equation}
	\widehat\cI=\frac{\ii\pi^2\kappa^3}{12G}\,C_{IJK}\left(\frac{(\frac{2\pi\ii}\kappa\,\etacoho_1)^3}{2\pi\ii\,\omega_1\omega_2}
		-\frac{(\frac{2\pi\ii}\kappa\,\etacoho_1-\flux_2\,\omega_2)^3}{2\pi\ii\,\omega_2\,(\omega_1+2\,\omega_2)}
		+\flux^I_2(\gaugeint^J_0\gaugeint^K_1\!+\!\gaugeint^J_0\gaugeint^K_2\!+\!\gaugeint^J_2\gaugeint^K_3)
		-2\gaugeint^I_0\gaugeint^J_1\gaugeint^K_2\right)\:,
\end{equation}
where we are using the shorthand notation $(\etacoho_1)^3\,\equiv\,\etacoho^I_1\etacoho^J_1\etacoho^K_1$, and similarly for the other cubic term.
If $\kappa$ satisfies \eqref{kappa_fixing}, the dependence on the arbitrary gauge choice integers $\gaugeint^I_a$ disappears from the on-shell action
modulo $2\pi\ii\,\bZ$, and we can write
\begin{equation}
\label{osa_black_lens}
	\widehat\cI=\frac{\pi\kappa^3}{24G}\,C_{IJK}\left(\frac{\varphi^I\varphi^J\varphi^K}{\omega_1\omega_2}
		-\frac{(\varphi^I -\flux^I\,\omega_2) (\varphi^J-\flux^J\,\omega_2) (\varphi^K-\flux^K\,\omega_2)}{\omega_2\,(\omega_1+2\,\omega_2)}\right)\:,
\end{equation}
where we have used \eqref{etacoho_vs_varphi} and omitted the patch indices in $\varphi^I\equiv\varphi^I_1$ and $\flux^I\equiv\flux^I_2$.
The above on-shell action generalizes the one computed in \cite{Cassani:2025iix}, which was derived for minimal ungauged supergravity.
The constraints \eqref{etacoho_vs_varphi2} and \eqref{flux_constraint_wba} are
\begin{equation}
\label{lens_constraints}
	3\kappa\,\V_I\,\varphi^I=\omega_1+\omega_2-2\pi\ii\:,\qquad\V_I\,\flux^I=0\:.
\end{equation}
In ungauged supergravity $\V_I=0$ and thus the second constraint is trivial.

\paragraph{Thermodynamics of the asymptotically-flat black lens.}

The entropy of asymptotically\texttt{-}flat black lenses can be obtained as the extremum the functional
\begin{equation}
\label{lens_extremand}
	\cS=-\widehat\cI-\varphi^IQ_I-\omega_1J_1-\omega_2J_2+\Lambda\,(\omega_1+\omega_2-2\pi\ii)\:.
\end{equation}
As usual, it is easy to verify that the entropy can be derived from the critical value of the Lagrange multiplier $\Lambda$ as
\begin{equation}
	S=\cS\big|_{\text{extremum}}=-2\pi\ii\,\Lambda\:.
\end{equation}
For simplicity let us start with the STU model, for which it is easy to check that a polynomial in $\Lambda$ that vanishes at its critical value is
given by
\begin{equation}
\begin{aligned}
\label{black_lens_polynomial_STU}
	0=\,&\frac{4G}{\pi\kappa^3}\,Q_1Q_2Q_3+(J_1-\Lambda)(J_2-\Lambda)-\frac12(J_2-\Lambda)^2
		-\frac12\,\flux^IQ_I\,(J_2-\Lambda)+\\
	&+\frac{\pi\kappa^3}{8G}\,\flux^1\flux^2\flux^3(J_1-\Lambda)
		-\frac12\,(\flux^1\flux^2Q_1Q_2+\flux^2\flux^3Q_2Q_3+\flux^3\flux^1Q_3Q_1)\:.
\end{aligned}
\end{equation}
The above quadratic equation in $\Lambda$ has a purely imaginary solution only if the charges obey the constraint
\begin{equation}
	2J_1+\frac{\pi\kappa^3}{4G}\flux^1\flux^2\flux^3-(\flux^1Q_1+\flux^2Q_2+\flux^3Q_3)=0\:,
\end{equation}
and the solution for $\Lambda$ leads to the following value for the entropy:
\begin{equation}
	S=4\pi^2\sqrt{2G\left(\frac{Q_1}{\pi\kappa}-\kappa^2\frac{\flux^2\flux^3}{8G}\right)
		\left(\frac{Q_2}{\pi\kappa}-\kappa^2\frac{\flux^3\flux^1}{8G}\right)
		\left(\frac{Q_3}{\pi\kappa}-\kappa^2\frac{\flux^1\flux^2}{8G}\right)
		-\frac14\left(\frac{J_2}\pi+\kappa^3\frac{\flux^1\flux^2\flux^3}{8G}\right)^2}\:,
\end{equation}
matching the result of \cite{Kunduri:2016xbo}.%
\footnote{In order to match their conventions, set $G=1$, $(Q_I)_{\text{our}}=\kappa(Q_i)_{\text{their}}$, $(\flux^I)_{\text{our}}=(q_i)_{\text{their}}$,
		$\kappa=-2$, $J_1\equiv J_\phi-J_\psi$, $J_2\equiv J_\phi+J_\psi$.}

For a more general supergravity, assuming that there exist a symmetric tensor $C^{IJK}$ satisfying \eqref{CIJK_inverse},
with some algebra it is possible to show that the identity \eqref{black_lens_polynomial_STU} can be generalized as follows:
\begin{equation}
\begin{aligned}
\label{black_lens_polynomial}
	0=\,&\frac{2G}{3\pi\kappa^3}\,C^{IJK}Q_IQ_JQ_K+(J_1-\Lambda)(J_2-\Lambda)-\frac12(J_2-\Lambda)^2
		-\frac12\,\flux^IQ_I\,(J_2-\Lambda)+\\
	&+\frac{\pi\kappa^3}{48G}\,\flux^I\flux^J\flux^K(J_1-\Lambda)
		-\frac18C^{IJK}C_{KLM}Q_IQ_J\flux^L\flux^M\:.
\end{aligned}
\end{equation}
Solving the above quadratic equation in $\Lambda$, one can then easily extract the black lens entropy $S=-2\pi\ii\Lambda$ and a constraint
among the charges corresponding to Im$\,S=0$.

\paragraph{Thermodynamics of asymptotically-AdS black lenses.}

Let us discuss the thermodynamics of a hypothetical black lens solution in AdS with the same fan as the known asymptotically-flat black lens solutions,
\eqref{black_lens_fan}.
The entropy would be found as the extremum of the functional
\begin{equation}
\label{lens_AdS_extremand}
	\cS=-\widehat\cI-\varphi^IQ_I-\omega_1J_1-\omega_2J_2+\Lambda\left(\omega_1+\omega_2-3\kappa\,\V_I\,\varphi^I-2\pi\ii\right)\:,
\end{equation}
where $\widehat\cI$ is the on-shell action \eqref{osa_black_lens}, formally equal to its asymptotically flat counterpart,
and we have imposed the constraint \eqref{lens_constraints}, where now the Fayet-Iliopoulos parameters $V_I$ are no longer zero.
Notice that \eqref{lens_constraints} constrains also the fluxes, $\V_I\flux^I=0$. Similarly to what we did in the case of AdS black rings,
we can observe that the extremand \eqref{lens_AdS_extremand} can formally be obtained from \eqref{lens_extremand} by shifting
$Q_I\to Q_I+3\kappa\V_I\Lambda$, and in particular the identity \eqref{black_lens_polynomial} can be generalized for gauged supergravities
by shifting $Q_I$ in the same manner. The resulting identity is a cubic polynomial in the critical value of the Lagrange multiplier $\Lambda$,
from which we can extract the AdS black lens entropy and supersymmetric constraint among the charges in the same manner as
\eqref{cubic_real_solutions}, finding
\begin{equation}
	S=2\pi\sqrt{\frac{3\ell^3}{2\kappa^2}\,C^{IJK}\V_IQ_JQ_K-\frac{\pi\ell^3}{4G}\left(J_1-\frac12\flux^IQ_I+
		\frac{3\kappa}{4}C^{IJK}C_{KLM}V_IQ_J\flux^L\flux^M+\frac{\pi\kappa^3}{48G}\,C_{IJK}\flux^I\flux^J\flux^K\right)}\:,
\end{equation}
and the constraint (which we write in terms of the above $S$ for simplicity)
\begin{align}
	{}&\bigg(\bigg(\frac{S}{2\pi}\bigg)^2-\frac{3\ell^3}{2\kappa^2}\,C^{IJK}\V_IQ_JQ_K\bigg)\left(\frac12+\frac{18G}{\pi\kappa}C^{IJK}V_IV_JQ_K 
		-\frac{9\kappa^2}8\right)=\\\nonumber
	{}&=J_1J_2-\frac12J_2^2-\frac12J_2\,\flux^IQ_I+\frac{\pi\kappa^3}{48G}J_1\,C_{IJK}\flux^I\flux^J\flux^K
		+\frac{2G}{3\pi\kappa^3}\,C^{IJK}Q_IQ_JQ_K-\frac18C^{IJK}C_{KLM}Q_IQ_J\flux^L\flux^M\:.
\end{align}

An asymptotically-AdS black lens with $L(2,1)$ horizon and a single bubble may also have a fan that is different from \eqref{black_lens_fan}. For example, it may have the following fan:
\begin{equation}
\label{black_lens_fan_convex}
	V^0=(0,0,1)\:,\qquad V^1=(1,0,0)\:,\qquad V^2=(0,2,1)\:,\qquad V^3=(0,1,0)\:.
\end{equation}
A similar computation as the one at the beginning of section \ref{black lens section} leads to the following on-shell action:
\begin{equation}
\label{osa_black_lens_convex}
	\widehat\cI=\frac{\pi\kappa^3}{24G}\,C_{IJK}\left(\frac{\varphi^I\varphi^J\varphi^K}{\omega_1\omega_2}
		-\frac{(\varphi^I+\flux^I\,\omega_2)(\varphi^J+\flux^J\,\omega_2)(\varphi^K+\flux^K\,\omega_2)}{\omega_2\,(\omega_1-2\,\omega_2)}\right)\:.
\end{equation}
We also have the constraints
\begin{equation}
\label{lens_constraints}
	3\kappa\,\V_I\,\varphi^I=\omega_1+\omega_2-2\pi\ii\:,\qquad\V_I\,\flux^I=-\frac2{3\kappa}\:.
\end{equation}
We will not do the full thermodynamic analysis of this case, since the entropy and constraints can be found in a very similar manner as before,
with the only changes being some signs and some extra terms due to $\V_I\flux^I$ no longer being zero.

\subsection{Topological solitons with an arbitrary number of bubbles}
\label{sect:example top sol}

Topological solitons are nontrivial solutions without horizons. Many such solutions are known in asymptotically-flat space-time,
such as \cite{Bena:2005va,Berglund:2005vb,Bena:2007kg,Gibbons:2013tqa} and more,
and in the context of the fuzzball program they are expected to be microstate geometries for black holes, see \cite{Mathur:2005zp}.
Solitonic solutions are also known in AdS space-time \cite{Chong:2005hr,Cassani:2015upa,Durgut:2021bpk}.
In this section we will use the main formula \eqref{OSA_localized} to compute the on-shell action of supersymmetric solitons with U(1)$^3$ symmetry
and discuss their physics.

Let us focus our attention on solitonic solutions whose boundary topology is $S^1\times S^3$.
In section \ref{Solitonic solutions} we have described the key features of such solutions and given a parametrization \eqref{gauge_parametrization_soliton1}
of the gauge fields in each patch.
This parametrization contains arbitrary gauge choice parameters $\gaugeint^I_a\in\bZ$, but the on-shell action does not depend (modulo $2\pi\ii\,\bZ$)
on them, as we prove in appendix \ref{Algebraic proof of gauge invariace}. We may therefore set the $\gaugeint^I_a$ to zero without loss of generality,
and rewrite the parametrization of the gauges simply as
\begin{equation}
\label{gauge_parametrization_soliton2}
	\left(\gauge^I_a\right)_i=h_a\,\kappa\Big[(e_i,V^a,W)\,\flux^I_{a-1}-(e_i,V^{a-1},W)\,\flux^I_a+(e_i,V^{a-1},V^a)\,\gaugereal^I\Big]\:,
\end{equation}
where we have also set $a_1=d$, $a_2=0$ and $\etacoho^I_0=0=\etacoho^I_d$.
We remind that $\flux^I_a\in\bZ$ are magnetic fluxes, whereas $\gaugereal^I\in\bR/\bZ$ is the flat connection parameter.
Without loss of generality we can set $W=(1,0,0)$. 
It is convenient to reabsorb the signs $h_a=-(V^{a-1},V^a,W)$ by defining
\begin{equation}
	\hat\flux^I_a\,\equiv\,\fks_a\flux^I_a=-\frac{\fks_a}{\kappa}\,\etacoho^I_a\:,\qquad V^a\,\equiv\,(0,\,\fks_a\,v^a)\:,\quad v^a\in\bZ^2\:,
\end{equation}
where the $\sigma_a$ are signs that we choose so that%
\footnote{Note that $h_0$ never appears in \eqref{gauge_parametrization_soliton2} since $\hat\flux^I_0=0=\hat\flux^I_d$.}
\begin{equation}
\label{soliton_signs}
	\fks_{a-1}\,\fks_a=h_a\qquad\forall a\ne0\:,\qquad\fks_0=1\:.
\end{equation}
Then \eqref{gauge_parametrization_soliton2} becomes
\begin{equation}
\label{gauge_parametrization_soliton3}
	\left(\gauge^I_a\right)_i=\kappa\Big[(e_i,v^a)\,\hat\flux^I_{a-1}-(e_i,v^{a-1})\,\hat\flux^I_a-\delta_{i,0}\,\gaugereal^I\Big]\:,
\end{equation}
where we are using the notation
\begin{equation}
	(X,v^a)\,\equiv\,\det\begin{pmatrix}X_1&\:X_2\\v^a_1&v^a_2\end{pmatrix}=v^a_2\,X_1-v^a_1\,X_2\:,\quad\forall\text{ vector }
		X\text{ with components }X_0,X_1,X_2\:.
\end{equation}
We also observe that the linear relation \eqref{etacoho_linear_relation} can be translated to%
\footnote{\label{footnote_spinc3}Note that this is the analogue of the relation \eqref{flux_constraint_wba} under $V^{\wb a}\to W$,
		which is a special case of the mapping \eqref{soliton_mapping}. In particular, as per the comment in footnote \ref{footnote_spinc2},
		we expect that $\int_{\cC_a}w_2=\fks_a-v^a_1-v^a_2\mod2$, where
		$\cC_a$ is the two-cycle that supports the flux $\flux^I_a$ and $w_2$ is the second Stiefel-Whitney class. For the simpler case
		where the signs $\sigma_a$ are all +1 we can use standard techniques for four-dimensional toric manifolds to verify that this is indeed correct.
		Writing $M=S^1\times M_4$, where $S^1$ is the time-circle parametrized by $\phi_0$,
		then $M_4$ has two-cycles $D_a$ corresponding to its divisors, which can be related to the $\cC_a$ as $D_a=\sum_{b}D_{ab}\,\cC_b$,
		where $D_{ab}=\int_{M_4}c_1(\cL_a)\,c_1(\cL_b)$ are the intersection numbers, $D_{aa}=-(v^{a-1},v^{a+1})$, $D_{a-1,a}=1$,
		and $D_{ab}=0$ if $|a-b|>1$. In terms of the $D_a$, $\int_{D_a}w_2=\int_{D_a}\sum_bc_1(\cL_b)\mod2=\sum_bD_{ab}\mod2$,
		and it is easy to verify that the formula $\int_{\cC_a}w_2=1-v^a_1-v^a_2\mod2$ gives the same result using $\sum_bD_{ab}v^b=0$.}
\begin{equation}
\label{hat_flux_constraint}
	\V_I\,\hat\flux^I_a=\frac{1}{3\kappa}(\fks_a-v^a_1-v^a_2)\:,
\end{equation}
which for $a=0$ and $a=d\,$ is automatically verified since $\hat\flux^I_0=0=\hat\flux^I_d$ and the right hand side vanishes as well.

If we plug the parametrization \eqref{gauge_parametrization_soliton3} in the localized formula for the on-shell action \eqref{OSA_localized},
after some algebra we obtain
\begin{align}
\label{soliton_OSA}
	\widehat\cI=\frac{\ii\pi^2\kappa^3}{12G}\,C_{IJK}\,\sum_{a=0}^d
		&\Big(\hat\flux^I_{a-1}\,\hat\flux^J_a+\hat\flux^I_a\,\hat\flux^J_{a+1}+(v^{a-1},v^{a+1})\,\hat\flux^I_a\,\hat\flux^J_a\Big)\cdot\\\nonumber
	\cdot&\Big(3M^K+\xi_0^{-1}(\xi,v^a)(\hat\flux_{a+1}^K-\hat\flux_{a-1}^K)+\xi_0^{-1}(\xi,v^{a-1}-v^{a+1})\,\hat\flux^K_a\Big)\:.
\end{align}
An important observation to make is that the above on-shell action is linear in the flat connection parameter $\gaugereal^I$ and the components
$\xi_1,\xi_2$ of the supersymmetric Killing vector. This is relevant considering that $\xi_{1,2}$ are related to the chemical potentials
$\omega_{1,2}$ conjugated to the angular momenta $J_{1,2}$ by the usual
\begin{equation}
	\xi=\frac1\beta\left(2\pi\ii\,\partial_{\phi_0}-\omega_1\partial_{\phi_1}-\omega_2\partial_{\phi_2}\right)\:,
\end{equation}
and it is also sensible to identify $\gaugereal^I$ with the chemical potential $\varphi^I$ conjugate to the electric charge $Q_I$,
\begin{equation}
	\varphi^I\,\equiv\,-2\pi\ii\:\gaugereal^I\:.
\end{equation}
Indeed, identifying $\varphi^I$ with the holonomy along the asymptotic $S^1$ is natural in the context of holography;
for example, in the case of single-horizon solutions we have $\left(\gauge^I_a\right)_0=\alpha_{\wb a}^I=-\frac\gamma\beta\,\varphi^I$,
which should be compared with $\left(\gauge^I_a\right)_0=-\kappa\gaugereal^I$ (modulo $\kappa\bZ$) for the solitons.
In particular, the on-shell action \eqref{soliton_OSA} is linear in the chemical potentials.
A fully rigorous thermodynamic analysis for the topological solitons is beyond the scope of this paper,
nonetheless we will now argue that the charges of the solitonic solution should be found as
\begin{equation}
  \label{soliton_proposal}
  Q_I=-\partial_{\varphi^I}\,\widehat\cI\:,\qquad J_i=-\partial_{\omega_i}\,\widehat\cI\:,
\end{equation}
and will also verify it by matching with the explicit AdS soliton solution of \cite{Durgut:2021bpk}.
First, let us notice that relations analogous to \eqref{soliton_proposal} have been derived for asymptotically-flat single-horizon solutions
(which may also include bubbles) using the supersymmetric quantum statistical relation together with the Smarr formula of \cite{Kunduri:2013vka},
and it is reasonable to think that a similar formula could still be valid in the absence of the horizon.
Furthermore, in analogy with the extremization principle discussed in the previous section, we may propose to extremize the functional
\begin{equation}
	\cS=-\widehat\cI-\varphi^IQ_I-\omega_1J_1-\omega_2J_2\:,
\end{equation}
whose extremization equations admit solutions only if the charges take the values
\begin{equation}
\label{soliton_charges}
\begin{aligned}
	Q_I=&-\partial_{\varphi^I}\,\widehat\cI=\frac{\pi\kappa^3}{8G}\,C_{IJK}\,\sum_{a=0}^d
		\Big(\hat\flux^J_{a-1}\,\hat\flux^K_a+\hat\flux^J_a\,\hat\flux^K_{a+1}+(v^{a-1},v^{a+1})\,\hat\flux^J_a\,\hat\flux^K_a\Big)\:,\\
	J_i=&-\partial_{\omega_i}\,\widehat\cI=\frac{\pi\kappa^3}{24G}\,C_{IJK}\,\sum_{a=0}^d
		\Big(\hat\flux^I_{a-1}\,\hat\flux^J_a+\hat\flux^I_a\,\hat\flux^J_{a+1}+(v^{a-1},v^{a+1})\,\hat\flux^I_a\,\hat\flux^J_a\Big)\cdot\\
	&\qquad\qquad\qquad\qquad\quad\:\:\cdot\Big((e_i,v^a)(\hat\flux_{a+1}^K-\hat\flux_{a-1}^K)+(e_i,v^{a-1}-v^{a+1})\,\hat\flux^K_a\Big)\:,
\end{aligned}
\end{equation}
which makes the extremand identically zero, $\cS(\varphi^I,\omega_1,\omega_2)=0$.
This is in line with the expectation that the solitonic solutions are not thermal ensembles and thus
their entropy must be zero. Notice that we did not introduced any Lagrange multiplier imposing a constraint involving $\varphi^I$ and $\omega_{1,2}$,
because no such constraint is implied by \eqref{etacoho_vs_varphi2} for soliton solutions.
These constraints usually originate from the linear relation \eqref{etacoho_linear_relation}
among the $\etacoho^I_a$, but for solitonic solutions $\varphi^I$ is determined by $\gaugereal^I$ which is unrelated to the $\etacoho^I_a$.
Rather, the linear relation \eqref{etacoho_linear_relation} only leads to the constraints \eqref{hat_flux_constraint} on the fluxes $\hat\flux^I_a$.

Equation \eqref{soliton_charges} is a prediction for the charges of solitonic solutions that we can compare with the charges of the
explicitly known solutions in the literature.

\subsubsection{Multicharge solitons in AdS}

In \cite{Durgut:2021bpk} asymptotically-AdS multicharge topological solitons with a single bubble were constructed,
generalizing the known solutions in minimal supergravity \cite{Chong:2005hr} (see also \cite{Cassani:2015upa}).
We can describe the topology of these solutions with the fan
\begin{equation}
	V^0=(0,0,1)\:,\qquad V^1=(0,1,1)\:,\qquad V^2=(0,1,0)\:.
\end{equation}
The bubble $\cD_1$ where the Killing vector $V^1$ vanishes has the topology of $S^2\times S^1$ and can support the magnetic flux
$\hat\flux^I_1=\flux^I_1\,\equiv\,\flux^I$.%
\footnote{For this solution the signs defined in \eqref{soliton_signs} are $\sigma_0=\sigma_1=\sigma_2=1$.}
These fluxes must obey the constraint \eqref{hat_flux_constraint} which now reads
\begin{equation}
\label{soliton_flux_constraint}
	\V_I\flux^I=-\frac1{3\kappa}\:.
\end{equation}
Overall the (Euclidean) solitons described by this fan have topology $S^1\times(\cO(-1)\to S^2)$.
The integers $\flux^I$ are identifiable with the fluxes supported on this $S^2$.
Then the formula \eqref{soliton_charges} predicts the following value for the charges:
\begin{equation}
\label{soliton_charges2}
	Q_I=-\frac{\pi\kappa^3}{8G}\,C_{IJK}\,\flux^J\flux^K\:,\qquad
		J_1=J_2=-\frac{\pi\kappa^3}{24G}\,C_{IJK}\,\flux^I\flux^J\flux^K\:.
\end{equation}
We can now compare the above values with the ones computed in \cite{Durgut:2021bpk} from their explicit solution.
For concreteness, let us do the comparison for the STU model first, whose $\V_I$ and $C_{IJK}$ are specified by \eqref{STU}.
The charges \eqref{soliton_charges2} are then
\begin{equation}
\label{soliton_charges_STU}
	Q_I=-\frac{\pi\kappa^3}{4G}\,\frac{\flux^1\flux^2\flux^3}{\flux^I}\:,\qquad
		J_1=J_2=-\frac{\pi\kappa^3}{4G}\,\flux^1\flux^2\flux^3\:,
\end{equation}
and the constraint \eqref{soliton_flux_constraint} becomes
\begin{equation}
\label{soliton_flux_constraint_STU}
	\flux^1+\flux^2+\flux^3=-\frac\ell\kappa\:,
\end{equation}
where the quantity $\ell/\kappa$ must be a non-zero integer  since the $\flux^I$ are integers and $\ell\ne0$.
This might put constraints on the internal manifold of the ten-dimensional uplift. Orbifolding the internal manifold by $\mathbb{Z}_n$
usually helps in avoiding such problems since $\kappa$ is replaced by $\kappa/n$. The fact that there are limits on the possible values on $n$ was already observed in \cite{Cassani:2015upa} in the context of minimal supergravity
and generalized in \cite{Durgut:2021bpk} for the multicharge case; even if any
supersymmetric solution of the STU can locally be uplifted to type IIB supergravity on $S^5/\bZ_n$, there may be topological obstructions that prevent the continuation
of local solutions to smooth global ones.
Our equations \eqref{soliton_charges_STU} and \eqref{soliton_flux_constraint_STU} perfectly match equations
(84-85-86) of \cite{Durgut:2021bpk}, with $G=1$ and the identifications
\begin{equation}
\label{soliton_matching}
	\flux^I\,=\,-\frac1{\kappa}\cD^I\:,\quad(Q_I)_{\text{our}}=\kappa\,(Q_I)_{\text{their}}\:,\quad (J_1+J_2)_{\text{our}}=(J)_{\text{their}}\:,
		\quad (J_1-J_2)_{\text{our}}=(J_\phi)_{\text{their}}\:.
\end{equation}

For a more general supegravity model associated to a symmetric scalar coset, we can also match our results with the explicit solutions of \cite{Durgut:2021bpk}.
Using the relation \eqref{CIJK_inverse}, we find a good match between \eqref{soliton_charges} and equations (65-73-75) of \cite{Durgut:2021bpk}
by setting $G=1$ and using the same identifications as in \eqref{soliton_matching}.

\subsection{A general class of (multi-center) black holes}\label{sec:general}

In this section we consider a general class of Euclidean supersymmetric non-extremal geometries that could arise as continuation  of Lorentzian multi-center black holes.
Using the same terminology of the previous sections, we take a fan of primitive vectors of the form
\begin{equation}\label{Lorfan}
	V^a = (n_a,x_a,y_a)\:,
	\end{equation}
where $n_a=1$ for a horizon and $n_a=0$ for a bubble,  and where we assume that all horizons are surrounded  by bubbles, as it happens in the case of Lorentzian multi-center black holes \cite{Cassani:2025iix}.  We also assume that the boundary is $S^1\times S^3$ and there are no flat connections which can be achieved by choosing 
\begin{equation}
\label{quotiented_BH}
	V^0=(0,0,1)\:,\qquad V^{\bar a}=(1,0,0)\:,\qquad V^d=(0,1,0)\:,
\end{equation}
for some $\bar a$. These choices and the assumption of regularity will allow to simplify the formula \eqref{OSA_localized} by eliminating the dependence on the auxiliary vectors $w^a$. The requirement that $n_a=1$ for an horizon can be easily relaxed to $n_a=\pm 1$ by observing that the formula \eqref{OSA_localized} is invariant under $v^a \rightarrow \sigma_a V^a$ and $\fn_a\rightarrow \sigma_a \fn_a$. On the other hand, to generalize the results of this section to the case $|n_a|\ne 0,1$ or to general Euclidean solutions with adjacent non-vanishing $n_a$ would require a non-trivial work. 
 Notice that our choice contains the fan  \eqref{flat_fan} of the  most general  asymptotically flat Lorentzian black hole  
and the corresponding supersymmetric non extremal geometries considered in \cite{Cassani:2025iix}.\footnote{ Notice however that a generic black saddle in \cite{Cassani:2025iix} can have $|n_a|\ne 0,1$.}
 
  Regularity requires that $V^{a-1}$ and $V^a$ can be completed to a basis of $\mathbb{Z}^3$, which in turn is equivalent to the existence of a vector $w^a$ with $(V^{a-1},V^a,w^a)=1$. In particular, this implies that, for consecutive bubbles,
\begin{equation}\label{consbubbles}
x^{a-1} y^a - y^{a-1} x^a =-h^a \, ,
\end{equation}
where $h^a=\pm 1$.

With the previous assumptions, the fan \eqref{Lorfan} allows one to make a general choice for  the auxiliary vectors $w^a$. 
For each bubble $V^a=(0,x_a,y_a)$ we define the dual vector with integer entries $\tilde V^a=(0,\tilde x_a,\tilde y_a)$  where $x_a \tilde y_a-y_a \tilde x_a=1$, which exists by Bezout lemma since $V^a$  is primitive. Then we can consistently
satisfy the identity $(V^{a-1},V^a,w^a)=1$ by choosing
\begin{itemize}
\item if both $V^a$ and $V^{a-1}$ are bubbles: $w^a= - h^a (1,0,0)$ where $h_a$ is defined in \eqref{consbubbles};
\item  if $V^{a}$ is a bubble and $V^{a-1}$ is a horizon: $w^a=\tilde V^a$. If $V^{a+1}$ is also a bubble, we can equivalently choose $w^a =-h^{a+1} V^{a+1}$ because of  \eqref{consbubbles};
\item  if $V^{a}$ is a horizon and $V^{a-1}$ is a bubble: $w^a=-\tilde V^{a-1}$. If $V^{a-2}$ is also a bubble, we can equivalently choose $w^a =-h^{a-1} V^{a-2}$ because of  \eqref{consbubbles}.
\end{itemize}

With these ingredients and a lot of algebra we can simplify formula \eqref{OSA_localized}.  Reporting all the steps would be too cumbersome and we simply quote the final result. This  is expressed in terms of the variables $\alpha^I_{\bar a}$ and $\fn^I_a$ defined in  \eqref{special_quantization}  which can be physically interpreted as the electric chemical potential and the quantized magnetic fluxes of the black hole. The on-shell action can be written as a sum of four terms
\begin{equation}
\widehat\cI = \frac{\ii\pi^2}{12G} C_{IJK} \sum_{k=0}^3 \cI^{(k)}_{IJK}\, ,\end{equation}
where $\cI^{k}_{IJK}$ is homogeneous of degree $k$ in $\fn_a^I$ and of degree $3-k$ in  $\alpha^I_{\bar a}$. To save space, we will use the abbreviation $(a,b,c)=(V^a,V^b,V^c)$  and other similar ones in the following.
The first three terms can be compactly written as
\begin{align}
\label{OSAk}
	\cI^{(0)}_{IJK} &=\:\,  \alpha_{\bar a}^I\alpha_{\bar a}^J\alpha_{\bar a}^K (\xi, d, 0)^2\sum_{a=0}^d \frac{(a,d,0)(a\!-\!1, a, a\!+\!1)}
		{(\xi, a\!-\!1, a)(\xi, a, a\!+\!1)}\; ,  \nonumber \\
		\cI^{(1)}_{IJK} &=\:\,-3 \kappa	\alpha_{\bar a}^I\alpha_{\bar a}^J (\xi, d, 0)^2\sum_{a=0}^d  \frac{\fn_a^K (a\!-\!1, a, a\!+\!1)}
		{(\xi, a\!-\!1, a)(\xi, a, a\!+\!1)}\; , \\
		\cI^{(2)}_{IJK} &=\:\, 3  	\kappa^2 \alpha_{\bar a}^I (\xi, d, 0) \sum_{a=0}^d \Bigg(
		\frac{\fn^J_{a\!-\!1}\:\fn^K_a}{(\xi, a\!-\!1, a)}+\frac{\fn^J_a\:\fn^K_{a\!+\!1}}{(\xi, a, a\!+\!1)}
	-\frac{\fn^J_a\:\fn^K_a\:(\xi, a\!-\!1, a\!+\!1)}{(\xi, a\!-\!1, a)(\xi, a, a\!+\!1)}\Bigg)\nonumber \:,
		\end{align}
where we notice that $\cI^{(0)}$  is a generalization to non Calabi-Yau fans of the Sasakian volume defined in \cite{Martelli:2005tp,Martelli:2006yb},
$\cI^{(1)}$ contains informations about the Sasakian volume of divisors \cite{Martelli:2005tp,Martelli:2006yb}, and
		$\cI^{(2)}$ is the master volume defined in \cite{Gauntlett:2018dpc} with $\fn_a$ playing
 the role of K\"ahler parameters. The final term $\cI^{(3)}$ is more complicated and reads

\begin{align}
\label{OSA3}
&\cI^{(3)}_{IJK} = -\kappa^3\sum_{a=0}^d \Bigg[
   \frac{\delta_{(a-1, a, a+1)= 0}}{\xi_0} \Bigg(
      3 \fn_a^I \fn^J_a \fn^K_{a\!+\!1} (\xi, a\!+\!1, \bar a)
      - 3 \fn_a^I \fn^J_a \fn^K_{a\!-\!1} (\xi, a\!-\!1, \bar a)  \\
&\qquad\qquad
      + \fn_a^I \fn^J_a \fn^K_a (a\!-\!1, a\!+\!1, \bar a)
        \Big[(a, a\!+\!1, \bar a)(\xi, a\!-\!1, \bar a)
        - (a\!-\!1, a, \bar a)(\xi, a\!+\!1, \bar a)\Big]
   \Bigg) \nonumber \\
&
   + \frac{\delta_{(a\!-\!1, a, a\!+\!1)\ne 0}}{(a\!-\!1, a, a\!+\!1)} \Bigg(
      \frac{\fn_a^I \fn^J_a \fn^K_a (\xi, a\!-\!1, a\!+\!1)^2}
           {(\xi, a\!-\!1, a)(\xi, a, a\!+\!1)}
      - \frac{3 \fn_a^I \fn^J_a \fn^K_{a\!-\!1} (\xi, a\!-\!1, a\!+\!1)}
             {(\xi, a\!-\!1, a)} 
             %\nonumber \\
%&\qquad\qquad
      - \frac{3 \fn_a^I \fn^J_a \fn^K_{a\!+\!1} (\xi, a\!-\!1, a\!+\!1)}
             {(\xi, a, a\!+\!1)}
   \Bigg)\nonumber \\
&
+\delta_{V_0^{a-1}=V_0^{a+1}\ne 0} \, \Bigg ( \fn_a^I \fn^J_a \fn^K_a \frac{(a\!-\!1, \tilde a, a\!+\!1)^2}{(a\!-\!1, a, a\!+\!1)}   -3 \fn_a^I \fn^J_a \fn^K_{a\!-\!1}\frac{(a\!-\!1, \tilde a, a\!+\!1)}{(a\!-\!1, a, a\!+\!1)}  + 3 \fn_a^I \fn^J_a \fn^K_{a\!+\!1}\frac{(a\!-\!1, \tilde a, a\!+\!1)}{(a\!-\!1, a, a\!+\!1)} \Bigg )
\Bigg] \nonumber
\end{align}
where $\tilde a$ stands for $\tilde V^a$, $\bar a$ for $V^{\bar a}=(1,0,0)$ and we need to distinguish the  contributions of triples of consecutive coplanar vectors, $(a\!-\!1, a, a\!+\!1)= 0$,  from  generic triples with $(a\!-\!1, a, a\!+\!1)\ne 0$ and add a specific contribution every time a bubble is surrounded by two horizons. One can check that the coplanar contribution is consistent with what we found in section \ref{sect:example top sol}. The result for $\cI^{(0)}$, $\cI^{(1)}$ and $\cI^{(2)}$ are valid in general, while the result for $\cI^{(3)}$ has been checked for the class of geometries \eqref{Lorfan} only. In deriving the previous identities we used the fact that $\fn_a^I$ are integers and discarded $2\pi \ii\, \mathbb{Z}$  contributions to the on-shell action.  As in section \ref{sec:lensesandrings},   we can extrapolate our analysis   to cover solutions in the ungauged supergravity too. 

Altogether the first three terms $\cI^{(0)}$, $\cI^{(1)}$ and $\cI^{(2)}$  are formally in correspondence with the expansion up to quadratic order in the K\"ahler parameters of the equivariant volume of the cone over the five dimensional geometry, as discussed in \cite{Martelli:2023oqk}.  $\cI^{(3)}$ has also an interpretation in terms of the cubic term in the equivariant volume, but it is more complicated. We will  discuss all these issues in details  in section \ref{sec:equivariant}. 

We can further simplify the on-shell action for solutions in minimal gauged supergravity.  We obtain minimal gauged supergravity from the STU model by setting $V^I=\frac{1}{3\ell}$ and identifying the three gauge fields. There the relations \eqref{etacoho_vs_varphi}  and  \eqref{etacoho_vs_varphi2}
\begin{equation}
	\alpha_a\:=\frac{\ell}{3}\,\frac{(\xi,V^0,V^a)+(\xi,V^a,V^d)-(\xi,V^0,V^d)}{(\xi,V^0,V^d)}\:,
\end{equation}
fixes $\alpha_{\bar a}$ and the fluxes $\fn_a$ completely in terms of $\xi$ and $V^a$. Explicitly, 
\begin{align}\label{constr}
\alpha_{\bar a}&=-\frac{\ell}{3} \frac{\xi_0+\xi_1+\xi_2}{\xi_0} = -\frac{\ell}{3}\frac{\eta_i \xi_i}{\xi_0}\, , \\ \label{constr2}
\fn_a &=  \frac{\ell}{3\kappa} (1-x_a-y_a-n_a) =   \frac{\ell}{3\kappa}(1-\eta_i V_i^a)\, ,
\end{align} 
where  we introduced the vector $\eta=(1,1,1)$. Notice that, similarly to the case of the topological solitons discussed in section \ref{sect:example top sol},  the formula for the fluxes generically puts  constraints on the internal manifold of the ten-dimensional type IIB uplift, which can be usually solved by performing a suitable orbifold projection  $\mathbb{Z}_n$ under which  $\kappa$ is replaced by $\kappa/n$. 
Setting as usual 
\begin{equation} \xi =\frac{1}{\beta}(2 \pi \ii, -\omega_1,-\omega_2) \end{equation}
we see that the expression for $\alpha_{\bar a}$  is nothing else that the linear constraint that relates  the thermodynamical potential $\varphi =\frac{2\pi \ii} {\kappa}  \alpha_{\bar a}$ to the angular momentum chemical potentials
\begin{equation}
\varphi=\omega_1 +\omega_2 - 2\pi \ii \, .
\end{equation}
With some algebra and a lot of patience, using $\alpha_{\bar a} (\xi, d,0) = - \frac{\ell}{3} \eta_i \xi_i$, one can check that for all $\fn_a$
\begin{equation}\label{rels}
 \sum_{a=0}^d \eta_i V_i^a \frac{\partial \cI^{(k)}}{\partial \fn_a} = (4-k)  \frac{3\kappa}{\ell} \cI^{(k-1)}\, ,
\end{equation}
where the relation with $k=3$ holds only up to integer terms.\footnote{More precisely, the relation holds up to terms $\kappa^3 A_{ab} \fn_a \fn_b$ where $A_{ab}$ are $\xi$-independent and integer. Such terms only contribute $2\pi \ii \, \mathbb{Z}$ shifts to the result \eqref{gauged0} for the on-shell action.}   
It easily follows from the relations \eqref{rels} that $\widehat\cI$ is a function of the combination $ \fn_a +  \frac{\ell}{3\kappa} \eta_i V_i^a$ only (modulo $2\pi \ii \, \mathbb{Z}$). Indeed, it follows from Taylor expansion that 
\begin{equation}\label{gauged0}
\widehat\cI(\fn_a) = \frac{\ii\pi^2}{2G}   \cI^{(3)}(\fn_a +  \frac{\ell}{3\kappa} \eta_i V_i^a)\, ,
\end{equation}
and, therefore, \eqref{constr2} gives
\begin{equation}\label{gauged}
\widehat\cI(\fn_a) = \frac{\ii\pi^2}{2G}    \cI^{(3)}\left ( \frac{\ell}{3\kappa}\right )\, .
\end{equation}
As already discussed in sections \ref{example BH} and \ref{sec:lensesandrings}, in  ungauged supergravity, the constraints \eqref{constr} and \eqref{constr2} are modified. $\varphi$ and $\fn_a$ are now independent variables and the right hand side of the relations \eqref{constr} and \eqref{constr2}  vanishes, imposing the linear constraint $\omega_1+\omega_2=2\pi \ii$ among the  angular momentum chemical potentials only and a condition on the fan vector that is equivalent to the CY condition. Therefore, \eqref{gauged} is valid only in gauged supergravity.

\subsection{Relation to the equivariant volume}\label{sec:equivariant}

In this section we discuss the relation of the form of the on-shell action discussed in section \ref{sec:general} to the equivariant volume of the  non-compact toric geometry of complex dimension three with fan $V^a$. It was observed in \cite{Martelli:2023oqk,Colombo:2023fhu} that many extremization problems in holography can be expressed in terms of the
equivariant volume of the internal manifold. Here we will show that a similar statement can be made for the equivariant volume of the six-dimensional geometry associated with a large class of non-compact asymptotically-AdS geometries. We will also make a connection between the odd-dimensional localization for a five-dimensional manifold and the more standard ABBV formula \cite{Duistermaat,berline1982classes,ATIYAH19841} for the cone over it.

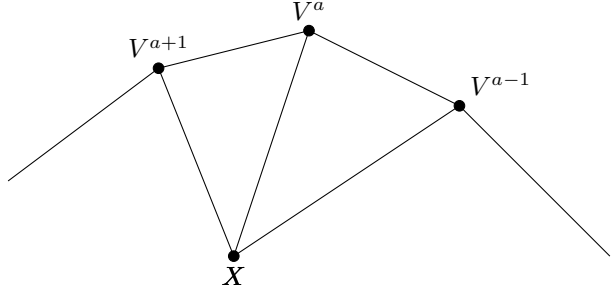
\begin{figure}[hhh!]
 \begin{center}
\begin{tikzpicture}[font = \footnotesize]
\draw (-4,0)--(-2,1.5)  node[above] {$V^{a+1}$} --(0,2) node[above] {$V^{a}$}--(2,1) node[above right] {$V^{a-1}$}--(4,-1);
\draw (-1,-1) node[below] {$X$}--(0,2);
\draw (-1,-1) node[below] {$X$}--(-2,1.5);
\draw (-1,-1) node[below] {$X$}--(2,1);
\filldraw (0,2) circle (2pt);
\filldraw (-2,1.5)  circle (2pt);
\filldraw (2,1)  circle (2pt);
\filldraw (-1,-1) circle (2pt);
%\node[draw=none] at (0.2, 0.8) {I}; 
%\node[draw=none] at (-1, 0.8) {II}; 
\end{tikzpicture}
\caption{The triangulation of the fan.}\label{fig:threefold}
\end{center}
\end{figure}

The equivariant volume for toric three-folds $M_6$ is defined as
\begin{equation}\label{evolH} \mathbb{V}(\lambda_a,\xi_i) = \frac{1}{(2 \pi)^3}\int_{M_{6}} e^{-H}  \frac{\omega^3}{3!} \, ,\end{equation}
where $\omega= \dd y_i\wedge \dd \phi_i$ is the symplectic form, $H=\xi_i y_i$ is the Hamiltonian  for the vector field $\xi$
and the geometry $M_6$ is a torus fibration over the three-dimensional polytope \begin{equation}\mathcal{P}=\{ y\in \mathbb{R}^3 \, : \, y_i V_i^a - \lambda_{a}  \ge 0\}\end{equation} whose shape is controlled by the K\"ahler parameters $\lambda_a$. We use the notations of \cite{Martelli:2023oqk} to which we refer for more details. We are interested in cases where the vectors of the fan span a cone in $\mathbb{R}^3$. The equivariant volume can be computed by the ABBV formula for even-dimensional geometry provided the fan is resolved to have at worst orbifold singularities. This can be done by triangulating the fan by adding an internal vector $X$ as in figure \ref{fig:threefold}. 
Each tetrahedron $(V^{a-1},V^a,X)$ corresponds to a fixed point and we can compute the equivariant volume as \cite{Martelli:2023oqk} 
\begin{equation}\label{eqfixed}
\mathbb{V}(\lambda_A,\xi_i) =  \sum_{a=0}^d \frac{d_a^2 e^{-[\lambda_X (a\!-\!1, a, \xi) +\lambda_{a-1} (a, X, \xi) +\lambda_a ( X, a\!-\!1, \xi)]/d_a}}{(a\!-\!1, a, \xi)  (a, X, \xi)  ( X, a\!-\!1, \xi)}\; , 
\end{equation}
where $\lambda_X$ is the K\"ahler parameter associated with the internal point $X$, and $d_a=(a\!-\!1, a, X)$. In writing this formula, we assumed that the cone is convex. For non-convex fans we take the formula \eqref{eqfixed} as the operational definition of the equivariant volume.  It is easy to check that
\begin{equation}
\mathbb{V}(\lambda_A + \beta_i V_i^A,\xi)= e^{-\beta_i \xi_i} \mathbb{V}(\lambda_A ,\xi)
\end{equation}
or, equivalently,
\begin{equation}\label{rec}
\sum_A V_i^A \frac{\partial \mathbb{V}}{\partial \lambda_A}=-\xi_i \mathbb{V}\,  ,
\end{equation}
where $\lambda_A$ collectively runs over $\lambda_a$ and $\lambda_X$. The equivariant volume can be expanded in a formal power series in the K\"ahler parameters
\begin{equation}
 \mathbb{V}(\lambda_A,\xi)=\sum_{k=0}^\infty \mathbb{V}^{(k)}(\lambda_A,\xi)\; , 
 \end{equation}
 where $\mathbb{V}^{(k)}$ is homogeneous of degree $k$ in $\lambda_A$. On  general grounds, for a threefold the terms up to the quadratic order are independent of both $X$ and $\lambda_X$ \cite{Martelli:2023oqk}. These terms explicitly read
 \begin{align}
\label{Vk}
	\mathbb{V}^{(0)} &=\:\, \sum_{a=0}^d \frac{ (X, a, a\!+\!1)^2}
		{(\xi, a\!+\!1, X)(\xi, X, a) (\xi, a, a\!+\!1)} = \frac{1}{\mu_i \xi_i}\sum_{a=0}^d \mu_i V_i^a \frac{ (a\!-\!1, a, a\!+\!1)}
		{(\xi, a\!-\!1, a)(\xi, a, a\!+\!1)} \; , \nonumber \\
		\mathbb{V}^{(1)} &=\:\,- \sum_{a=0}^d  \frac{\lambda_a (a\!-\!1, a, a\!+\!1)}
		{(\xi, a\!-\!1, a)(\xi, a, a\!+\!1)}\; , \\
		\mathbb{V}^{(2)} &=\:\, \frac12 \sum_{a=0}^d \Bigg(
		\frac{\lambda_{a\!-\!1}\:\lambda_a}{(\xi, a\!-\!1, a)}+\frac{\lambda_a\:\lambda_{a\!+\!1}}{(\xi, a, a\!+\!1)}
	-\frac{\lambda_a^2\:(\xi, a\!-\!1, a\!+\!1)}{(\xi, a\!-\!1, a)(\xi, a, a\!+\!1)}\Bigg)\nonumber \:,
		\end{align}
where the first identity holds for arbitrary $\mu_i\in \mathbb{R}$. 		
The last two equations are easy to check.\footnote{See for example appendix  B of \cite{Martelli:2023oqk}.}  One way to see that the two expressions in the first line are equal is to set $\lambda_a=0$ in  \eqref{rec} and contracting with $\mu_i$. Setting $\lambda_X=0$ in \eqref{rec} we also find
\begin{equation}\label{rec2}
\sum_{a=0}^d V_i^a \frac{\partial \mathbb{V}^{(k)}}{\partial \lambda_a}=-\xi_i \mathbb{V}^{(k-1)}\, ,\qquad\qquad k=1,2\ .
\end{equation} 
$\mathbb{V}^{(0)}$, $\mathbb{V}^{(1)}$ and $\mathbb{V}^{(2)}$ are a generalization of quantities defined  for Calabi-Yau cones: the Sasakian volume \cite{Martelli:2005tp,Martelli:2006yb},    the generating function for the volumes of divisors and   the master volume \cite{Gauntlett:2018dpc}, respectively.

We dubbed the vector $X$ by no coincidence. There is a connection between the odd-dimensional localization for a five-dimensional Sasakian manifold $Y_5$ with respect to the Killing vector $X$ and the even-dimensional localization  for the cone $M_6=C(Y_5)$ with resolution vector field $X$. One can easily convince oneself that  there is a one-to-one correspondence between the fixed points of the even-dimensional ABBV formula and the close orbits of the five-dimensional localization. Moreover  the corresponding contributions to the calculation of the Sasakian and master volume match exactly.

The cubic term in the equivariant volume is not universal and it depends on both $X$ and $\lambda_X$. For the class of geometries \eqref{Lorfan}, one can check that, for {\it integer} $\lambda_a$ and $\lambda_X=0$, 
\begin{equation}
 \mathbb{V}^{(3)}(\lambda_a,\xi,X)=  \hat{ \mathbb{V}}^{(3)}(\lambda_a,\xi) - \hat{\mathbb{V}}^{(3)}(\lambda_a,X) \qquad {\rm (mod}\,  \mathbb{Z}/6)
 \end{equation}
 where
 \begin{align}
\label{V3}
&\hat{ \mathbb{V}}^{(3)} = -\frac16\sum_{a=0}^d \Bigg[
   \frac{\delta_{(a-1, a, a+1)= 0}}{\xi_0} \Bigg(
      3 \lambda_a^2 \lambda_{a\!+\!1} (\xi, a\!+\!1, \bar a)
      - 3 \lambda_a^2 \lambda_{a\!-\!1} (\xi, a\!-\!1, \bar a)  \\
&\qquad\qquad
      + \lambda_a^3 (a\!-\!1, a\!+\!1, \bar a)
        \Big[(a, a\!+\!1, \bar a)(\xi, a\!-\!1, \bar a)
        - (a\!-\!1, a, \bar a)(\xi, a\!+\!1, \bar a)\Big]
   \Bigg) \nonumber \\
&
   + \frac{\delta_{(a\!-\!1, a, a\!+\!1)\ne 0}}{(a\!-\!1, a, a\!+\!1)} \Bigg(
      \frac{\lambda_a^3 (\xi, a\!-\!1, a\!+\!1)^2}
           {(\xi, a\!-\!1, a)(\xi, a, a\!+\!1)}
      - \frac{3 \lambda_a^2 \lambda_{a\!-\!1} (\xi, a\!-\!1, a\!+\!1)}
             {(\xi, a\!-\!1, a)} 
             %\nonumber \\
%&\qquad\qquad
      - \frac{3 \lambda_a^2 \lambda_{a\!+\!1} (\xi, a\!-\!1, a\!+\!1)}
             {(\xi, a, a\!+\!1)}
   \Bigg)\nonumber \\
&
+\delta_{V_0^{a-1}=V_0^{a+1}\ne 0} \, \Bigg ( \lambda_a^3 \frac{(a\!-\!1, \tilde a, a\!+\!1)^2}{(a\!-\!1, a, a\!+\!1)}   -3 \lambda_a^2 \lambda_{a\!-\!1}\frac{(a\!-\!1, \tilde a, a\!+\!1)}{(a\!-\!1, a, a\!+\!1)}  + 3 \lambda_a^2 \lambda_{a\!+\!1}\frac{(a\!-\!1, \tilde a, a\!+\!1)}{(a\!-\!1, a, a\!+\!1)} \Bigg )
\Bigg] \nonumber \, ,
\end{align}
  which satisfies
  \begin{equation}\label{rec3}
\sum_{a=0}^d V_i^a \frac{\partial \hat{\mathbb{V}}^{(3)}(\lambda_a,\xi)}{\partial \lambda_a}=-\xi_i \mathbb{V}^{(2)}(\lambda_a,\xi)\, \qquad {\rm (mod}\,  \mathbb{Z}/6) \, .
\end{equation} 
This last identity is related to \eqref{rec}. To use \eqref{rec} we need to be careful that $\mathbb{V}^{(3)}$ now depends on $\lambda_X$ and, although we set it to zero, we need to keep the derivatives with respect to it
\begin{equation}
\sum_{a=0}^d V_i^a \frac{\partial \mathbb{V}^{(3)}}{\partial \lambda_a} + X_i \frac{\partial \mathbb{V}^{(3)}}{\partial \lambda_X} =-\xi_i \mathbb{V}^{(2)}\, .
\end{equation} 
One easily computes $\frac{\partial \mathbb{V}^{(3)}}{\partial \lambda_X}=-\mathbb{V}^{(2)}(\xi=X)$ and therefore
\begin{equation}
\sum_{a=0}^d V_i^a \frac{\partial \hat{\mathbb{V}}^{(3)}(\xi)}{\partial \lambda_a} -\sum_{a=0}^d V_i^a \frac{\partial \hat{\mathbb{V}}^{(3)}(X)}{\partial \lambda_a} =-\xi_i \mathbb{V}^{(2)}(\xi) + X_i \mathbb{V}^{(2)}(\xi=X) \, ,
\end{equation} 
which is indeed compatible with \eqref{rec3}. It would be interesting to see if \eqref{V3} and \eqref{rec3} hold for a generic choice of fan.

The reader has certainly already appreciated the formal similarities between  the formulae for the equivariant volume and those appearing in the on-shell action for fans of the form \eqref{Lorfan}. The terms \eqref{OSAk} and \eqref{OSA3} are a multivariable generalizations of  the first three homogeneous pieces in the equivariant volume expansion in power of $\lambda_a$. To be more explicit, specializing to minimal gauged supergravity, we can write
\begin{equation}
\begin{aligned}
\widehat\cI = \frac{\ii\pi^2}{2G} \Bigg ( &(\alpha_{\bar a}(\xi,d,0))^3 \mathbb{V}^{(0)}(\fn_a,\xi)   + 3 \kappa(\alpha_{\bar a}(\xi,d,0))^2 \mathbb{V}^{(1)}(\fn_a,\xi) \\ &+ 6 \kappa^2 (\alpha_{\bar a}(\xi,d,0)) \mathbb{V}^{(2)}(\fn_a,\xi) +6 \kappa^3 \hat{\mathbb{V}}^{(3)}(\fn_a,\xi) \Bigg )\, .
\end{aligned}
\end{equation}
Moreover the identities \eqref{rels} follow from the properties \eqref{rec2} and \eqref{rec3}.

We see again that the equivariant volume  plays an important role in holography  as argued in \cite{Martelli:2023oqk,Colombo:2023fhu}. Our results points toward the existence of another physically relevant component  of the equivariant volume, beside the Sasakian and master volume, which appear at order three in the power series expansion in the K\'alher parameters. We denoted it with $\hat{\mathbb{V}}^{(3)}$ and we defined it only for the particular class of geometries \eqref{Lorfan}. It would be very interesting to understand better its role  and how to generalise its definition for a generic fan.

\section{Boundary analysis}
\label{Boundary analysis}
In this section we analyse the boundary contribution identified in \eqref{bdy_contribution}. We work in Lorentzian signature and we also set the AdS scale $\ell = 1$.  We will make two simplifying assumptions, making this section less general than the rest of the paper: first, we specialize to the STU model (see \cref{sect:sugra}, \ref{STU_model}), second, we will assume more isometry compared to the other sections, namely: $\mathbb{R} \times \text{SU}\qty(2) \times \text{U}\qty(1)$, where $\mathbb{R}$ denotes the time coordinate in Lorentzian signature.  Defining the left-invariant one forms
\begin{align}\label{eq:left-invariant}
  \begin{aligned}
    \sigma_1 ={}& \cos^{}{\psi} \dd[]{\theta} + \sin^{}{\theta} \sin^{}{\psi} \dd[]{\phi} \,, \\
    \sigma_2 ={}& - \sin^{}{\psi} \dd[]{\theta} + \sin^{}{\theta} \cos^{}{\psi} \dd[]{\phi} \,, \\
    \sigma_3 ={}& \dd[]{\psi} + \cos^{}{\theta} \dd[]{\phi} \,, 
  \end{aligned}
\end{align}
the most general field configuration that respects this isometry and is asymptotically locally AdS reads
\begin{align}
  \begin{aligned}
  \dd[]{s}^2_g ={}& \frac{1}{\rho^2} \qty[ \dd[]{\rho}^2 + g_{tt}(\rho) \dd[]{t}^2 + g_{\theta\theta}(\rho) \qty(\sigma_1^2 + \sigma_2^2) + g_{\psi\psi}(\rho) \sigma_3^2 + 2 g_{t\psi}(\rho) \dd[]{t} \sigma_3] \,, \\
  A^I_{(0)} ={}& A^I_t(\rho) \dd[]{t} + A^I_\psi(\rho) \sigma_3 \,, \\
  X^I ={}& X^I(\rho) \,. 
  \end{aligned}
\end{align}
Here $\rho$ is the Fefferman-Graham radial coordinate and the boundary is located at $\rho = 0$.
We can expand the functions of $\rho$ as
\begin{align}
  \begin{aligned}
    g_{ \bullet \bullet } ={}& g_{ \bullet \bullet }^{(0,0)} + g_{ \bullet \bullet }^{(2,0)} \rho^2 + \qty(g_{ \bullet \bullet }^{(4,0)} + g_{ \bullet \bullet }^{(4,1)} \log{\rho} + g_{ \bullet \bullet }^{(4,2)} (\log{\rho})^2) \rho^4 + \dots \,, \\[5pt]
    A^I_{ \bullet } ={}& A^{I(0,0)}_{ \bullet } + \qty(A^{I(2,0)}_{ \bullet } + A^{I(2,1)}_{ \bullet } \log{\rho}) \rho^2 + \qty(A^{I(4,0)}_{ \bullet } + A^{I(4,1)}_{ \bullet } \log{\rho} + A^{I(4,2)}_{ \bullet } (\log{\rho})^2) \rho^4 + \dots \,,  \\[5pt]
  X^I ={}& \bar{X}^I + \qty(X^{I(0,0)} + X^{I(0,1)} \log{\rho}) \rho^2 + \qty(X^{I(2,0)} + X^{I(2,1)} \log{\rho} + X^{I(2,2)} (\log{\rho})^2) \rho^4 + \dots \,, 
  \end{aligned}
\end{align}
where two bullets stand for $\qty(tt, \theta\theta, \psi\psi, t\psi)$ and a single bullet stands for $\qty(t, \psi)$, we have expanded to the relevant order in 5d and the $(\log{\rho})^2$ terms appear, since the scalars source more logarithmic terms compared to Einstein-Maxwell theories. The boundary data is specified by
\begin{align}
  \qty(g_{ \bullet \bullet }^{(0,0)} \,, A_{ \bullet }^{I(0,0)} \,, X^{I (0,0)}) \,. 
\end{align}
We fix the boundary metric to be
\begin{align}\label{eq:boundary_metric}
  \begin{aligned}
  \dd[]{s}^2_{g^{(0,0)}} ={}& - \dd[]{t}^2 + \frac{1}{4} \qty(\sigma_1^2 + \sigma_2^2 + v^2 \sigma_3^2 ) \\
  \implies& \quad g_{t t}^{(0,0)} = - 1 \,, \quad g_{\theta\theta}^{(0,0)} = \frac{1}{4} \,, \quad g_{\psi\psi}^{(0,0)} = \frac{v^2}{4} \,, \quad g_{t\psi}^{(0,0)} = 0 \,,
  \end{aligned}
\end{align}
where $ v > 0$ is a squashing parameter. The rationale behind this choice is that we want a simple, yet non-trivial, squashing of the $S^3$, but we want to eliminate twisting between $\mathbb{R}$ and the $S^3$ at infinity. Locally, such twisting can always be eliminated and it is in fact the preferred Fefferman-Graham coordinate system for holographic renormalization analysis for known rotating black holes.

\subsection{Asymptotic perturbative solutions}\label{Asymptotic perturbative solutions}
First, we impose the Einstein equations \eqref{eq:einstein} and the Maxwell equations \eqref{F_eq}. This fixes
\begin{align}\label{eq:orig_expansion}
  \begin{aligned}
  & g_{ \bullet \bullet }^{(2,0)} \,, & & g_{\psi\psi}^{(4,0)} \,, & & g_{ \bullet \bullet }^{(4,1)} \,, & & g_{ \bullet \bullet }^{(4,2)} \,, \\
  & A^{I(2,1)}_{ \bullet } \,, & & A^{I(4,0)}_{ \bullet } \,, & & A^{I(4,1)}_{ \bullet } \,, & & A^{I(4,2)}_{ \bullet } \,, \\
  & & & X^{I(2,0)} \,, & & X^{I(2,1)} \,, & & X^{I(2,2)} \,. 
  \end{aligned}
\end{align}
Then we demand supersymmetry. To achieve that we perform the coordinate transformation
\begin{align}
  t ={}& y \,, \quad \psi = \widehat{\psi} + c y \,,
\end{align}
and express the metric as
\begin{align}\label{eq:susy_metric}
  \dd[]{s}^2_g ={}& - f^2 \qty(\dd[]{y} + \frac{\omega_\psi}{\rho^2} \, \widehat{\sigma}_3)^2 + f^{-1} \qty(\frac{\gamma_{\rho\rho}^2}{\rho^2} \dd[]{\rho}^2 + \frac{\gamma_{\theta\theta}^2}{\rho^2} \qty(\widehat{\sigma}_1^2 + \widehat{\sigma}_2^2) + \frac{\gamma_{\psi\psi}^2}{\rho^4} \, \widehat{\sigma}_3^2) \,, 
\end{align}
where $\widehat{\sigma}_{1,2,3}$ are the same as \eqref{eq:left-invariant} but with $\psi$ replaced by $\widehat{\psi}$, $\qty(f, \omega_\phi, \gamma_{\rho\rho}, \gamma_{\theta\theta}, \gamma_{\psi\psi})$ are functions of $\rho$, and the additional factors of $\rho$ are introduced for convenience. The constant $c$ is fixed to
\begin{align}
  c ={}& \frac{2}{v} \,, 
\end{align}
such that $f$ doesn't diverge at infinity. Then its expansion takes the shape
\begin{align}\label{eq:f_expansion}
  f ={}& v + \qty(f^{(2,0)} + f^{(2,1)} \log{\rho}) \rho^2 + \qty(f^{(4,0)} + f^{(4,1)} \log{\rho} + f^{(4,2)} (\log{\rho})^2) \rho^4 + \dots \,, 
\end{align}
with the coefficients completely fixed in terms of the original expansion coefficients in \eqref{eq:orig_expansion}. The remaining functions are then determined through
\begin{align}\label{eq:sol_gamma}
  \gamma_{\rho\rho}^2 ={}& f \,, \quad \gamma_{\theta\theta}^2 = f g_{\theta\theta} \,, \quad \gamma_{\psi\psi}^2 = \frac{g_{t\psi}^2 - g_{t t} g_{\psi\psi}}{f} \,, \quad \omega_\psi = - \frac{g_{t\psi} + \frac{2}{ v} g_{\psi\psi}}{f^2} \,,
\end{align}
and the factors of $\rho$ in \eqref{eq:susy_metric} ensure that their expansions start at $\order{1}$. Next we demand that
\begin{align}\label{eq:susy_F}
  \dd[]{A^I} = F^I ={}& \dd[]{\qty[X^I f \qty(\dd[]{y} + \frac{\omega_\psi}{\rho^2} \, \widehat{\sigma}_3)]} + \text{basic} \,. 
\end{align}
This strong constraint fixes some of the non-boundary functions:
\begin{align}
  g_{t\psi}^{(4,0)} \,, \quad A^{1(2,0)}_{\psi} \,, \quad A^{2(2,0)}_{\psi} \,, \quad X^{I(0,1)} \,,
\end{align}
and provides the following constraint between the boundary functions
\begin{align}
  \bar{X}_I A^{I(0,0)}_{\psi} ={}& \frac{1}{3} (v^2 - 1) \,. 
\end{align}
To proceed we have to demand that the base
\begin{align}
  \dd[]{s}^2_\gamma ={}& \frac{\gamma_{\rho\rho}^2}{\rho^2} \dd[]{\rho}^2 + \frac{\gamma_{\theta\theta}^2}{\rho^2} \qty(\widehat{\sigma}_1^2 + \widehat{\sigma}_2^2) + \frac{\gamma_{\psi\psi}^2}{\rho^4} \, \widehat{\sigma}_3^2 \,, 
\end{align}
has the required structure that we have reviewed in section \ref{Supersymmetric solutions in the time-like class}. We take the following ansatz for the anti-self-dual two-forms that define the almost-hyper-K\"ahler structure:
\begin{align}
  \begin{aligned}
  \mathcal{X}^1 ={}& x^1(\rho) \qty[\dd[]{\rho} \wedge \widehat{\sigma}_3 + \frac{\rho \gamma_{\theta\theta}^2}{\gamma_{\rho\rho} \gamma_{\psi\psi}} \widehat{\sigma}_1 \wedge \widehat{\sigma}_2] \,, \\
  \mathcal{X}^2 ={}& x^2(\rho) \qty[\dd[]{\rho} \wedge \widehat{\sigma}_1 + \frac{\gamma_{\psi\psi}}{\rho \gamma_{\rho\rho}} \widehat{\sigma}_2 \wedge \widehat{\sigma}_3 ] \,, \\
  \mathcal{X}^3 ={}& x^3(\rho) \qty[\dd[]{\rho} \wedge \widehat{\sigma}_2 + \frac{\gamma_{\psi\psi}}{\rho \gamma_{\rho\rho}} \widehat{\sigma}_3 \wedge \widehat{\sigma}_1] \,. 
  \end{aligned}
\end{align}
Then, the algebra
\begin{align}
  \qty(\mathcal{X}^i)\indices{_m^p} \qty(\mathcal{X}^j)\indices{_p^n} ={}& - \delta^{ij} \delta\indices{_m^n} + \varepsilon^{ijk} \qty(\mathcal{X}^k)\indices{_m^n} \,, 
\end{align}
fixes\footnote{This fixing is up to signs, namely, $( -, -, -)$, $( -, +, +)$, $( +, -, +)$ are also allowed. The choice $( +, +, -)$ is consistent with our orientation choices for the total space and the base.}
\begin{align}
  x^1 ={}& \frac{\gamma_{\rho\rho} \gamma_{\psi\psi}}{\rho^3} \,, \quad x^2 = \frac{\gamma_{\rho\rho} \gamma_{\theta\theta}}{\rho^2} \,, \quad x^3 = - \frac{\gamma_{\rho\rho} \gamma_{\theta\theta}}{\rho^2} \,,
\end{align}
and ensures that
\begin{align}
  \star_\gamma ={}& - \frac{1}{2} \mathcal{X}^1 \wedge \mathcal{X}^1 = - \frac{1}{2} \mathcal{X}^2 \wedge \mathcal{X}^2 = - \frac{1}{2} \mathcal{X}^3 \wedge \mathcal{X}^3 \,. 
\end{align}
Demanding that $\mathcal{X}^1$ is closed leads to
\begin{align}
  \dd[]{\mathcal{X}^1} ={}& 0 \quad \implies \quad \rho \, \partial_\rho \gamma_{\theta\theta} = \gamma_{\theta\theta} - \frac{\gamma_{\rho\rho} \gamma_{\psi\psi}}{2 \gamma_{\theta\theta}} \,, 
\end{align}
however we observe that this relation is already satisfied after the equation of motion and \eqref{eq:susy_F} are imposed, leading to no new constraints. The potential for the Ricci form $\mathcal{R} = \dd[]{P}$ can be found as
\begin{align}
  \qty(\nabla_m + \iu P_m) \qty(\mathcal{X}^2 + \iu \mathcal{X}^3)_{np} ={}& 0 \,, 
\end{align}
which uniquely determines
\begin{align}
  P ={}& \qty( - 1 + \frac{\gamma_{\psi\psi}^2}{2 \rho^2 \gamma_{\theta\theta}^2} + \frac{2 \gamma_{\psi\psi} - \rho \partial_\rho \gamma_{\psi\psi}}{\rho^2 \gamma_{\rho\rho}}) \, \widehat{\sigma}_3 \,. 
\end{align}
We can then verify that
\begin{align}
  \mathcal{R}_{mn} ={}& \frac{1}{2} (R_\gamma)_{mnpq} \qty(\mathcal{X}^1)^{pq} \,, 
\end{align}
is indeed satisfied. The relation
\begin{align}
  f ={}& \frac{4}{R_\gamma} \mathcal{V} = - \frac{8}{R_\gamma} \qty(\frac{1}{X^1} + \frac{1}{X^2} + \frac{1}{X^3}) 
\end{align}
is automatically satisfied once the equations of motion and \eqref{eq:susy_F} are imposed, leading to no new constraints. To proceed we calculate
\begin{align}
  \mathcal{V}^I ={}& -9 f^{-1} C^{IJK} \bar{X}_J X_K \,, 
\end{align}
and we extract the basic self-dual two form $\Theta^I$ from
\begin{align}
  F^I ={}& \dd[]{\qty[X^I f \qty(\dd[]{y} + \frac{\omega_\psi}{\rho^2} \, \widehat{\sigma}_3)]} + \Theta^I + \mathcal{V}^I \mathcal{X}^1 \,. 
\end{align}
The last two supersymmetry relations that we have to impose are
\begin{align}
  \bar{X}_I \Theta^I ={}& \frac{1}{3} \qty(\mathcal{R} - \frac{1}{4} R_\gamma \mathcal{X}^1) \,, \quad X_I \Theta^I = - \frac{1}{3} f \qty(\dd[]{\omega} + \star_\gamma \dd[]{\omega}) \,, 
\end{align}
where in the current setup $\omega = \frac{\omega_\psi}{\rho^2} \, \widehat{\sigma}_3$. Both of these relations leads to a single new constraint on $g_{\theta\theta}^{(4,0)}$. 

In summary: Given the choice of boundary metric \eqref{eq:boundary_metric} the supersymmetry equations demand that the boundary data
\begin{align}\label{eq:bdy_data}
  \text{boundary data}: \quad \qty(A^{I(0,0)}_t \,, A^{I(0,0)}_\psi \,, X^{I(0,0)}) \,, 
\end{align}
is subject to the constraint\footnote{In addition to the STU model constraint $X^1 X^2 X^3 = 1$.}
\begin{align}
  \bar{X}_I A^{I(0,0)}_\psi = \frac{1}{3} ( v^2 - 1) \,. 
\end{align}
Among the non-boundary constants, after imposing the equations of motion and supersymmetry what remains is
\begin{align}\label{eq:non_bdy_data}
  \text{non-boundary data}: \quad \qty( g_{tt}^{(4,0)} \,, A^{3(2,0)} ) \,.
\end{align}
Notably, throughout the calculation we always work with the Fefferman-Graham radial coordinate $\rho$. First we impose the equations of motion using the Fefferman-Graham coordinates $\qty(t, \rho, \theta, \psi, \phi)$. Then, after the simple coordinate transformation
\begin{align}\label{eq:t_to_y}
  t ={}& y \,, \quad \psi = \widehat{\psi} + \frac{2}{ v} y \,, 
\end{align}
we express all the supersymmetric quantities (eg. $f, \omega, \Theta^I, \mathcal{X}^1, \mathcal{R}, \dots $) as direct $\rho$-expansions, whose coefficients are related to the coefficients in \eqref{eq:orig_expansion}, though \eqref{eq:f_expansion} and \eqref{eq:sol_gamma}.
In contrast, in \cite{Colombo:2025ihp} we began with a particular coordinate system which makes supersymmetry manifest, but is not Fefferman-Graham. Then we had to switch to a Fefferman-Graham coordinate system (equation (D.22)) in order to compute the boundary integrals.

\subsection{Holographic renormalization and background subtraction}

\noindent
Including only the standard (divergent) counterterms, the overall boundary contribution in \eqref{bdy_contribution} is the sum of\footnote{Notice that we are ignoring the terms in the third line of \eqref{bdy_contribution}; indeed, we can fix the redundancy of the cohomological parameters $\etacoho^I_a$ by setting $\etacoho^I_0=0=\etacoho^I_d$.}:
\begin{align}
  &\text{loc. formula}: & &- \frac{1}{6}C_{IJK} \, \eta_{(0)}^{I} \wedge \widetilde{\eta}^J \wedge \dd[]{\eta^K} \label{eq:from_loc_fla}\\
  &\text{action}: & &- e^0 \wedge \qty(f^{-2} X_I \star_\gamma \dd[]{\qty(f X^I)}) + \frac{1}{6} C_{IJK} X^I e^0 \wedge \eta_{(0)}^{J} \wedge \dd[]{\qty(X^K e^0 - 2\eta^K)} \label{eq:from_action} \\
  &\text{GHY}: & &+ 2 K \, \star_h \label{eq:from_GHY} \\
  &\text{divergent c-t's}: & &+ 2 \qty(- 3 - \frac{1}{4} R_h - \frac{3}{2} \frac{(\bar{X}_I X^I - 1)}{\log{\rho_c}} + \frac{1}{8} \mathcal{A} \log{\rho_c})\star_h \label{eq:from_divergent_ct} \,, 
\end{align}
integrated over $ \frac{1}{16\pi G_5} \int_{\partial M_{\rho_c}}$.  Here the induced metric is
\begin{align}
  \dd[]{s}^2_h ={}& \frac{1}{\rho^2} \qty[g_{tt}(\rho) \dd[]{t}^2 + g_{\theta\theta}(\rho) \qty(\sigma_1^2 + \sigma_2^2) + g_{\psi\psi}(\rho) \sigma_3^2 + 2 g_{t\psi}(\rho) \dd[]{t} \sigma_3] \,,
\end{align}
and the integral is performed on a spatial slice $\partial M_{\rho_c}$ at a finite cutoff $\rho = \rho_c$. Let us describe in more detail the terms \eqref{eq:from_loc_fla}---\eqref{eq:from_divergent_ct}. \eqref{eq:from_loc_fla} is the piece arising from the localisation formula \eqref{localization_formula}, which is manifestly independent of the vector $X$ used in \cref{The localization formula} to build to equivariant cohomology. Separating $X$ out of the boundary integral in \eqref{localization_formula} was discussed in \ref{Simplification of the boundary integral} and it constitutes a major advantage compared to our previous work \cite{Colombo:2025ihp}, allowing for a clean separation between bulk and boundary analysis. The pieces entering in \eqref{eq:from_loc_fla} are now discussed. First we have the regular one-form $\eta_{(0)}^I$ in the boundary patch:
\begin{align}
  \eta_{(0)}^I = X^I e^0 - A^I_{(0)}\,, \quad e^0 = f \qty(\dd[]{y} + \omega) \,.
\end{align}
Then the (non-regular) form $\widetilde{\eta}^I$ 
can be written as
\begin{align}
  \widetilde{\eta}^I ={}& \eta^I_{(0)} - \nu^I_{0}\cdot\dd\phi= \eta^I_{(0)} - (\xi\cdot\nu^I_{0})\,\dd y\:,
\end{align}
since in our chosen coordinate system
\begin{align}
  \xi = \partial_y \,, \quad V^0 = \partial_\phi + \partial_{\widehat{\psi}} \,, \quad V^d = \partial_\phi - \partial_{\widehat{\psi}} \,, 
\end{align}
and we have used the regularity condition \eqref{gauge_regularity} with $\etacoho^I_0=0=\etacoho^I_d$.
In particular we have
\begin{align}
  \xi\cdot\nu_0^I = \iota_\xi \eta^I ={}& v \bar{X}^I - A^{I(0,0)}_t - \frac{2}{ v} A^{I(0,0)}_\psi \,, 
\end{align}
\eqref{eq:from_action} is the explicit boundary term that arose from re-writing of the on-shell action in \cref{The on-shell Lagrangian}. We construct \eqref{eq:from_loc_fla} and \eqref{eq:from_action} in the $(y, \widehat{\psi})$ coordinates and at the end we coordinate transform the resulting four-form to the $(t,\psi)$ coordinates using \eqref{eq:t_to_y}. As a sanity check we report that \eqref{eq:from_loc_fla} + \eqref{eq:from_action} is proportional to $\star_h$. \eqref{eq:from_GHY} is the Gibbons-Hawking-York term where
\begin{align}
  K ={}& - \frac{\rho}{2} h^{-1} \cdot \partial_\rho h \,, 
\end{align}
is the extrinsic curvature. \eqref{eq:from_divergent_ct} is the set of standard (divergent) counterterms required for holographic renormalization of the STU model \cite{Cassani:2019mms, Ntokos:2021duk}, where
\begin{align}
  \mathcal{A} = \text{Ric}_h \cdot \text{Ric}_h - \frac{1}{3} R_h^2 - \delta_{IJ} F^{I} \cdot F^{J} \,, 
\end{align}
is related to the conformal anomaly, which is non-zero for a squashed boundary.

Now comes the crucial point. Taken separately, all of \eqref{eq:from_loc_fla}, \eqref{eq:from_action}, \eqref{eq:from_GHY}, \eqref{eq:from_divergent_ct} are divergent, with divergences appearing as $(\rho_c^{-4}, \rho_c^{-2}, \log{\rho_c})$, and all of them depend both on the boundary data: $(A^{I(0,0)}_t, A^{I(0,0)}_\psi, X^{I(0,0)})$ and on the non-boundary data: $( g_{tt}^{(4,0)}, A^{3(2,0)} )$.
Taking everything together the divergences cancel, and, \textit{remarkably}, the non-boundary data drops out. Further, the boundary values of the scalars also drop out leaving the following boundary contribution
\begin{align}\label{eq:bdy_fla}
  \begin{aligned}
    \mathcal{B} ={}& \frac{1}{16\pi G_5} \int_{\partial M} \qty( \frac{1}{6} C_{IJK} \qty(2 \bar{X}^I + \frac{\xi\cdot\nu_0^I}{v}) A^{J(0,0)}_\psi A^{K(0,0)}_\psi + \frac{1}{32} v^2(8 - 5 v^2)) \qty(8 \, \star_{g^{(0,0)}}) \,,
  \end{aligned}
\end{align}
where $8 \, \star_{g^{0,0}} = v \sin^{}{\theta} \dd[]{t} \wedge \dd[]{\theta} \wedge \dd[]{\psi} \wedge \dd[]{\phi}$, and we have the constraint $ \bar{X}_I A^{I(0,0)}_\psi = \frac{1}{3} ( v^2 - 1)$. 
This is the main result of this section. It shows that within the background subtraction scheme (implemented through holographic renormalization) the boundary contribution $\mathcal{B}$ in our main formula for the on-shell action \eqref{OSA_localized} can be dropped out. From \eqref{OSA_localized} we have
\begin{align}
  \mathcal{I}[M] ={}& \widehat \cI [M] + \mathcal{B}[M] \,, 
\end{align}
where $M$ is the manifold whose on-shell action we want to compute. As a subtraction manifold we take a manifold $N$ with $\partial M = \partial N$, the same boundary data, and only two vectors in the fan that are the same as $V^0$ and $V^d$ of $M$.\footnote{Such manifold exists and it is smooth when the boundary is topologically $S^1\times S^3$.} Practically, such a geometry is some possibly squashed $\text{AdS}_5$, whose on-shell action is
\begin{align}
  \mathcal{I}[N] ={}& \mathcal{B}[N] \,,
\end{align}
since the bulk contribution $\widehat \cI [N]$ in \eqref{OSA_localized} vanishes for a geometry with only two vectors in the fan. Then the background subtraction scheme gives
\begin{align}
  \mathcal{I}[M] - \mathcal{I}[N] ={}& \widehat \cI [M] + \mathcal{B}[M] - \mathcal{B}[N] = \widehat \cI [M]  \,,
\end{align}
since $\mathcal{B}[M]$ and $\mathcal{B}[N]$ only depend on boundary data and cancel out, provided that we take the same boundary values for \textit{all fields} on $M$ and $N$. In the concrete example analysed above, clearly, $A^{I(0,0)}_\psi[M] = A^{I(0,0)}_\psi[N]$.\footnote{The gauge field component $A_\psi^{I(0,0)}$ at the boundary is related directly to the field strength at the boundary through 
  $F^{I(0,0)} = \dd[]{A^{I(0,0)}} = A_{\psi}^{I(0,0)} \dd[]{\sigma_3}$. 
} The only question is whether it is consistent to fix the respective gauges, such that $A^{I(0,0)}_t[M] = A^{I(0,0)}_t[N]$. Regularity in the bulk  of $M$ constrains the value of  $A^{I(0,0)}_t[M]$. For $N$, there is no regularity condition on $A^{I(0,0)}_t[N]$ since the time circle never shrinks, so there is no obstruction in choosing it such that $A^{I(0,0)}_t[M] = A^{I(0,0)}_t[N]$. Practically, this means that the subtraction manifold assumes the gauge of the manifold in question, and that
typically leads to a time-dependent spinor for the (squashed) $\text{AdS}_5$ subtraction manifold.

It would be interesting to understand better the boundary term $\cB$ without resorting to the background subtraction scheme. In appendix \ref{app:renormalization} we will discuss some preliminary results and puzzles about this. 

\section{Discussion}
\label{sect:Discussion}

In this paper we have computed the on-shell action of supersymmetric solutions with U(1)$^3$ symmetry of arbitrary gauged (and also ungauged) supergravities
with vector multiplets. We have employed methods of equivariant localization introduced in \cite{Colombo:2025ihp}, adding  the substantial new ingredient of working in patches covering the geometry, which was crucial 
to deal with Chern-Simons terms in the supergravity action. 
 Our main result is summarized in section \ref{The formula for the localized on-shell action}, and in section \ref{sect:examples}
we have discussed in detail many examples of physical interest. In addition to black holes and  black saddles  \cite{Aharony:2021zkr},  for which explicit solutions are known in the literature, 
 we have investigated configurations such as black rings, black lenses, as well as a large class of multi-horizon black holes (or rings/lenses). In the ungauged limit, these kind of solutions are known to exist and have been revisited recently 
 in \cite{Cassani:2025iix,Boruch:2025sie}, where non-extremal supersymmetric deformations that are relevant for the computation of the on-shell action \cite{Cabo-Bizet:2018ehj} are being considered. In these cases our methods reproduce the results obtained with the explicit solutions, only using information about their topology. In the gauged case, instead, the  construction of these kind of solutions, with AdS asymptotics, has remained an open problem, despite the absence of no-go theorems  ruling out their existence\footnote{Existing no-go theorems typically do not apply for non-minimal supergravity, complex Euclidean saddles or combination thereof}. 
 It is therefore astounding that our approach can be used to unlock the physics of a plethora of gravitational configurations, whose corresponding solutions are unlikely to ever be found explicitly.
We have also discussed so-called topological solitons, for which only a simple example of explicit solution is known.

For this large class of solutions we have also related our result to the equivariant volume;
it was observed in \cite{Martelli:2023oqk,Colombo:2023fhu} that many extremization problems in holography can be expressed in terms of the
equivariant volume of the internal manifold, and in this paper we have shown that a similar statement can be made for 
the equivariant volume of the non-compact asymptotically-AdS geometries that we considered.

Our computation has some restrictions  which could be relaxed in future works. First, we only consider geometries with U(1)$^3$ symmetry,
despite the fact that generically supersymmetry only ensures a U(1) symmetry. An analogue of the localization formula \eqref{localization_formula}
that just assumes U(1)$^2$ symmetry could be derived in a similar manner,
but at present it is not clear to us what would be the strategy to obtain a formula that only uses U(1) symmetry.
Furthermore, the division of the solution in patches would be much more complicated without U(1)$^3$ symmetry,
and it would be desirable to find a more elegant way to deal with this.
The alternative approach of \cite{BenettiGenolini:2025icr} can in principle cover solutions with only U(1)$^2$ symmetry,
provided that they can be described with a single global gauge patch; again, it is not clear how the symmetry assumption could be relaxed to U(1).
Another current limitation of our approach is that we do not cover solutions in the null class, that is solutions for which $\lVert\xi\rVert^2=0$,
and also solutions with $T^2\times S^2$  
boundary\footnote{Or more generally $T^2\times \Sigma_g$, that would be relevant as supergravity dual geometries to field theories in the background used 
for the four dimensional topologically twisted index.}, or solutions with orbifold singularities.
Other possible extensions of this work could feature the inclusion of hypermultiplets and/or higher-derivative corrections.

In section \ref{Boundary analysis} we have discussed the issue of holographic renormalization (making the additional simplifying assumption of cohomogeneity one of the boundary metric), combining this with the background subtraction method. 
Without emplyoing any background subtraction one has to deal with a boundary contribution that should be interpreted as a Casimir energy, that is notoriously scheme-dependent and therefore impossible to match precisely with a field theory
 computation. We also pointed out that  emplyoing the supersymmetric holographic renormalization scheme developed in \cite{BenettiGenolini:2016tsn} leads to a contribution that does not match the supersymmetric 
 Casimir
  energy \cite{Assel:2014paa}, 
due to a subtle difference for the holonomy  of the background R-symmetry gauge field along the $S^1$.  This puzzle is further discussed in appendix  \ref{app:renormalization}.

Lastly, it would be interesting to understand the relation with the equivariant volume at a more conceptual level.
Currently, expressing our result in terms of the equivariant volume requires a lot of heavy algebraic manipulations,
and we have done it only for a subclass of the possible solutions with U(1)$^3$ symmetry.

\section*{Acknowledgments}

We thank Davide Cassani, Roberto Emparan, Ohad Mamroud, Noppadol Mekareeya, Ioannis Papadimitriou, Jaeha Park, Alejandro Ruip\'erez, and Enrico Turetta for useful discussions.
We  acknowledge partial support by the INFN.  EC, VD and DM are partially  supported by a grant Trapezio (2023) of the Fondazione Compagnia di San Paolo. AZ is partially supported by the INFN, and the MIUR-PRIN grant  2022NY2MXY (finanziato dall'Unione europea - Next Generation EU, Missione 4 Componente 1 CUP H53D23001080006).

\appendix

\section{Gauge invariance}\label{Gauge invariance}

\subsection{Interface integrals for a well defined Chern-Simons action}
\label{Interface integrals for a well defined Chern-Simons action}

The Chern-Simons action is not just equal to the sum of the Chern-Simons actions in each patch.
Suppose that there exist a 6D manifold $\wb M$ such that $\partial\wb M=M$. For this discussion let us ignore the presence of the boundary $\partial M$;
if the conditions at $\partial M$ are Dirichlet, it should make $M$ effectively compact, at least at the level of the Chern-Simons action.
If the boundary conditions are not Dirichlet, there are subtleties that we can ignore if we are primarily interested in gauge transformations in the bulk.
Let us also assume that it is possible to choose patches $\wb U_{(a)}$ on $\wb M$ such that
\begin{center}
\begin{tabular}{lll}
	Patches in $\wb M$ (6D) & $\longrightarrow\qquad \wb U_{(a)}\:,$ & $\quad\partial \wb U_{(a)}=U_{(a)}-(\wb U_{(a,a+1)}-\wb U_{(a,a-1)})\:,$\\
	Interfaces in $\wb M$ (5D) & $\longrightarrow\qquad \wb U_{(a,a+1)}\:,$ & $\quad\partial\wb U_{(a,a+1)}=U_{(a,a+1)}-\wb U_{(0,\ldots,d)}\:,$\\
	Interface intersection in $\wb M$ (4D) & $\longrightarrow\qquad \wb U_{(0,\ldots,d)}\:,$ & $\quad\partial \wb U_{(0,\ldots,d)}=U_{(0,\ldots,d)}\:.$
\end{tabular}
\end{center}
\vspace{3mm}
Notice that these conventions, together with the conventions outlined at the beginning of section \ref{Patch-wise localization of the on-shell action},
are consistent with $\partial^2=0$ (if we ignore $\partial M$). Then we can define the Chern-Simons action as
\begin{equation}
	16\pi\ii G\cdot\text{(Chern-Simons)}=-\frac16C_{IJK}\int_{\wb M}\wb F^I\wedge\wb F^J\wedge\wb F^K\:,
\end{equation}
where we have extended the gauge fields $A^I_{(a)}$ in $U_{(a)}$ to gauge fields $\wb A^I_{(a)}$ in $\wb U_{(a)}$
and set $\wb F^I=\dd\wb A^I_{(a)}$. We can then express everything in terms of quantities on $M$ by repeatedly using Stokes:
\begin{equation}
\begin{aligned}
	\,&16\pi\ii G\cdot\text{(Chern-Simons)}=-\frac16C_{IJK}\sum_{a=0}^d\int_{\wb U_{(a)}}\dd\left(\wb A_{(a)}^I\wedge\wb F^J\wedge\wb F^K\right)=\\
	&\qquad=\,-\frac16C_{IJK}\sum_{a=0}^d\int_{U_{(a)}} A_{(a)}^I\wedge F^J\wedge F^K
		+\frac16C_{IJK}\sum_{a=0}^d\int_{\wb U_{(a,a+1)}}\!\!\!\!\!\!\!\!\!\!\dd\wb\Lambda^I_{a,a+1}\wedge\wb F^J\wedge\wb F^K\:,
\end{aligned}
\end{equation}
where we have defined
\begin{equation}
	\dd\wb\Lambda^I_{a,b}=\wb A_{(a)}^I-\wb A_{(b)}^I\:,\qquad\dd\Lambda^I_{a,b}=A_{(a)}^I-A_{(b)}^I\:.
\end{equation}
We can use Stokes again by noticing that
\begin{equation}
\begin{aligned}
	\,&\sum_{a=0}^d\int_{\wb U_{(a,a+1)}}\!\!\!\!\!\!\!\!\!\!\dd\wb\Lambda^I_{a,a+1}\wedge\wb F^J\wedge\wb F^K=
		\sum_{a=0}^d\int_{\wb U_{(a,a+1)}}\!\!\!\!\!\!\!\!\!\!\dd\left(\wb A^J_{(a)}\wedge\dd\wb\Lambda^I_{a,a+1}\wedge\wb F^K\right)=\\
	&\qquad=\sum_{a=0}^d\int_{U_{(a,a+1)}}\!\!\!\!\!\!\!\!\!\! A^J_{(a)}\wedge\dd\Lambda^I_{a,a+1}\wedge F^K-
		\sum_{a=0}^d\int_{\wb U_{(0,\ldots,d)}}\!\!\!\!\!\!\!\!\!\!\wb A^J_{(a)}\wedge\dd\wb\Lambda^I_{a,a+1}\wedge\wb F^K\:.
\end{aligned}
\end{equation}
Then by writing $\wb A^J_{(a)}=\wb A^J_{(0)}+\dd\wb\Lambda^J_{(a,0)}$ and using that $\sum_{a=0}^d\dd\wb\Lambda_{a,a+1}^I=0$ we have
\begin{equation}
\begin{aligned}
	\,&\sum_{a=0}^d\int_{\wb U_{(0,\ldots,d)}}\!\!\!\!\!\!\!\!\!\!\wb A^J_{(a)}\wedge\dd\wb\Lambda^I_{a,a+1}\wedge\wb F^K=
		\sum_{a=0}^d\int_{\wb U_{(0,\ldots,d)}}\!\!\!\!\!\!\!\!\!\!\wb \dd\Lambda^J_{(a,0)}\wedge\dd\wb\Lambda^I_{a,a+1}\wedge\wb F^K=\\
	&\qquad=\sum_{a=0}^d\int_{\wb U_{(0,\ldots,d)}}\!\!\!\!\!\!\!\!\!\!
		\dd\left(\wb\dd\Lambda^J_{(a,0)}\wedge\dd\wb\Lambda^I_{a,a+1}\wedge\wb A^K_{(a+1)}\right)=
		\sum_{a=0}^d\int_{U_{(0,\ldots,d)}}\!\!\!\!\!\!\!\!\!\!\dd\Lambda^J_{(a,0)}\wedge\dd\Lambda^I_{a,a+1}\wedge A^K_{(a+1)}\:.
\end{aligned}
\end{equation}
Putting everything together we have
\begin{align}\nonumber
\label{CS_interfaces_appendix}
	16\pi\ii G\cdot\text{(Chern-Simons)}=-\frac16C_{IJK}\sum_{a=0}^d\Bigg[&\int_{U_{(a)}} A_{(a)}^I\wedge F^J\wedge F^K
		+\int_{U_{(a,a+1)}}\!\!\!\!\!\!\!\!\!\!\dd\Lambda^I_{a,a+1}\wedge A^J_{(a)}\wedge F^K-\\
	&-\int_{U_{(0,\ldots,d)}}\!\!\!\!\!\!\!\!\!\!\dd\Lambda^J_{(a,a+1)}\wedge\dd\Lambda^I_{a,0}\wedge A^K_{(a+1)}\Bigg]\:.
\end{align}
The above Chern-Simons action is entirely written in terms of quantities on the 5D manifold $M$, but it also include interface integrals.

\subsection{Algebraic proof of the gauge invariace of the on-shell action}
\label{Algebraic proof of gauge invariace}

In this appendix we provide an algebraic proof of the identity \eqref{gauge_invariance_formula}, which shows the gauge invariance (modulo $2\pi\ii\,\bZ$)
of the on-shell action $\widehat\cI$, as given by the main formula \eqref{OSA_localized},
focusing on the case of solutions whose boundary geometry is homeomorphic to $S^1\times S^3$.
For the class of solutions with a trivial fundamental group  discussed in section \ref{Solutions without flat connections},
we can use the parametrization \eqref{gauge_parametrization2} for the gauges $\gauge^I_a$.
The case of solitonic solutions only requires minimal changes, as explained at the end of section \ref{Solitonic solutions}.
We will not consider the case of solutions whose fundamental group has torsion here.

Our starting point will be the parametrization \eqref{gauge_parametrization2} for the gauges $\gauge^I_a$.
Our objective is then to show that the on-shell action is independent of $\gaugeint^I_a$ (modulo $2\pi\ii\,\bZ$).
Let us begin by massaging the term
\begin{equation}
\label{torusint_def}
	\torusint\equiv C_{IJK}\sum_{a=0}^d\left(\gauge^I_0,\gauge^J_a,\gauge^K_{a+1}\right)\:,
\end{equation}
which appears in the formula \eqref{OSA_localized}, multipled by the numerical factor $\ii\pi^2/(12G)$.
By induction it can be shown to be equal to
\begin{equation}
	\torusint=C_{IJK}\underset{a\ne2}{\sum_{a=1}^d}\left(\gauge^I_0,\gauge^J_a-\gauge^J_{a-1},\gauge^K_{a+1}-\gauge^K_a\right)\:.
\end{equation}
Since the two differences $\gauge^J_a-\gauge^J_{a-1}$ and $\gauge^K_{a+1}-\gauge^K_a$ have components belonging in $\kappa\,\bZ$,
all the terms in $\gauge_0^I$ that also belong in $\kappa\,\bZ$ will contribute a total of $2\pi\ii$-integer
to the on-shell action. We can therefore make the substitution
\begin{equation}
	\left(\gauge_0^I\right)_i\quad\longrightarrow\quad(e_i,V^d,V^0)\,\etacoho^I_{\wb a}
\end{equation}
in \eqref{torusint_def}, finding
\begin{align}
\label{torusint_rewritten}\nonumber
	\torusint=\kappa^2\,C_{IJK}\,\etacoho^I_{\wb a}\,&\sum_{a=0}^d\epsilon_{ijk}(e_i,V^d,V^0)\Big((e_j,V^a,w^a)\,\flux^J_{a-1}-
		(e_j,V^{a-1},w^a)\,\flux^J_a+(e_j,V^{a-1},V^a)\,\gaugeint^J_a\Big)\cdot\\
	&\cdot\Big((e_k,V^{a+1},w^{a+1})\,\flux^K_a-
		(e_k,V^{a},w^{a+1})\,\flux^J_{a+1}+(e_k,V^a,V^{a+1})\,\gaugeint^K_{a+1}\Big)\:,
\end{align}
where we have also substituted the expression \eqref{gauge_parametrization2} for $\gauge^J_a$ and $\gauge^K_{a+1}$,
and used the fact that $\epsilon_{ijk}(e_i,V^d,V^0)(e_j,V^d,V^0)=0$.

Let us focus on the terms that are quadratic in $\gaugeint^I_a$ first. From \eqref{torusint_rewritten} we find the quadratic terms
\begin{equation}
\label{torusint_quadratic}
	\torusint=\kappa^2\,C_{IJK}\,\etacoho^I_{\wb a}\sum_{a=1}^d\gaugeint^J_a\,\gaugeint^K_{a+1}\,
		(V^a,V^d,V^0)(V^{a-1},V^a,V^{a+1})+\cO(\gaugeint)\:,
\end{equation}
where we have used the relation
\begin{equation}
\label{det_of_det}
	\epsilon_{ijk}(e_i,v^1,v^2)(e_j,v^3,v^4)(e_k,v^5,v^6)=(v^1,v^3,v^4)(v^2,v^5,v^6)-(v^2,v^3,v^4)(v^1,v^5,v^6)\:.
\end{equation}
In the on-shell\ action \eqref{OSA_localized} the other terms quadratic in the $\gaugeint^I_a$ are the one coming from the
$(\gauge^I_{a}\cdot\xi)(\gauge^J_{a+1}\cdot\xi)$ terms,
which using \eqref{gauge_parametrization2} are (ignoring the overall numerical factor $\ii\pi^2/(12G)$)
\begin{equation}
\begin{aligned}
	\,&-C_{IJK}\sum_{a=0}^d\frac{(\gauge^I_{a}\cdot\xi)(\gauge^J_{a+1}\cdot\xi)\:\etacoho^K_a\:(V^{a-1},V^a,V^{a+1})}
		{(\xi,V^{a-1},V^a)(\xi,V^a,V^{a+1})}=\\
	&\qquad=-\kappa^2\,C_{IJK}\,\sum_{a=1}^d\etacoho^I_{a}\,\gaugeint^J_a\,\gaugeint^K_{a+1}\,
		(V^{a-1},V^a,V^{a+1})+\cO(\gaugeint)\:.
\end{aligned}
\end{equation}
Because of \eqref{special_quantization}, the above term can be combined with \eqref{torusint_quadratic} to form a $\kappa^3\,C_{IJK}\,$(integers) term, 
which only contributes to the on-shell action as a $2\pi\ii\,$(integer), once the overall numerical factor of $\ii\pi^2/(12G)$ is accounted,
provided that $\kappa$ is fixed as in \eqref{kappa_fixing}.

We have thus shown that terms that are quadratic in $\gaugeint^I_a$ cancel from the on-shell action, modulo $2\pi\ii\,\bZ$; in order to complete the
proof of gauge invariance of the on-shell action we need to do the same for the linear terms.
Using \eqref{det_of_det}, we find that the terms in $\torusint$ that are linear in $\gaugeint^I_a$ can be written as
\begin{equation}
\begin{aligned}
	\torusint\big|_{\text{linear in }\gaugeint^I_a}=&\,\kappa^2\,C_{IJK}\,\etacoho^I_{\wb a}\sum_{a=0}^d\bigg\{
		\gaugeint^K_{a+1}\bigg[\flux_{a-1}^J(V^a,V^d,V^0)(V^{a+1},V^a,w^a)-\\
	&-\flux_a^J\Big((V^a,V^d,V^0)(V^{a+1},V^{a-1},w^a)+(V^{a+1},V^d,V^0)\Big)\bigg]+\\
	&+\gaugeint^J_a\bigg[\flux^K_a\Big((V^a,V^d,V^0)(V^{a-1},V^{a+1},w^{a+1})-(V^{a-1},V^d,V^0)\Big)-\\
	&-\flux^K_{a+1}(V^a,V^d,V^0)(V^{a-1},V^a,w^{a+1})\bigg]\bigg\}\:,
\end{aligned}
\end{equation}
and after imposing the quantization condition \eqref{special_quantization} it simplifies down to
\begin{align}\nonumber
\label{torusint_linear}
	\torusint\big|_{\text{linear in }\gaugeint^I_a}=&\,\kappa^2\,C_{IJK}\sum_{a=0}^d\bigg\{
		\gaugeint^K_{a+1}\bigg[\flux_{a-1}^J\,\etacoho^I_{a}(V^{a+1},V^a,w^a)
		-\flux_a^J\Big(\etacoho^I_{a}(V^{a+1},V^{a-1},w^a)+\etacoho^I_{a+1}\Big)\bigg]+\\\nonumber
	&+\gaugeint^J_a\bigg[\flux^K_a\Big(\etacoho^I_{a}(V^{a-1},V^{a+1},w^{a+1})-\etacoho^I_{a-1}\Big)
		-\flux^K_{a+1}\,\etacoho^I_{a}(V^{a-1},V^a,w^{a+1})\bigg]\bigg\}+\\[2mm]
	&+\kappa^3\,C_{IJK}\,\big(\text{integers}\big)\:.
\end{align}
In order to simplify the terms coming from the rest of the on-shell action we need the relation
\begin{equation}
\label{det_prod}
	(v^1,v^2,v^3)(v^4,v^5,v^6)=(v^2,v^3,v^4)(v^1,v^5,v^6)-(v^1,v^3,v^4)(v^2,v^5,v^6)+(v^1,v^2,v^4)(v^3,v^5,v^6)\:,
\end{equation}
which can be found from \eqref{det_of_det} by taking its left hand side, permuting $i,j$ before applying the same identity \eqref{det_of_det},
and then comparing with the right hand side of \eqref{det_of_det}.
The terms
\begin{align}
\label{cc_linear}\nonumber
	\,&-C_{IJK}\sum_{a=0}^d\frac{(\gauge^I_{a}\cdot\xi)(\gauge^J_{a+1}\cdot\xi)\:\etacoho^K_a\:(V^{a-1},V^a,V^{a+1})}
		{(\xi,V^{a-1},V^a)(\xi,V^a,V^{a+1})}\bigg|_{\text{linear in }\gaugeint^I_a}=\\
	&\qquad=\kappa\,C_{IJK}\sum_{a=0}^d\frac{\etacoho^K_a\:(V^{a-1},V^a,V^{a+1})}{(\xi,V^{a-1},V^a)(\xi,V^a,V^{a+1})}\cdot\\\nonumber
	&\qquad\qquad\cdot\bigg[\gaugeint_a^I(\xi,V^{a-1},V^a)\Big((\xi,V^d,V^0)\,\etacoho^J_{\wb a}-\kappa\,\flux^J_{a}(\xi,V^{a+1},w^{a+1})+
		\kappa\,\flux^J_{a+1}(\xi,V^{a},w^{a+1})\Big)+\\\nonumber
	&\qquad\qquad+\gaugeint_{a+1}^I(\xi,V^a,V^{a+1})\Big((\xi,V^d,V^0)\,\etacoho^J_{\wb a}-\kappa\,\flux^J_{a-1}(\xi,V^a,w^a)+
		\kappa\,\flux^J_a(\xi,V^{a-1},w^a)\Big)\bigg]
\end{align}
can be simplified by using \eqref{det_prod} to write
\begin{equation}
\begin{aligned}
\label{gauge_invariance_det1}
	\,&(V^{a-1},V^a,V^{a+1})(\xi,V^{a},w^{a})=(\xi,V^{a},V^{a+1})-(V^{a},V^{a+1},w^{a})(\xi,V^{a-1},V^{a})\:,\\
	\,&(V^{a-1},V^a,V^{a+1})(\xi,V^{a-1},w^{a})=(\xi,V^{a-1},V^{a+1})-(V^{a-1},V^{a+1},w^{a})(\xi,V^{a-1},V^{a})\:,\\
	\,&(V^{a-1},V^a,V^{a+1})(\xi,V^{a+1},w^{a+1})=-(\xi,V^{a-1},V^{a+1})+(V^{a-1},V^{a+1},w^{a+1})(\xi,V^{a},V^{a+1})\:,\\
	\,&(V^{a-1},V^a,V^{a+1})(\xi,V^{a},w^{a+1})=-(\xi,V^{a-1},V^{a})+(V^{a-1},V^{a},w^{a+1})(\xi,V^{a},V^{a+1})\:,
\end{aligned}
\end{equation}
and
\begin{align}\nonumber
\label{gauge_invariance_det2}
	(V^{a-1},V^a,V^{a+1})(\xi,V^d,V^0)=\,&(V^{a-1},V^d,V^0)(\xi,V^a,V^{a+1})-(V^{a},V^d,V^0)(\xi,V^{a-1},V^{a+1})+\\
	&+(V^{a+1},V^d,V^0)(\xi,V^{a-1},V^{a})\:,
\end{align}
which can be combined with \eqref{special_quantization} to give
\begin{align}\nonumber
\label{gauge_invariance_wba}
	(V^{a-1},V^a,V^{a+1})(\xi,V^d,V^0)\,\etacoho^J_{\wb a}=\,&\left(\etacoho^J_{a-1}+\kappa\,\flux^J_{a-1}\right)(\xi,V^a,V^{a+1})
		-\left(\etacoho^J_a+\kappa\,\flux^J_a\right)(\xi,V^{a-1},V^{a+1})+\\
	&+\left(\etacoho^J_{a+1}+\kappa\,\flux^J_{a+1}\right)(\xi,V^{a-1},V^{a})\:.
\end{align}
Let us proceed in steps: first we combine \eqref{torusint_linear} with \eqref{cc_linear} and substitute \eqref{gauge_invariance_det1}, obtaining
\begin{align}\nonumber
\label{cc_linear2}
	\,&\eqref{torusint_linear}+\eqref{cc_linear}\big|_{\eqref{gauge_invariance_det1}}=
		\kappa\,C_{IJK}\sum_{a=0}^d\Bigg[-\gaugeint^K_{a+1}\,\flux^J_a\,\etacoho^I_{a+1}-\gaugeint^J_a\,\flux^K_a\,\etacoho^I_{a-1}+\\\nonumber
	&\quad+\frac{\etacoho^K_a\gaugeint_a^I\Big((V^{a-1},V^a,V^{a+1})(\xi,V^d,V^0)\,\etacoho^J_{\wb a}+\kappa\,\flux^J_{a}(\xi,V^{a-1},V^{a+1})-
		\kappa\,\flux^J_{a+1}(\xi,V^{a-1},V^{a})\Big)}{(\xi,V^a,V^{a+1})}+\\\nonumber
	&\quad+\frac{\etacoho^K_a\gaugeint_{a+1}^I\Big((V^{a-1},V^a,V^{a+1})(\xi,V^d,V^0)\,\etacoho^J_{\wb a}
		-\kappa\,\flux^J_{a-1}(\xi,V^{a},V^{a+1})+\kappa\,\flux^J_a(\xi,V^{a-1},V^{a+1})\Big)}{(\xi,V^{a-1},V^a)}\Bigg]+\\[2mm]
	&\quad+\kappa^3\,C_{IJK}\,\big(\text{integers}\big)\:,
\end{align}
then we substitute \eqref{gauge_invariance_wba} as well
\begin{align}\nonumber
\label{cc_linear3}
	\eqref{cc_linear2}\big|_{\eqref{gauge_invariance_wba}}=
		\kappa\,C_{IJK}\sum_{a=0}^d&\,\Big(\etacoho^J_{a-1}\,(\xi,V^a,V^{a+1})
		-\etacoho^J_a\,(\xi,V^{a-1},V^{a+1})+\etacoho^J_{a+1}\,(\xi,V^{a-1},V^{a})\Big)\cdot\\
	&\cdot\Bigg[\frac{\etacoho^K_a\,\gaugeint_a^I}{(\xi,V^a,V^{a+1})}+\frac{\etacoho^K_a\,\gaugeint_{a+1}^I}{(\xi,V^{a-1},V^a)}\Bigg]
		+\kappa^3\,C_{IJK}\,\big(\text{integers}\big)\:.
\end{align}
Notice that in \eqref{cc_linear2} all the terms involving $w^a$ canceled, while in \eqref{cc_linear3} all the terms involving $\flux^J_a$ canceled;
in the latter case we had to invoke the cyclicity of the sum over $a$ in order to cancel some terms.
Considering that
\begin{equation}
	\xi\cdot\big(\gauge_a^I+\gauge_{a+1}^I\big)\big|_{\text{linear in }\gaugeint^I_a}=-\kappa\,\gaugeint^I_a\,(\xi,V^{a-1},V^{a})
		-\kappa\,\gaugeint^I_{a+1}\,(\xi,V^{a},V^{a+1})\:,
\end{equation}
it is easy to see that the terms that are linear in $\gaugeint^I_a$ from the $\xi\cdot\big(\gauge_a^I+\gauge_{a+1}^I\big)$ term in the on-shell action \eqref{OSA_localized}
cancel perfectly with \eqref{cc_linear3}, and there are no other terms linear in $\gaugeint^I_a$ to account for.

In conclusion we have shown that
\begin{equation}
	\widehat\cI\:=\:\widehat\cI\big|_{\gaugeint^I_a=0}\,+\,\frac{\ii\pi^2\kappa^3}{12G}\,C_{IJK}\,\big(\text{integers}\big)\:,
\end{equation}
and thus the on-shell action is invariant under gauge transformations up to $2\pi\ii\,\bZ$ terms, as long as $\kappa$ is chosen as in \eqref{kappa_fixing}.

\section{Comments on the supersymmetric Casimir energy}\label{app:renormalization}
In this appendix we ask whether the boundary contribution can acquire physical meaning on its own without resorting to the background subtraction scheme. Reference \cite{BenettiGenolini:2016tsn} studies this problem (in the case of minimal supergravity) in great detail\footnote{The geometries considered in \cite{BenettiGenolini:2016tsn} manifestly have no contributions from the interior and the main quantity calculated there $S_{\text{susy}}$ is in fact equivalent to our $\mathcal{B}$.}. 
The main conclusion of that paper is that it is not possible to add finite counterterms which are simultaneously
\begin{enumerate}[(i)]
  \item diffeomorphism invariant, 
  \item gauge invariant,
  \item supersymmetry invariant,
\end{enumerate}
such that the supersymmetric Ward identities are satisfied. Instead they look for finite counterterms that ensure the validity of the supersymmetric Ward identities at the expense of (i) and (ii). To that end they introduce the following coordinate system and R-symmetry gauge field at the boundary\footnote{The overall prefactor of $A^{R(0,0)}$ is to ensure matching with our normalization of the gauge field.} 
\begin{align}\label{eq:their_boundary_metric}
  \begin{aligned}
  \dd[]{s}^2_{g^{(0,0)}} ={}& - \dd[]{t}^2 + \qty(\dd[]{\psi} + a)^2 + 4 \eu^w \dd[]{z} \dd[]{\bar{z}} \,, \quad a = a_z \dd[]{z} + \overline{a_z} \dd[]{\bar{z}} \, , \\
  A^{R(0,0)} ={}& - \frac{2}{3}\qty(\frac{u}{4} \qty(\dd[]{\psi} + a) + \frac{\iu}{4} \qty(\dd[]{\bar{z}} \partial_{\bar{z}} w - \dd[]{z} \partial_z w) + \gamma \dd[]{\psi} + \dd[]{\lambda} - \frac{u}{8} \dd[]{t} + \gamma' \dd[]{t}) \,, 
  \end{aligned}
\end{align}
where $a_z = a_z(z, \bar{z}), \, \overline{a_z} = \overline{a_z}(z, \bar{z}), \, w = w(z, \bar{z}), \, u = u(z, \bar{z}), \, \lambda = \lambda(z, \bar{z})$ with the flux constraint
\begin{align}
  \dd[]{a} ={}& \iu \, u \, \eu^w \dd[]{z} \wedge \dd[]{\bar{z}} \,. 
\end{align}
Then they show that $\mathcal{B}' \equiv \mathcal{B} + \mathcal{B}_{\text{new}}$ satisfies the supersymmetric Ward identities where  
\begin{align}
  \begin{aligned}
  \mathcal{B}_{\text{new}} ={}& \frac{1}{8\pi G_5} \int_{\partial M} \qty(A^{R(0,0)} \wedge \Phi + \Psi) \,, \\
  \Phi ={}& - \frac{1}{2^4 3^2} \qty(u^3 - 4 u R_{2d}) \iu \eu^w \dd[]{z} \wedge \dd[]{\bar{z}} \wedge (2 \dd[]{\psi} - \dd[]{t}) \,, \\
  \Psi ={}& \frac{1}{2^{11} 3^2} (19 u^4 - 48 u^2 R_{2d}) \, \star_{g^{(0,0)}} \,,
  \end{aligned}
\end{align}
and $R_{2d} = - \eu^{-w} \partial_z \partial_{\bar{z}} w$ is the Ricci scalar of $4 \eu^w \dd[]{z} \dd[]{\bar{z}}$. The configuration \eqref{eq:their_boundary_metric} admits more general squashing compared what we analysed in \cref{Asymptotic perturbative solutions}. In particular, it allows for a boundary with $\mathbb{R} \times \text{U}\qty(1)^2$ isometry. It also captures our $\mathbb{R} \times \text{SU}\qty(2) \times \text{U}\qty(1)$ boundary \eqref{eq:boundary_metric} as a special case. To see this one can perform the coordinate transformation
\begin{align}
  t \rightarrow {}& - t \,, \quad \psi \rightarrow \frac{v}{2} \psi \,, \quad z \rightarrow \eu^{-\iu \phi} \cot{\frac{\theta}{2}} \,, \quad \bar{z} \rightarrow \eu^{\iu \phi} \cot{\frac{\theta}{2}} \,, 
\end{align}
and identify
\begin{align}
  \begin{aligned}
  u ={}& - 4 v \,, \\
  a_z ={}& \frac{\iu v (z \bar{z} - 1)}{4 z (1 + z \bar{z})} = \frac{\iu}{4} \eu^{\iu \phi} v \cos^{}{\theta} \cot{\frac{\theta}{2}} \,, \\
  w ={}& \log{\qty(\frac{1}{4(1 + z \bar{z})^2})} = \log{\qty(\frac{1}{4} \sin^{4}{\frac{\theta}{2}})} \,. \\
  \end{aligned}
\end{align}
Upon this the boundary metric in \eqref{eq:their_boundary_metric} is mapped to ours \eqref{eq:boundary_metric}. Further, the R-symmetry gauge field in \eqref{eq:their_boundary_metric} is mapped to ours
\begin{align}
  A^{R(0,0)} ={}& \bar{X}_I \qty(A^{I(0,0)}_t \dd[]{t} + A^{I(0,0)}_\psi \sigma_3) \,,  
\end{align}
provided that
\begin{align}
  \bar{X}_I A^{I(0,0)}_\psi ={}& \frac{1}{3}(v^2 - 1) \,, \quad \bar{X}_I A^{I(0,0)}_t = \frac{1}{3} (v + 2 \gamma') \,, \quad \gamma = \frac{1}{v} \,, \quad \lambda = - \frac{\phi}{2} \,.
\end{align}
Finally, for our simple squashing we have
\begin{align}
  R_{2d} ={}& 8 \,. 
\end{align}
Then we obtain $\mathcal{B}_{\text{new}}$ as
\begin{align}\label{eq:B_new}
  \mathcal{B}_{\text{new}} ={}& \frac{1}{16\pi G_5} \int \qty( - \frac{2}{27} + \frac{7}{36} v^2 - \frac{89}{864} v^4 + \frac{1}{9} v( v^2 - 2) \bar{X}_I A^{I(0,0)}_t) \qty(8 \, \star_{g^{(0,0)}}) \,. 
\end{align}
Assuming that there is no modification to $\mathcal{B}_{\text{new}}$ for the STU model compared to minimal supergravity, except identifying $A^{R(0,0)} \equiv \bar{X}_I A^{I(0,0)}$, we can now write the full answer $\mathcal{B}' = \mathcal{B} + \mathcal{B}_{\text{new}}$. Using the relations
\begin{align}
  \xi\cdot\nu_{0}^I ={}& v \bar{X}^I - A^{I(0,0)}_t - \frac{2}{ v} A^{I(0,0)}_\psi \,, \quad \bar{X}_I A^{I(0,0)}_\psi = \frac{1}{3}(v^2 - 1) \,, 
\end{align}
there are many equivalent ways of writing the final answer. We find the following one convenient
\begin{align}\label{eq:Bprime_final}
  \mathcal{B}' ={}& \frac{1}{16\pi G_5} \int \qty[\frac{\xi\cdot\nu_{0}^I}{9 v}\,\bar{X}_I + \frac{1}{6} \qty(C_{IJK} - \bar{X}_I \bar{X}_J \bar{X}_K) A_\psi^{I(0,0)} A^{J(0,0)}_{\psi} \qty(2 \bar{X}^K + \frac{\xi\cdot\nu_{0}^K}{v})] \qty(8 \, \star_{g^{(0,0)}}) \,. 
\end{align}
The special case of minimal supergravity is reached as
\begin{align}
  A^{I(0,0)}_\psi = \frac{1}{3} \qty( v^2 - 1) \,, \quad A^{I(0,0)}_t = \frac{1}{3} \qty( v + 2 \gamma') \,, \quad \xi\cdot\nu_{0}^I = \frac{2}{3} \qty(\frac{1}{ v} - \gamma') \,, \quad \forall I \,. 
\end{align}
Upon this the second term in the round bracket in \eqref{eq:Bprime_final} vanishes and we can bootstrap our special case of $\mathbb{R} \times \text{SU}\qty(2) \times \text{U}\qty(1)$ squashing to the more general case of $\mathbb{R} \times \text{U}\qty(1)^2$ squashing as
\begin{align}
  8 \rightarrow {}& R_{2d} \,, \quad \frac{1}{v} \rightarrow \gamma \,,
\end{align}
and obtain agreement with (4.49) of \cite{BenettiGenolini:2016tsn}
\begin{align}
  \mathcal{B}'_{\text{minimal}} ={}& \frac{\gamma(\gamma - \gamma')}{27 \qty(8\pi G_5)} \int \dd[]{t} \int R_{2d} \, \text{vol}_3 \,, 
\end{align}
For the case of a Hopf sufrace $\mathbb{R} \times S^3_{b_1, b_2}$ discussed in section 5.4 of \cite{BenettiGenolini:2016tsn} one obtains
\begin{align}
  \int R_{2d} \, \text{vol}_3 ={}& 8\pi^2 \frac{b_1 + b_2}{b_1 b_2} \,, \quad \gamma = \frac{1}{2} (b_1 + b_2) \,, 
\end{align}
and then, upon Wick rotation $t \rightarrow - \iu \tau$ to a periodic time coordinate $\tau \sim \tau + \beta$, the integral evaluates to
\begin{align}\label{eq:final_res_susy_casimir}
  \mathcal{B}'_{\text{minimal}} ={}& \frac{2\pi^2 \beta}{27 (8\pi G_5)}  \frac{(b_1 + b_2)^3}{b_1 b_2} - \frac{4\pi^2 \beta}{27(8\pi G_5)} \frac{(b_1 + b_2)^2}{b_1 b_2} \gamma' \,. 
\end{align}
When the solution in question is an $\mathbb{R} \times \text{U}\qty(1)^2$-squashed $\text{AdS}_5$, one can consistently choose a gauge where $\gamma' = 0$\footnote{This ensures that the spinor is time-independent.} and recover the supersymmetric Casimir energy \cite{Assel:2015nca, Bobev:2015kza, Martelli:2015kuk}. However, when the solution in questions is an $\mathbb{R} \times \text{U}\qty(1)^2$-squashed black hole, or a more general geometry, $\gamma'$ is fixed by regularity of $A^{R}$ in the bulk.
In the latter case, $\gamma'$ is generically non-zero.
Namely, it is such that in Euclidean signature the spinor takes the form $\varepsilon \propto \eu^{\pi \iu \phi_0}$, where $\phi_0$ is an honest $2\pi$ periodic coordinate, whose periodicity is decoupled from the periodicities of the remaining $\text{U}\qty(1)$'s. Translated back to the Lorentzian $\qty(t,\phi, \psi)$ coordinates used here one obtains $\gamma' \neq 0$. In the classic papers computing the supersymmetric Casimir energy in field theory, eg. starting from \cite{Assel:2014paa} and leading to \cite{Assel:2015nca}, effectively the gauge choice $\gamma' = 0$ is being made even if not explicitly stated.
It would be interesting to revisit these calculations in a more general gauge (compatible with bulk regularity conditions) and test if one obtains agreement with \eqref{eq:final_res_susy_casimir}.

Another open problem that we leave for future work is to bootstrap the second term in the round bracket of \eqref{eq:Bprime_final} to the more general $\mathbb{R} \times \text{U}\qty(1)^2$ squashing and re-write it in the form (4.19) of \cite{Bobev:2015kza} as predicted by field theory\footnote{Note that it is likely that the two open problems are correlated, and that our assumption just below \eqref{eq:B_new} is open for scrutiny.}.
For the time being we simply note that the background subtraction scheme, implemented in this paper through an explicit holographic renormalization calculation, makes 
the subtracted on-shell action manifestly independent of any mysterious boundary contribution 
and is in good thermodynamic agreement with known 
geometries in AdS and flat space.

\newpage

\bibliographystyle{ytphys}
\bibliography{equivodd}

\end{document}